\documentclass[preprint2]{proto} 
\usepackage{times}
\usepackage{xcolor}
\usepackage{graphicx}
\usepackage[nointegrals]{wasysym}
\usepackage{float}
\usepackage{dblfloatfix}
\usepackage{widetext}
\usepackage{amsmath}		
\usepackage{soul}

\usepackage{microtype}

\usepackage[utf8]{inputenc}
\usepackage{hyperref}
\usepackage{aas_macros}
\usepackage{chemformula}    

\graphicspath{{./}{Figures/}}

\usepackage{todonotes}

\begin{document}

\title{\textbf{\LARGE Solar wind interaction with a comet: evolution, variability, and implication}}

\author {\textbf{\large C. Götz}}
\affil{\small\em European Space Agency, ESTEC, The Netherlands}

\author {\textbf{\large J. Deca}}
\affil{\small\em Laboratory for Atmospheric and Space Physics (LASP), University of Colorado Boulder, Boulder, Colorado 80303, USA.}

\author {\textbf{\large K. Mandt}}
\affil{\small\em Johns Hopkins Applied Physics Laboratory, Laurel, MD, USA}

\author {\textbf{\large M. Volwerk}}
\affil{\small\em Space Research Institute, Austrian Academy of Sciences, Graz, Austria}

\begin{abstract}

\begin{list}{ } {\rightmargin 1in}
\baselineskip = 11pt
\parindent=1pc
{\small 
Once a cometary plasma cloud has been created through ionisation of the cometary neutrals, it presents an obstacle to the solar wind and the magnetic field within it. The acceleration and incorporation of the cometary plasma by the solar wind is a complex process that shapes the cometary plasma environment and is responsible for the creation of boundaries such as a bow shock and diamagnetic cavity boundary. It also gives rise to waves and electric fields which in turn contribute to the acceleration of the plasma. This chapter aims to provide an overview of how the solar wind is modified by the presence of the cometary plasma, and how the cometary plasma is incorporated into the solar wind. We will also discuss models and techniques widely used in the investigation of the plasma environment in the context of recent findings by \emph{Rosetta}. In particular, this chapter highlights the richness of the processes and regions within this environment and how processes on small scales can shape boundaries on large scales. It has been fifteen years since the last book on Comets was published and since then we have made great advances in the field of cometary research. But many open questions remain which are listed and discussed with particular emphasis on how to advance the field of cometary plasma science through future space missions.\\~\\~\\~}
\end{list}
\end{abstract}  



{\bf Keywords}:
\begin{enumerate}
    \item[5] Magnetic Field; Draping; Magnetotail; Magnetic Reconnection; Plasma Boundaries; 
\end{enumerate}

\section{Introduction}
In the preceding chapter, the reader is introduced to the sources and losses of the plasma at a comet. In the following, we will describe what happens to the cometary plasma cloud as it interacts with the solar wind.

Several works have summarized and reviewed the extensive work done in the field of cometary plasma environments since the 1980s with the first artificial comet experiments and the 1P/Halley flyby. The reader is for example referred to summaries by \cite{Gombosi2015} for an overview or \cite{Szego2000} for the details of the physics in these plasmas.
Most new discoveries since the last book \citep[Comets II,][]{Combi2004} was written, are due to the European Space Agency's \emph{Rosetta} mission, we therefore highlight some peculiarities of this mission here. 
The spacecraft arrived at comet 67P/Churyumov-Gerasimenko (67P) in August 2014 and explored the plasma environment for over two years until end of mission at the end of September 2016. This is in stark contrast to all previous missions to comets that were equipped with plasma instruments, which were all single flybys with different closest approach distances (for more details see Chapter 6). The particular advantage of the long lasting rendezvous is that the observations cover a large range of activity levels. This allows us to study the cometary environment and its interaction with the solar wind at different stages and observe how the increase and decrease in gas production rate and consequently the cometary ion density affects the solar wind. Rosetta also covered much lower activity levels than other missions to comets and therefore expands our knowledge significantly. On the other hand, this means that comparisons with previous results always need to take into account the different situation and are usually non-trivial.

\emph{Rosetta's} instruments were able to measure the magnetic field vector, the plasma density, the electron temperature, the ion energy distribution with mass resolution and the electron energy distribution. 
For a few days in November 2014, data by the lander Philae were available for two-point measurements of the magnetic field. 
For more details, the reader is referred to the documentation for the Rosetta Plasma Consortium (RPC)\footnote{\url{https://www.cosmos.esa.int/web/psa/rosetta}}. There, one can also find an extensive list of available data products and the limits and error sources associated with them.

\begin{figure}[t]
    \centering
    \includegraphics[width=\columnwidth]{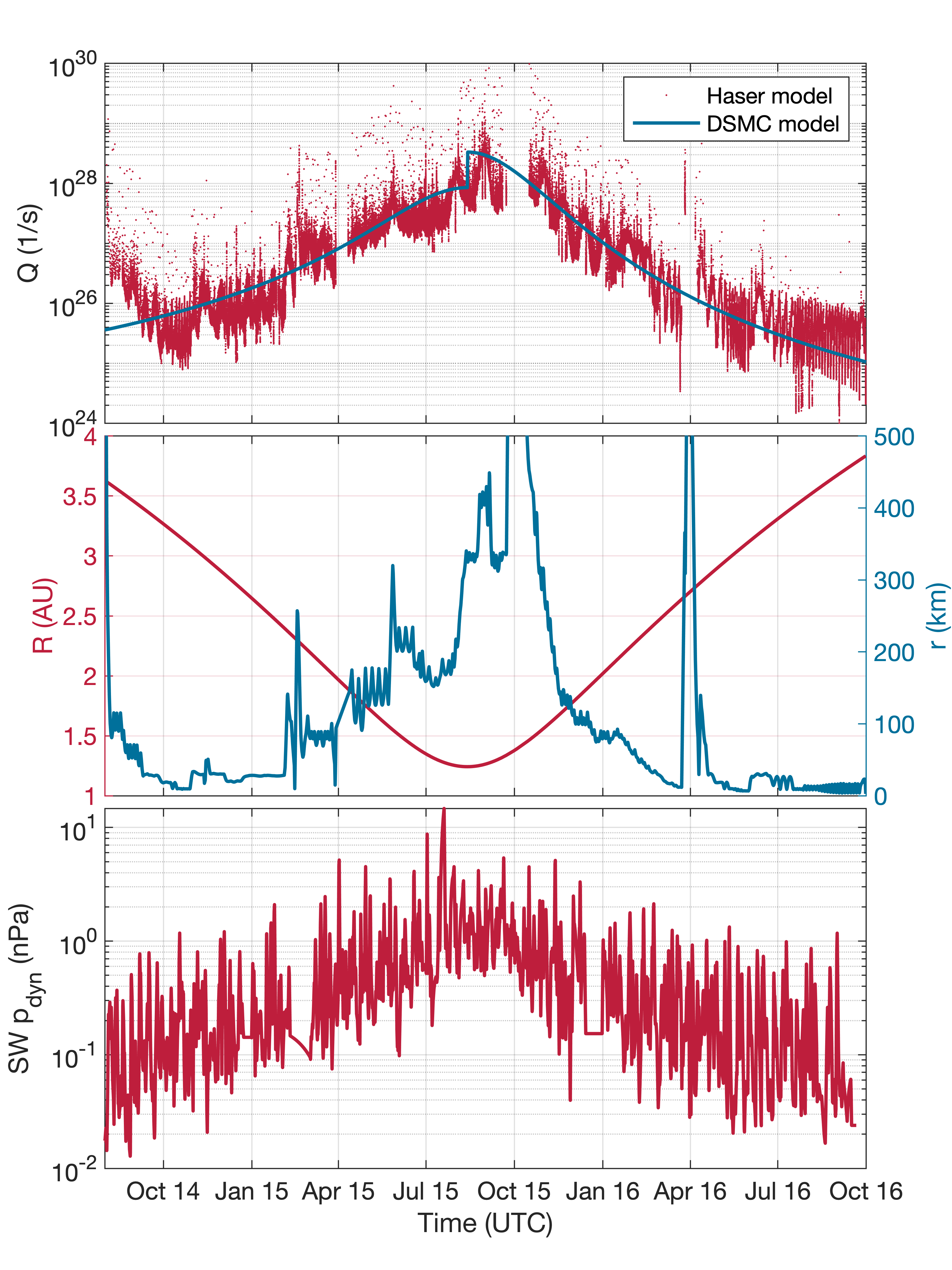}
    \caption{Top: Gas production rate of comet 67P derived from in-situ measurments using the Haser model (red) and averaged gas production rate from a DSMC model (blue). Middle: heliocentric distance of comet 67P (red) and cometocentric distance of \emph{Rosetta} at 67P (blue). The dayside excursion in October 2015 and the nightside excursion in April 2015 went out to $1500\,$km and $1000\,$km respectively. Bottom: solar wind dynamic pressure at 67P derived from Earth based data and a Tao model \citep{Tao2005}.} 
    \label{fig:Q_sw}
\end{figure}

As already described in Chapter 16, the primary source of the plasma at the comet is the neutral gas that is produced by the sublimation of the ices on the comet's surface. Processes like photoionisation, charge exchange, and electron impact ionisation produce ions and electrons from those neutrals. The resulting ion coma can extend millions of kilometres from the nucleus, interacting with the plasma and magnetic field of the solar wind in which it mostly resides. This forms the plasma environment of the comet.

As comet 67P journeys through the solar system, the heliocentric distance decreases and increases and therefore the gas production rate increases and decreases as well. This is shown in the top and middle panels of Fig. \ref{fig:Q_sw}. At the same time, the solar wind parameters change as well. The bottom panel of Fig. \ref{fig:Q_sw} shows the estimated dynamic pressure of the solar wind at 67P from a model to illustrate this. The closer the comet is to the Sun the higher the solar wind density (and therefore the dynamic pressure) and magnetic field strength. On shorter timescales such as days and hours, solar wind transients like Corotating Interaction Regions (CIRs) and Interplanetary Coronal Mass Ejections (ICMEs) can also cause the solar wind parameters to change significantly. Unfortunately, no solar wind monitor was included in the \emph{Rosetta} mission and models need to suffice to infer the solar wind parameters at the comet. They usually utilise some sort of propagation model and either observations of the Sun or in-situ measurements at other solar system locations to estimate the solar wind conditions with spatial and temporal uncertainties. This is one of the major drawbacks of a single-spacecraft mission and should be taken into consideration for future missions \citep{Goetz2019wp}.

Therefore, depending on the gas production rate and solar wind conditions, the comet influenced plasma environment may extend from as little as $100\,$km up to millions of km from the nucleus. As a consequence, the collisionality of the plasma, the gyroradii of the particles within it and the presence of different electron and ion populations also vary significantly and systematically over time.

In the following, the reader is given an overview of the solar wind -- comet interaction, starting with a description of the induced cometosphere, the magnetic field structure near the comet and in the comet plasma tail, followed by a list of the boundaries that form. At the end, various wave phenomena are discussed. 

\section{Induced Cometosphere}
\label{sec:cometosphere}
\subsection{Introduction}
Whereas the cometary environment might appear quite uneventful near aphelion, as a comet travels closer to the Sun, the increasing availability of neutral gas leads to the formation of a coma, and dust and plasma tails. Observations of the latter phenomena revealed the existence of the solar wind and the interplanetary magnetic field \citep{Biermann1967}. Important is the realisation that the presence of neutral gas in the solar wind leads to mass-loaded plasmas of various natures \citep{Szego2000,Gombosi2015}. One such example is an induced cometosphere, a term analogous to a magnetosphere at planets, that forms when the magnetized solar wind plasma interacts with the cometary plasma environment \citep{Goetz2021b}.

Mass-loading describes the process whereby slow, heavy ions like water, carbondioxide or oxygen are accelerated and incorporated into the fast solar wind flow. It is therefore an energy and momentum exchange process between different plasma populations \citep{Galeev1985,Huddleston1993}.
What distinguishes the cometary neutral environment from that of e.g. Venus or Mars, is the lack of significant gravitational acceleration of the central body: while Venus and Mars have an atmosphere that is for the most part gravitationally bound to the planet, the neutral exosphere of a comet is radially expanding into space without much gravitational influence. As a consequence, neutral gas can be ionised anywhere in the environment, even millions of km upstream of the source body. This then gives a size of the cometosphere which is dependent on the outgassing and ionisation conditions. For example, the first signatures of a cometosphere at comet 1P/Halley at about 1 AU, which had a gas production rate of $Q = 10^{30}\,$s$^{-1}$ and a water photoionisation rate of $\nu_{H_2O} = 10^{-6}\,$s$^{-1}$, were found at $2\times10^{6}\,$km upstream of the nucleus \citep{Neubauer1986}. On the other end of the scale, we find that the cometosphere at comet 67P at 3.4 AU only expands about $10^2\,$km at a gas production rate of $4\times10^{25}\,$s$^{-1}$ and a water photoionisation rate of $0.5\times10^{-7}\,$s$^{-1}$ \citep{Richter2015}. This impressive difference in scale of four orders of magnitude demonstrates the richness of the cometary plasma environment. Many of the processes and structures in the cometary plasma environment depend on the scale size, and therefore the outgassing rate of the comet, which is why it is useful to talk about different activity stages of a comet. 
Following \cite{Goetz2021b}, we will use the following definition: weakly active for $Q < 10^{26}\,$s$^{-1}$, intermediately active for $ 10^{26}\leq Q <  5\times10^{27}\,$s$^{-1}$, and strongly active for $Q \geq 5 \times 10^{27}\,$s$^{-1}$. These values are useful guidelines, but should not be taken too strictly, as the transition between stages is smooth.

In addition to the outgassing rate, the state of the solar wind plasma is an important parameter shaping the cometosphere. Fig. \ref{fig:Q_sw}, bottom panel, shows the solar wind dynamic pressure at comet 67P over two years, at different heliocentric distances.
The closer the comet is to the Sun, the higher the dynamic pressure $\rho v_{SW}^2$, because the density $\rho$ decreases approximately as $r_h^{-2}$, where $r_h$ is the heliocentric distance. The interplanetary magnetic field magnitude is also higher closer to the Sun, so that, as a comet follows its orbit, the solar wind parameters vary on timescales of weeks to months due to the heliocentric distance change.
Without knowing the exact solar wind conditions that a comet encounters, it is difficult (but not impossible!) to distinguish between intrinsic changes of the plasma environment and changes due to external triggers. 
The large timescale changes are therefore usually absorbed in the definitions of the comet's activity stages, as they happen on the same timescale. The shorter timescales are then discussed on a case by case basis.

While most of this chapter will concern itself with the interaction of different plasma populations with each other, the solar wind can also interact with the nucleus directly. This is only possible if the activity of the nucleus is negligible, because if there is a significant atmosphere, the solar wind will not be able to reach the nucleus. 
Solar wind sputtering causes dust particles to be freed from the nucleus. This process is useful to investigate surface composition without having a surface probe, but so far only comet 67P was suitable to observe sputtering \citep{Wurz2015}. 
There, silicon was the most abundant species, with traces of sodium, calcium and potassium all detected at 67P. Models showed that indeed the observed signal was due to sputtering, not gas or dust release due to thermal input into the surface. The ratio of sodium-to-potassium was of the same order as that of the lunar surface, the solar system and meteorites.

\subsection{Shaping the induced cometosphere}
When intercepting comet 67P at about 3.5 AU ($Q \simeq 10^{26}\,$s$^{-1}$), the \emph{Rosetta} plasma instruments reported the first signatures of a plasma environment significantly affected by cometary matter at a few hundreds of km from the cometary nucleus \citep{Clark2015a,Yang2016}. At this time, the outgassing rate of 67P was estimated to be four orders of magnitude smaller than comet 1P/Halley's outgassing rate \citep{Combi_Feldman_Icarus_1993,Hansen2016}. Hence, even during its weakly outgassing phases, 67P was actively mass-loading the solar wind plasma and maintaining an induced cometosphere \citep{Nilsson2015a,Richter2015,Deca2017}.

\subsubsection{Plasma sources and sinks}
Fundamentally, there are two populations of particles that contribute to shaping the near-cometary environment, or better, the induced cometosphere: the incoming solar wind plasma and the radially-expanding neutral gas coma that is ionised through a variety of processes, such as photo-ionization, charge exchange, and electron-impact ionization (see Chapter 16).

67P’s neutral coma consists primarily of $H_2O$, $CO_2$, $CO$ and $O_2$
(see Chapter 16). Historically, the Haser model, which assumes a spherically-symmetric homogeneous outgassing for each species at a constant neutral radial velocity, has been applied to estimate the production rates of cometary volatiles \citep{Haser1957}. While the model can include different species and loss through ionisation, a reasonable approximation of the overall, momentary gas production rate Q can be obtained by 
\begin{equation}
    Q = 4 \pi n_n u_n r^2,
\end{equation}
where $n_n$ is the local, measured neutral density, $u_n$ is the neutral velocity of $500 - 1000\,$m/s and $r$ is the cometocentric distance of the measurement point. Within a couple of km of the comet a more accurate model should be used, but for the distances covered by \emph{Rosetta} during most of its mission, this approximation is reasonable to disentangle the effects of the radial variation of the trajectory from that of the temporal variation. 

\emph{Rosetta's} advanced instruments were able to capture for the first time the details of the non-uniform time-varying volatile distribution in which the cometary nucleus is embedded \citep{Marshall2019}. The complexity of the latter is caused by the rotation and orbital motion of the comet. In other words, \emph{Rosetta's} high temporal and spatial resolution measurements allowed to develop an accurate shape model that allows solar illumination and shadowing effects to be taken into account to predict the evolution of the production rates as a function of heliocentric distance \citep{Bieler_etal_SPIE_2016,Fougere2016,Fougere2016MNRAS,Hansen2016}. 

The cometary plasma density is determined by a continuous interplay of five main processes. (1) When a neutral molecule absorbs sunlight, a positive ion and photoelectron may be created. Photoelectrons are believed to be a significant contributor to the warm electron population (see below).
(2) At larger heliocentric distances/lower outgassing rates, accelerated solar wind electrons (see Sect. \ref{4fluidsystem}) lead to elevated electron-impact ionisation rates, because the electron ionisation thresholds for most cometary species are significantly higher than the typical local photoelectron energies \citep{ItikawaMason2005,bodewits2016changes,Deca2017,Heritier2018,Divin2020}. (3) At very large cometocentric distances, fast and light solar wind H$^+$ and He$^{2+}$ ions may capture electrons from slow heavy neutral species through charge-exchange \citep{Fuselier1991,Bodewits2004,SimonWedlund2019a}. Although this process does not produce any net ionisation, momentum is transferred from the fast solar wind ions to the slow neutral coma and mass-loads the cometary environment \citep{SimonWedlund2017}. (4) Fast solar wind ions that impact cometary neutrals can provide an additional source of charged particles to the cometary plasma density. 
(5) Finally, the primary process causing a loss of plasma in the induced cometosphere is electron-ion dissociative recombination. When the plasma number density is sufficiently high and contains cold electrons, ions and electrons can merge back together and with the excess energy create a neutral fragment \citep[see also Chapter 16 and ][]{SimonWedlund2019c}. 

\subsubsection{Ion Dynamics}\label{iondynamics}
\begin{figure}
    \centering
    \includegraphics[width=0.99\columnwidth]{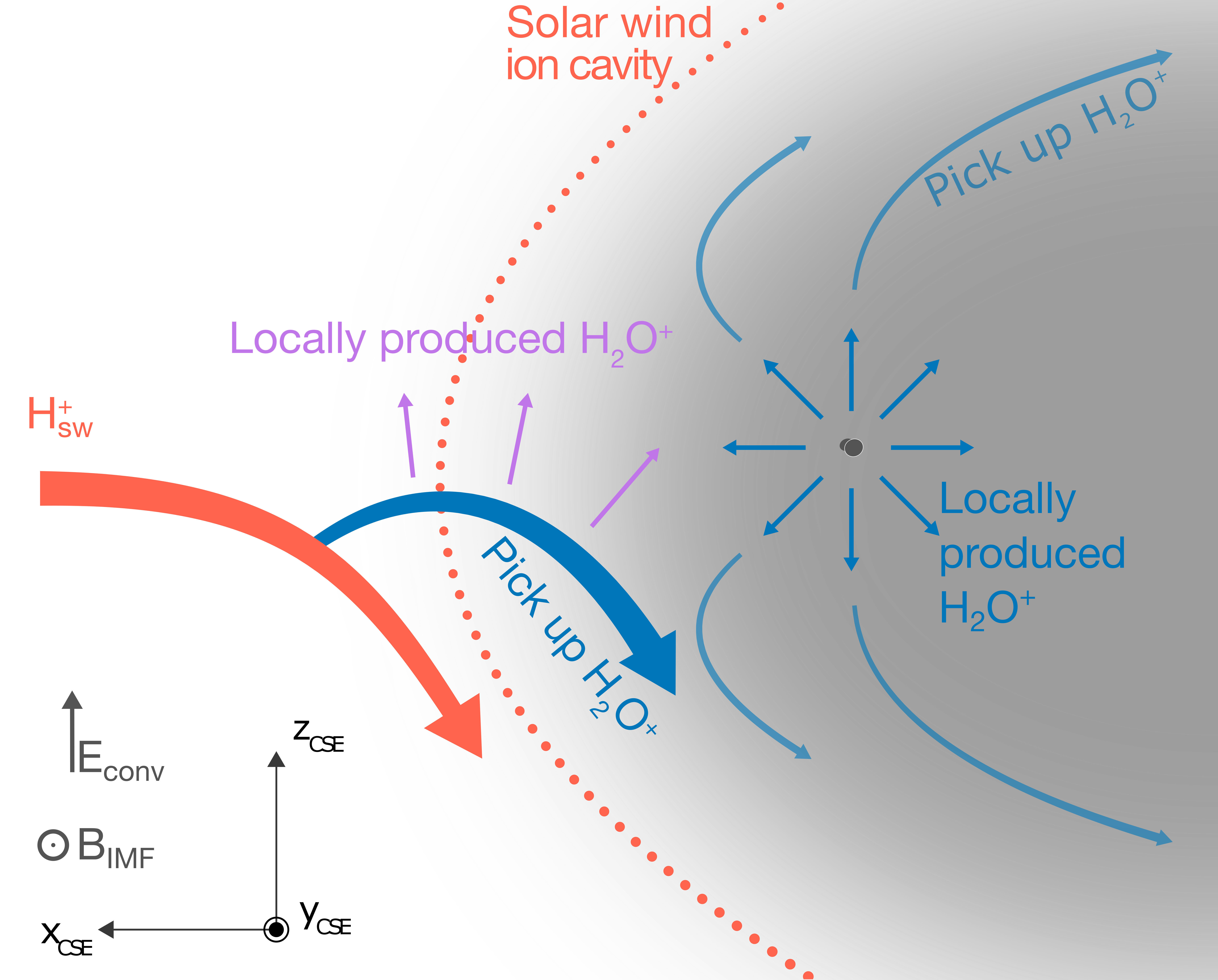}
    \caption{Sketch of the bulk ion motion at an intermediately active comet. The Sun is to the left. The solar wind protons (red) are deflected to conserve momentum as the locally produced H$_2$O$^+$ (purple) is accelerated along the solar wind convective electric field $E_\text{conv}$. Pick up ions (dark blue) that were produced far upstream eventually take over the role of the solar wind protons in the flow towards the comet. Within the solar wind ion cavity, the convective electric field is shielded. Within this region, the locally produced water ions (light blue) expand radially and are eventually accelerated and picked up by various processes (electric fields, wave-particle interaction). Adapted from \cite{Nilsson2020}.}
    \label{fig:ions}
\end{figure}

In previous books on cometary plasma processes, the discussion on the behaviour of ions was often based on a (single-)fluid dynamics approach. This had two reasons: 1. simulations in an MHD regime were the predominant simulation method, as hybrid and fully kinetic simulations required computational resources out of reach of the previous generations of supercomputers, and 2. for the comets visited until then, a fluid dynamics approach was in reasonable agreement with the measurements (partially because fine-scale particle dynamics could not be resolved by the onboard instruments). 
However, even then it was clear that weakly active comets necessitate a different approach, because ratios between the relevant plasma scales are different from more active comets. 
A useful parameter that highlights these differences is the particle gyroradius:
\begin{equation}
r_g = \frac{m v_\perp}{\left|q\right| B},
\end{equation}
where $m$ and $q$ are the particle mass and charge, $B$ is the magnetic field strength and $v_\perp$ is the particle velocity component that is perpendicular to the field. For $B=5\,$nT and a water group ion with a relative velocity to the magnetic field of $400\,$km/s, this results in a gyroradius of $\sim 15000\,$km. Compared to the scale size of comet 1P/Halley's interaction region ($2\times10^{6}\,$km) this is small and thus gyroradius effects on the large scale structure of the environment should be negligible. On the other hand, the interaction region at 67P only extends $100\,$km (at low activity) to $\sim 7000\,$km (high activity) and thus the cometary ion gyroradius is larger and cannot be neglected.

These gyroradius effects are observable in the data collected by \emph{Rosetta's} instruments. As was speculated in the last book, the cometary ions are virtually at rest in the comet's frame of reference, which means they move at $-v_{sw}$ in the solar wind frame. Therefore, they are subject to the solar wind convective electric field:
\begin{equation}
\vec{E}_\text{conv} = - \vec{v}_\text{sw}\times \vec{B}
\end{equation}
and are accelerated along the field. This is illustrated in Fig. \ref{fig:ions} by the purple arrows. \citet{Rubin2014} showed that these gyroradius effects can be modelled reasonably well with a hybrid as well as a multi-fluid MHD approach. The inclusion of the cometary ions as a separate fluid was shown to be a good method to simulate the large scale gyroradius structures, expanding the applicability of the MHD models. 
If all cometary ions are moving in one direction, momentum conservation dictates that the solar wind ions have to move in the opposite direction, they are deflected away from the comet-Sun line. This has been observed at low activity at 67P as well as in the artifical comet experiments, AMPTE \citep{Coates2015,Behar2016}. 
This in turn modifies the plasma flow, as the addition of the heavy cometary ions leads to a reduction in overall flow velocity. Incoming solar wind ions are thus faster than the bulk flow and start to gyrate in the magnetic field, similar to the cometary ions.
This leads to an additional component to the solar wind ion deflection that can add up to $180^\circ$ in the denser part of the coma, where the magnetic field is higher. In effect this means that under certain outgassing conditions, solar wind ions can flow Sunward (see also blue and red boxes in Fig. \ref{deflection}). 
At this point, the solar wind ions are so deflected, that they cannot flow towards the inner coma anymore (red arrow in Fig. \ref{fig:ions}). This is the approximate position of the solar wind ion cavity. Within that region, no solar wind ions can be observed (see Sect. \ref{sec:boundaries}). 
In the aftermath of the Halley flybys, the region outside of the solar wind ion cavity was often referred to as the cometosheath, i.e. the region where cometary and solar wind ions are mixed and flow around the inner coma \citep{Mandt2016}.

For any investigation of the ion flow pattern (as e.g. shown in Fig. \ref{fig:ions}) at low gas production rates, or far from the nucleus, it can be advantageous to transform the data into a Cometocentric Solar Electric (CSE) coordinate system. In this system the comet is at the origin, and the x-axis points toward the Sun. The y-axis is aligned with the solar wind convective electric field and the z-axis completes the triad.

Ignoring the electron dynamics for now (see section \ref{4fluidsystem}), an ion plasma flow towards the nucleus remains, as also evidenced by the presence of an interplanetary magnetic field inside the solar wind ion cavity. This means that accelerated (pick-up) cometary ions must have taken over the role of the solar wind ions in this flow. Indeed, observations show two distinct populations of cometary ions at this stage \citep{Bercic2018}: one accelerated population flowing approximately anti-Sunward (dark blue arrows in Fig. \ref{fig:ions}) and a radially expanding, slow moving population (blue locally produced H$_2$O$^+$ in Fig. \ref{fig:ions}). This flow pattern is consistently observed in the innermost coma, where accelerated ions are often observed in pulses, while the slower, radially expanding population is ubiquitous. An open question remains: How do the kinetic aspects of the ion dynamics around a low-to-medium activity comet continuously evolve towards this more fluid-like behaviour at high outgassing activity, close to the Sun? And, related, what is the speed of the bulk of those low energy ions accelerated near the nucleus? What is the role of the ambipolar, polarisation and Hall-electric field for these acceleration processes?

This transfer of energy and momentum from the solar wind to the cometary ions can also be surveyed in terms of momentum flux.
The momentum flux in a cometary environment is the sum of the momentum flux of the cometary ions, solar wind ions, electron pressure and magnetic pressure. The higher the gas production rate (and the solar wind dynamic pressure) the higher the total momentum flux measured by the ion instruments. 
From recent observations it becomes clear that the electron pressure gradient is the dominant source of momentum at all activity levels, except those that are very high (close to perihelion of 67P). While the solar wind ions carry some momentum into the coma at low activity, this momentum is taken up by the cometary ions within the solar wind ion cavity \citep{Williamson2020}. This indicates an efficient momentum and energy transfer from the solar wind ions to the pick-up cometary ions via the convective electric field.

Within the inner coma, the solar wind convective electric field is not present anymore, instead other fields such as an ambipolar or a polarization electric field gain importance and lead to more complicated flow patterns \citep{Nilsson2018,Gunell2019,Deca2019}. Close to the nucleus, where collisions are important, the dynamics of the ions are coupled to the radially expanding neutral molecules \citep{Nicolaou2017,Bercic2018,Nilsson2020}.
Closer to the Sun, more complex flow patterns were observed around the diamagnetic cavity as collisional effects become important in a greater area of the interaction region. In general, simulations and observations show that the accelerated ions have a flow pattern that guides them around the diamagnetic cavity most of the time, but in some instances the accelerated ions can go through the boundary and flow tailward also within the diamagnetic cavity \citep{Masunaga2019}.

\subsubsection{A four-fluid coupled system}\label{4fluidsystem}
Close to the nucleus of comet 67P, \emph{Rosetta's} Langmuir probe measured a 1/r decay of the plasma density with cometocentric distance \citep{Edberg2015,Galand2016}. Consequently, it indicates that the ratio between the cometary and solar wind plasma density, and by extension the full set of plasma parameters throughout 67P's plasma environment, continuously changes within a comet's plasma interaction region \citep{Nilsson2017,Eriksson2017,Bercic2018}. In other words, the global configuration of the induced cometosphere is driven by changes in the locally dominating physical processes and the relative ratios of the local plasma scales, such as the electron/ion-neutral collisional mean-free-path and the inertial lengths of the plasma species. A variety of permanent and transient boundaries can be defined \citep[Sect. \ref{sec:boundaries} and][]{Mandt2016}. In addition, (the existence of) these boundaries evolve(s) with heliocentric distance, or better, with the activity regime of the comet. 

During weakly outgassing regimes, i.e., well before a bow shock or diamagnetic cavity forms upstream of the cometary nucleus, ionization of the outflowing neutral gas from the nucleus compresses the incoming solar wind. A magnetic pileup region forms and the magnetic field lines drape around the nucleus \citep[see Sect. \ref{sec:magneticfield} and][]{Koenders2013,Koenders2015,Koenders2016,Behar2016aa,Behar2016,Volwerk2016}. 

\begin{figure*}
    \centering
    \includegraphics[width=0.9\textwidth]{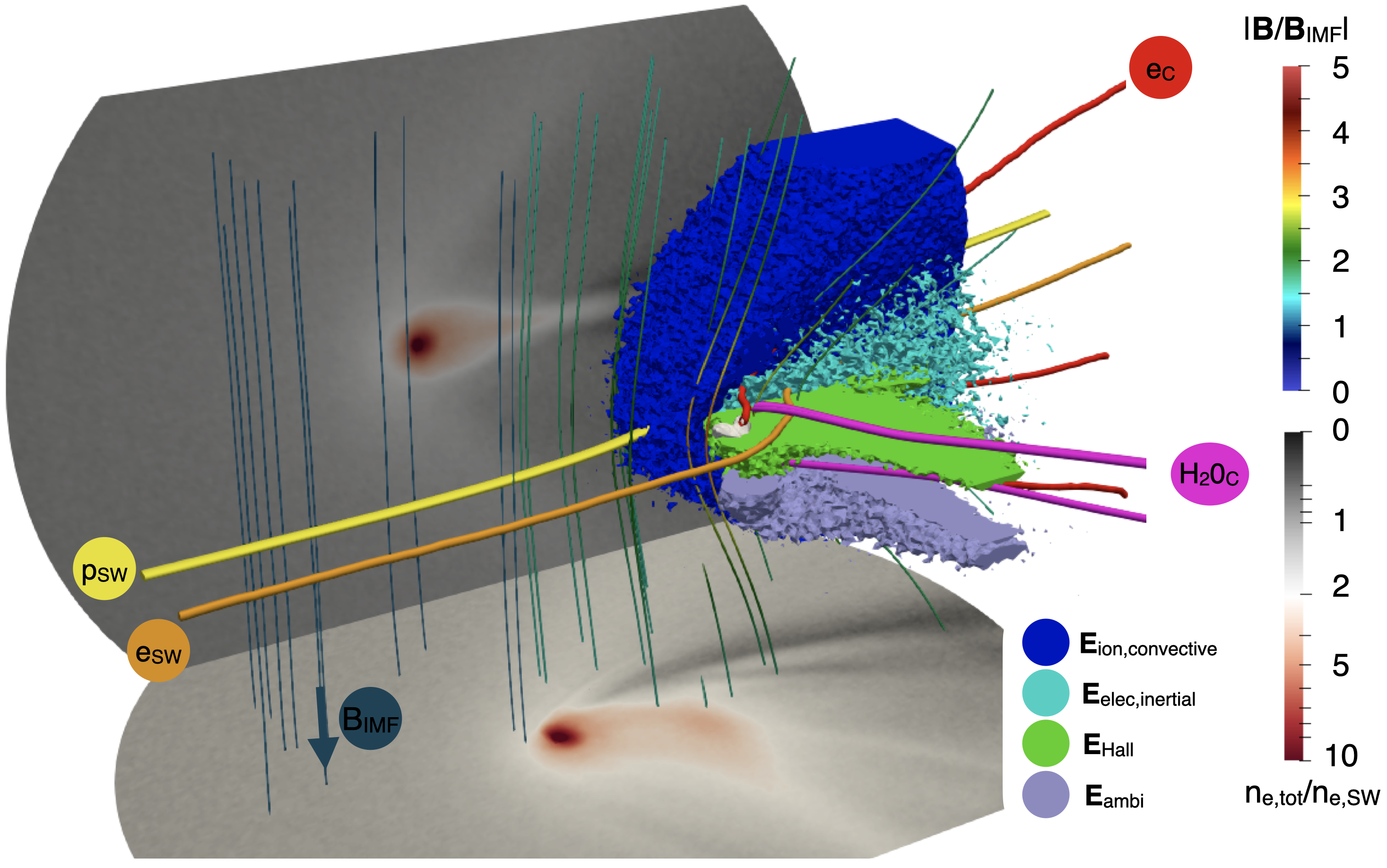}
    \caption{Illustration of the four-fluid coupled system, including various isovolumes that indicate the significant components of a Generalised Ohm's law. See text for details. Figure adapted from \citep{Deca2019}.}
    \label{fig:my_labelA}
\end{figure*}

In addition to section \ref{iondynamics}, here we include the electron dynamics. The discussion is based on several fully kinetic particle-in-cell simulations that model self-consistently four plasma populations: solar wind protons and electrons, cometary water ions and cometary electrons \citep{Deca2017, Deca2019, Sishtla2019, Divin2020}. Assuming a collisionless plasma, the dynamical interaction that determines the general structure of the induced cometosphere during the weakly outgassing phases can be interpreted as a four-fluid coupled system, where the solar wind electrons move to neutralize the cometary ions, and the cometary electrons organize themselves to neutralize the solar wind ions (illustrated in Fig. \ref{fig:my_labelA}) \citep{Deca2017}. More precisely, ions of cometary origin accelerate along the local solar wind convective electric field in the cross-magnetic field direction, whereas electrons of cometary origin are initially accelerated in the opposite direction. This results in a net (Hall) current that locally decouples the solar wind protons and electrons as the solar wind plasma becomes more and more massloaded by cometary ions. It is important to note that the convective electric field carries an opposite sign in the solar wind and cometary ion reference frame and transfers momentum between the two species. As the solar wind protons are deflected and not any longer frozen-in, the interplanetary magnetic field is carried close to the comet through the solar wind (and cometary) electrons. Note that a crucial component of the mechanism described above is provided by the electron dynamics. One of the revelations of the \emph{Rosetta} mission and subsequent numerical modelling is that key interaction mechanisms based on fluid concepts, such as momentum and energy transfer between the solar wind and cometary plasma can be explained by including self-consistently the electron dynamics. For example, the deflection of cometary electrons creates the electron current that induces the magnetic pileup region as the incoming solar wind is compressed \citep{Deca2019}. Figure \ref{fig:my_labelA} presents an overview of the four-fluid behaviour of the solar wind interaction with a weakly outgassing comet.

Although modelling both the electron and ion species is a necessity to unveil the interaction discussed above, less complete approaches (from an electron point-of-view) often present different benefits.
For example, more complex ion-interaction models can be implemented as less computational resources are needed when keeping the electron populations as a massless charge-neutralising fluid as it is done in a hybrid approach \citep{SimonWedlund2019c, Koenders2016,alho_hybrid_2020}. This has the advantage of reducing simulation time while retaining the ion kinetics.
To resolve larger interaction regions, simulate longer time scales and for parameter studies that require a short turnover time, a multi-fluid approach can be useful \citep[e.g.][]{Huang2018}. However, this comes at the cost of losing the electron and ion kinetics and many small scale structures. Compared to fully kinetic treatments, hybrid and multi-fluid (Hall-) MHD models often assume a a suitable electron closure relation combined with a generalised Ohm's law (GOL) to approximate the electric field in the system \citep{Huang2018,Deca2019}: 
\begin{equation}\label{eq2}
    \mathbf{E} = - (\mathbf{u}_{\rm i}\times\mathbf{B}) + \frac{1}{en}(\mathbf{j}\times\mathbf{B}) - \frac{1}{en}\nabla\cdot\mathbf{\Pi}_{\rm e}
    + \frac{m_e}{e}\nabla\cdot(\mathbf{u}_e\mathbf{u}_e),
\end{equation}
where $e$ and $m_e$ are the electron electric charge and mass, respectively, $\mathbf{B}$ the magnetic field, $\mathbf{u}_{\rm i}$ the ion mean velocity, $\mathbf{j}$ the current density, $n$ the plasma total number density defined as the sum of the solar wind and cometary densities, $n=n_{\rm sw} + n_{\rm c}$ and $\mathbf{\Pi}_{\rm e}$ the electron pressure tensor derived from the electron momentum equation. The various terms decompose the total electric field in (from left to right) the convective electric field, the Hall electric field, the ambipolar electric field and the contribution associated with the electron inertia. Using the results from their fully kinetic model, \citet{Deca2019} were able to compute the various contributions using the output particle data from the simulation, in this way identifying where an isotropic single-electron fluid Ohm’s law approximation can be adopted, and where it cannot. To illustrate, Figure \ref{fig:my_labelA} includes various isovolumes that indicate the regions where specific terms of the GOL described above are significant. Consistent with \emph{Rosetta}'s measurements, they conclude that near the nucleus the electron pressure gradient is dominant, and that at spatial scales smaller than the ion plasma scales the total electric field is primarily a combination of the solar wind convective electric field and the ambipolar electric field. Interestingly, throughout most of the region the electron inertial term was found negligible. This also implies that multi-fluid and hybrid simulations with an electron pressure equation are valid most of the time, except when a distinction between electron populations is necessary. 
While cometary missions and simulations have helped us understand the complex interplay of these four fluids to a greater degree, the exact plasma distribution with respect to cometocentric distance (in 3D), and how it varies in time is still unclear. In addition, 
it remains to be investigated what role the (quiet) solar wind dynamics (e.g. turbulence) play in the inner coma dynamics, where the plasma is shielded from direct influence of solar wind variation, but could be affected indirectly, via pick-up ion dynamics.

\subsubsection{Electron populations and the ambipolar electric field}
The electron distributions observed in the inner coma of 67P can be traced back to two main categories: cometary electrons and solar wind electrons. The cometary electron population originates from ionising neutral molecules of cometary origin through solar extreme ultraviolet radiation \citep[photoelectrons;][]{VigrenGaland2013} or secondary particle processes as described above \citep{SimonWedlund2017,Heritier2018}. The undisturbed solar wind electron population typically consists of a thermal core (E $<$ 50 eV) and a suprathermal tail (70 $<$ E $<$ 1000 eV) \citep{Pierrard2016}.

A variety of physical mechanisms in the cometary plasma environment can affect both these source populations, such as ambipolar electric fields \citep{Madanian_etal_2016,Deca2019,Divin2020}, adiabatic compression and expansion \citep{Nemeth2016,Broiles2016,Deca2019,Madanian2020}, collisional cooling \citep{Engelhardt2018b} and wave-particle interactions \citep{Goldstein2019a,Lavorenti_etal_2021}.

In reality, it is not feasible to tell apart which group an electron belongs to without additional information, e.g., from numerical modelling. Most generally, three distinct electron populations were observed: a warm (5-10 eV), cold (0.01-1 eV) and hot (or suprathermal) population (few 10s to 100 eV).

In order to understand the origins of these three distinct populations, we need to have a closer look at the electromagnetic environment surrounding the cometary nucleus. Whereas ionisation of the cometary gas results in electrons that have thermal speeds on the order of $\rm 1000\,km\,s^{-1}$ \citep{VigrenGaland2013}, the much heavier ion counterparts tend to retain the speed of their parent neutral molecule (typically on the order of $\rm 1\,km\,s^{-1}$). A strong electron pressure gradient forms, which results in a potential well and ambipolar electric field that surrounds the cometary nucleus \citep{Madanian_etal_2016,Deca2019,Divin2020}. Note that efficient electron cooling during high outgassing phases may neutralize the ambipolar electric field component, such as was the case for the \emph{Giotto} flyby of comet 1P/Halley \citep{GanCravens1990}, but the importance of the ambipolar electric field for low-activity comets has been one of the most intriguing discoveries of the \emph{Rosetta} mission. For example, \cite{Galand2020} found that this acceleration of the electrons in an ambipolar field can produce EUV emissions, similar to those associated to auroral emissions at other solar system bodies. 

The ambipolar electric field and its associated potential well can temporarily trap cometary electrons and accelerate solar wind electrons near the cometary nucleus, leading to the observed warm and hot electron populations \citep{Deca2017}. Studying the trajectories of trapped electrons in the ambipolar potential well surrounding the nucleus, a clear boundary in velocity space can be defined that separates temporarily trapped cometary electrons from passing solar wind electrons \citep{Sishtla2019}. It shows that electrons may stay much longer in the region of dense neutral gas around the nucleus than previously believed, which may also lead to increased cooling of electrons in non-collisionless plasma regimes \citep{Engelhardt2018b,Divin2020}. Using fully kinetic particle-in-cell simulations it was found that cometary electrons exhibit an apparent isotropic and almost isothermal behaviour, whereas the solar wind electrons, on the other hand, exhibit an anisotropic and apparent polytropic behaviour \citep{Deca2019}. In addition, at a few tens of kilometres from the nucleus, the electron sensor observed a highly variable bi-Maxwellian electron distribution with energies ranging from tens up to several hundred eV \citep{Clark2015a}, consistent with a dense warm and a rarefied hot electron population \citep{Madanian_etal_2016,Deca2017}. \cite{Broiles2016} constrained the kappa values of the warm population ranging from 10-1000, suggesting the warm electrons are near-thermal equilibrium (consistent with \citealt{Deca2019}), whereas values between 1-10 were found for the hot electron component. The warm population showed a 1/r density dependence up to about 900\,km from the nucleus and a $r^{-0.7}$ dependence farther out \citep{Edberg2015,Myllys2019}, where $r$ is the cometocentric distance. In addition, \cite{Myllys2019} corroborated on a $1/r^2$ dependence with density for the hot electrons.

Taking into account the radial cometocentric dependence of the warm population, both the warm and hot components manifest an increase in density with decreasing heliocentric distance, consistent with the observed increase in neutral gas density and photoionisation frequency. The density of the warm and hot population increased from about 30 to 100 $\rm cm^{-3}$ and 0.1 to 3  $\rm cm^{-3}$  between 3.5 and 1.3 AU, respectively. The hot electron population showed a 33\% increase in temperature with heliocentric distance, whereas the warm population seemed not affected by the latter parameter. \cite{Myllys2019} concluded as well that the ambipolar acceleration process could only explain part of the observed hot electron distribution, suggesting that part of the population may have its origin in the already accelerated suprathermal component of the solar wind electron distribution \citep{Broiles2016}. 

Quite intriguingly, the accelerated hot electron population was found to be correlated with Far-UV emissions observed near comet 67P, showing for the first time that cometary aurora can be driven by the interaction of the solar wind with the local environment \citep{Galand2016}.

\begin{figure*}
    \centering
    \includegraphics[width=0.95\textwidth]{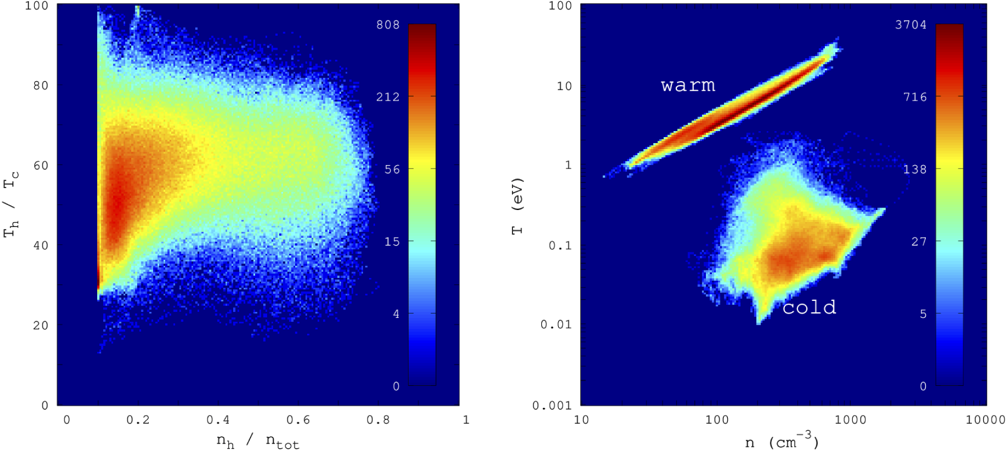}
    \caption{Distribution of $\rm T_h/T_c$ versus $\rm n_h/n_{tot}$ (left) and of the electron temperatures versus the electron densities (right) estimated at the \emph{Rosetta} orbiter locations in the cometary plasma of comet 67P from January to September 2016 (570 000 samples), where the subscripts $h$, $c$ and $tot$ indicate hot cold and total, respectively. Colors correspond to the number of responses (the colour map is logarithmically scaled in both panels). Figure adapted from \citet{Wattieaux2020}.}
    \label{fig:Wattieaux_etal_AA_2020_fig13}
\end{figure*}

Although \emph{Giotto} could not directly measure cold electrons, it was assumed that the dense inner coma of 1P/Halley could efficiently cool electrons through collisions with the neutral gas \citep{Feldman_etal_JGR_1975,Ip75,Eberhardt1995}. Electron cooling is thought to be most efficient within the electron exobase (or collisionopause, see Chapter 16). However, \emph{Rosetta’s} Langmuir Probe and Mutual Impedance Probe instruments detected cold electron signatures significantly outside the electron exobase, suggesting that once cooled in the inner coma, there is no mechanism that is able to reheat cold electrons when they travel away from the nucleus \citep{Eriksson2017}. At this point, it is not known what processes contribute to cooling newborn, warm electrons.
The cold electron temperature observed ranged between 0.05 to 0.3 eV \citep{Wattieaux2020}. In addition, cold electrons measurements were only significant in regions dominated by (warm) cometary plasma (Figure \ref{fig:Wattieaux_etal_AA_2020_fig13}). It was shown that there is a clear correlation between electron cooling and the observations of the diamagnetic cavity, suggesting that both have to be in close proximity \citep{Henri2017,Odelstad2018}. Finally, cold electrons have as of yet not been observed without the presence of a significant warm electron population \citep{Eriksson2017,Myllys2019,Gilet2020a,Wattieaux2020}. So far it is unclear what the energy distribution of electrons below the spacecraft potential is.

\subsubsection{Cometary outbursts}

Up to here, we have focussed on a steady-state solar wind plasma interaction with the cometary environment. Throughout the \emph{Rosetta} mission, however, at least one internal outburst was detected that significantly affected the near-cometary plasma environment \citep{Gruen2016}. During the outburst, the neutral density doubled, but interestingly, the plasma density quadrupled, which was attributed to an increased efficiency of the electron impact ionisation process as well as more efficient ion-neutral coupling in the denser neutral gas which reduces the ionisation length scales \citep{Hajra2017}. \emph{Rosetta} also measured locally a decreased suprathermal electron flux and solar wind ion density, as well as a reconfigured magnetic field and a change in the singing comet wave activity \citep[see Sect. \ref{sec:waves} and][]{Breuillard2019}. 

\subsection{External drivers to the induced cometosphere}
\label{sec:cometosphere_external}

\begin{figure}[t]
    \centering
    \includegraphics[width=0.95\columnwidth]{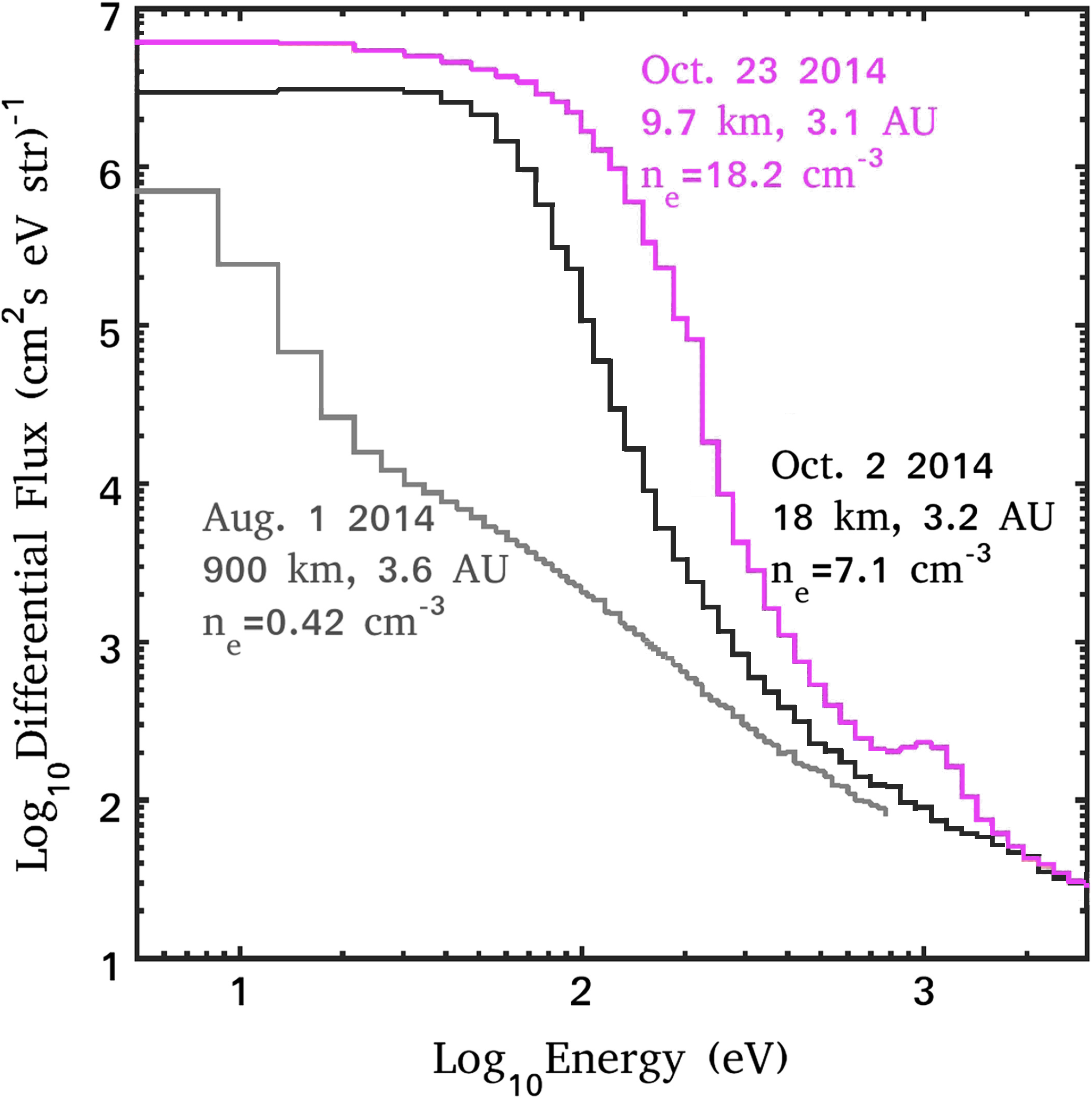}
    \caption{The daily averaged electron differential flux as measured by \emph{Rosetta}'s Ion and Electron Spectrometer. The spectrum from 1 August 2014 shows a typical solar wind type electron flux. The spectrum from 2 October 2014 and 23 October 2014 show a typical quiet and ICME-impacted electron flux profile near the comet, respectively.}
    \label{fig:Madanian_etal_JGR_2016_Figure_3_v0}
\end{figure}

Above we discussed the major internal processes that shape the near-cometary plasma environment. Solar wind (SW) transients, flares, CIRs, and ICMEs, on the other hand, are external drivers to the induced cometosphere. Transient phenomena change the solar wind parameters on relatively short time scales and, as a result, have the potential to significantly reconfigure/disturb the near-cometary plasma environment. The effects can be extremely variable and differ from event to event, depending on the strength and type of the transient and the heliocentric location and outgassing rate of the comet. Nevertheless, several characteristics can be identified that are consistently observed for specific classes of transient phenomena.

Solar flares are short-lived and temporarily increase the local extreme-ultraviolet flux. However, during the weakly outgassing stages of a comet, photoionisation is only a minor contributing plasma source. During the strongly outgassing stages, on the other hand, the induced cometosphere is inherently very variable even without external drivers. As a result, solar flares are not observed to significantly disturb the plasma interactions that shape the induced cometosphere \citep{Edberg2019flare}.  

CIRs, on the other hand, spiral outward in the solar system and are often accompanied by a forward shock at their leading edge \citep[and a reverse shock at the trailing edge, ][]{Balogh_etal_SSR_1999}. This has been observed to lead to an increase in the local solar wind velocity, proton and cometary plasma density, temperature, magnetic field strength, and suprathermal (hot) electron flux \citep{Edberg2016CIR,Hajra2018}. The latter results in an increased electron-impact ionisation rate, which, on its turn, may become the largest contributor to the total plasma density, rather than increased compression or acceleration due to the ion pickup process \citep{Galand2016,Heritier2018,Hajra2018}. The magnetic compression resulting from the CIR impact on the induced cometosphere, on the other hand, is responsible for the local increase in proton density and magnetic field strength.

Specific for 67P, \cite{Hajra2018} observed greater increases on the southern (summer) hemisphere of the comet versus its northern (winter) hemisphere. The effects of CIRs are typically observed for about 24h, with clear signatures of the forward and reverse shock wave visible even in the coma region. 
Langmuir waves can be excited near the forward and reverse shocks of a CIR in the solar wind, where electrons can be accelerated such that they counterstream in the solar wind flow \citep{Pierrard2016}.
Locally generated Langmuir waves have been observed during times of CIR (and ICME) interaction with the cometary environment, indicating that backstreaming electrons exist even in the coma of the comet.
However, the exact mechanism that causes the electrons to backstream along the field lines from the shock interface is not yet understood \citep{Myllys2021}.

ICMEs comprise the most variable class of external drivers to the induced cometosphere. However, they tend to accelerate most within 1 AU, after which they gradually decelerate travelling outward into the solar system \citep{Zhao_etal_ApJ_2017}. Their interaction with the cometosphere can be characterised by an increase in magnetic field and plasma density, an increase in energetic electron flux (e.g. Figure~\ref{fig:Madanian_etal_JGR_2016_Figure_3_v0}), and a rapid (Forbush) decrease in the observed galactic cosmic ray intensity \citep{Witasse2017}. ICMEs may also merge with other transients and form more complex structures \citep{Witasse2017,Wellbrock_etal_EPSC_2018}. Although the \emph{Rosetta} spacecraft was not often in an ideal position to characterise the impacts of ICMEs throughout the induced cometosphere, an ICME impact detected close to perihelion lead at 170\,km from the cometary nucleus to the highest magnetic field strength ($300\,$nT) measured through the mission \citep{Goetz2019}.

\subsection{Magnetic Field Pile-up}\label{sec:magneticfield}
\begin{figure}[h]
    \centering
    \includegraphics[width=0.99\columnwidth]{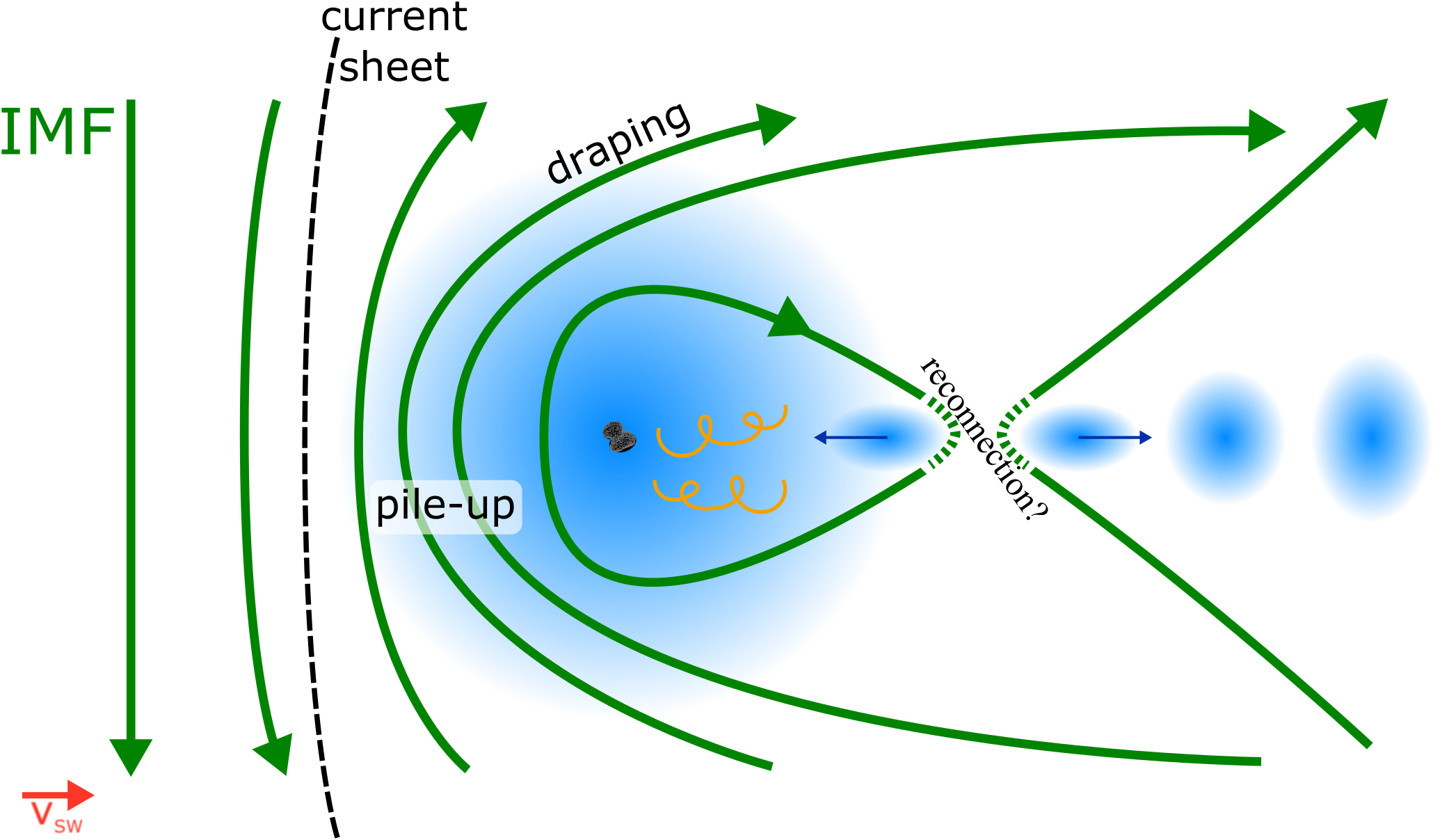}
    \caption{An illustration of the large scale magnetic field at a comet, including processes in the tail. The Solar wind with the IMF approaches from the left. 
    The magnetic field lines are shown in green, and a heliospheric current sheet is marked by the black dashed line. In the near comet tail there are spiralling magnetic field waves in orange. A possible reconnection region in the tail produces plasma density enhancements (blue) that travel towards and away form the reconnection region. For further information and scale sizes see text. }
    \label{fig:magnetic_field}
\end{figure}
\label{ssec:magnetic_structure}
The interaction of the interplanetary magnetic field (IMF, sometimes also referred to as the heliospheric magnetic field) carried by the solar wind and an outgassing comet leads to the formation of an induced magnetosphere with magnetic field pile-up and field line draping, an overview of which is shown in Figure \ref{fig:magnetic_field}. This happens as soon as the comet gets active, usually when it enters Jupiter's orbit, i.e.\ $\sim 5$ au radial distance from the Sun. The convective electric field of the solar wind will accelerate the new-born cometary ions, and these start to gyrate around the magnetic field lines. This is called ion pick-up or mass-loading of the solar wind. As most of the earlier work on cometary magnetospheres is discussed in \citet{Ip2004}, this section will mainly concentrate on work after 2004.

Because of conservation of momentum the mass-loaded solar wind will have to decelerate. \citet{Biermann1967} presented a simple 1-D MHD model of this solar wind - comet interaction by combining the continuity, momentum and energy equations. The solar wind velocity $u_{\rm x}(x)$ is in the $x$-direction and the IMF $B_{\rm y}$ is in the $y$-direction. This results in a solar wind velocity:

\begin{widetext}
\begin{align}
&u_{\rm x}(x) = \left( 2 (f + 1) (M(x) + \rho_{\infty} u_{\infty}) \right)  \nonumber \\
&\left[ (f+2) \rho_{\infty} u_{\infty}^2 \pm \sqrt{(f+2)\rho_{\infty}^2 u_{\infty}^4 - 4(f+1) (M(x) + \rho_{\infty} u_{\infty}) \rho_{\infty} u_{\infty}^3 } \right],
\label{Eq:uxx}
\end{align}
\end{widetext}
 
\noindent
where $\rho_{\infty}$ and $u_{\infty}$ are the undisturbed solar wind density and velocity, $f$ is the number of degrees of freedom of the ions, and $M(x)$ is the mass source. Using e.g.\ the \citet{Haser1957} model for cometary outgassing one can solve the continuity equation for the magnetic flux, which leads to $B_{\rm y}(x) \rightarrow \infty$ at the location where the bow shock is formed. This can be solved by adding cooling of the plasma through charge exchange processes, to obtain finite field strengths downstream of the bow shock \citep[for details see][]{Galeev1985, Goetz2017}. However, there are some limiting assumptions to this model: the solar wind magnetic field should be perpendicular to the plasma flow direction (but is in reality on average at the Parker spiral angle); single fluid MHD is assumed (although the ion gyro radius can be larger than the interaction region); and the model is only valid on the Sun-Comet line. Nevertheless, a comparison of the modeled magnetic field and the measured field at comet 67P gives reasonable agreement \citep{Goetz2017}. 

\subsection{Magnetic Field Draping}
\label{sec:draping}

\begin{figure*}
    \centering
    \includegraphics[width=0.4\textwidth]{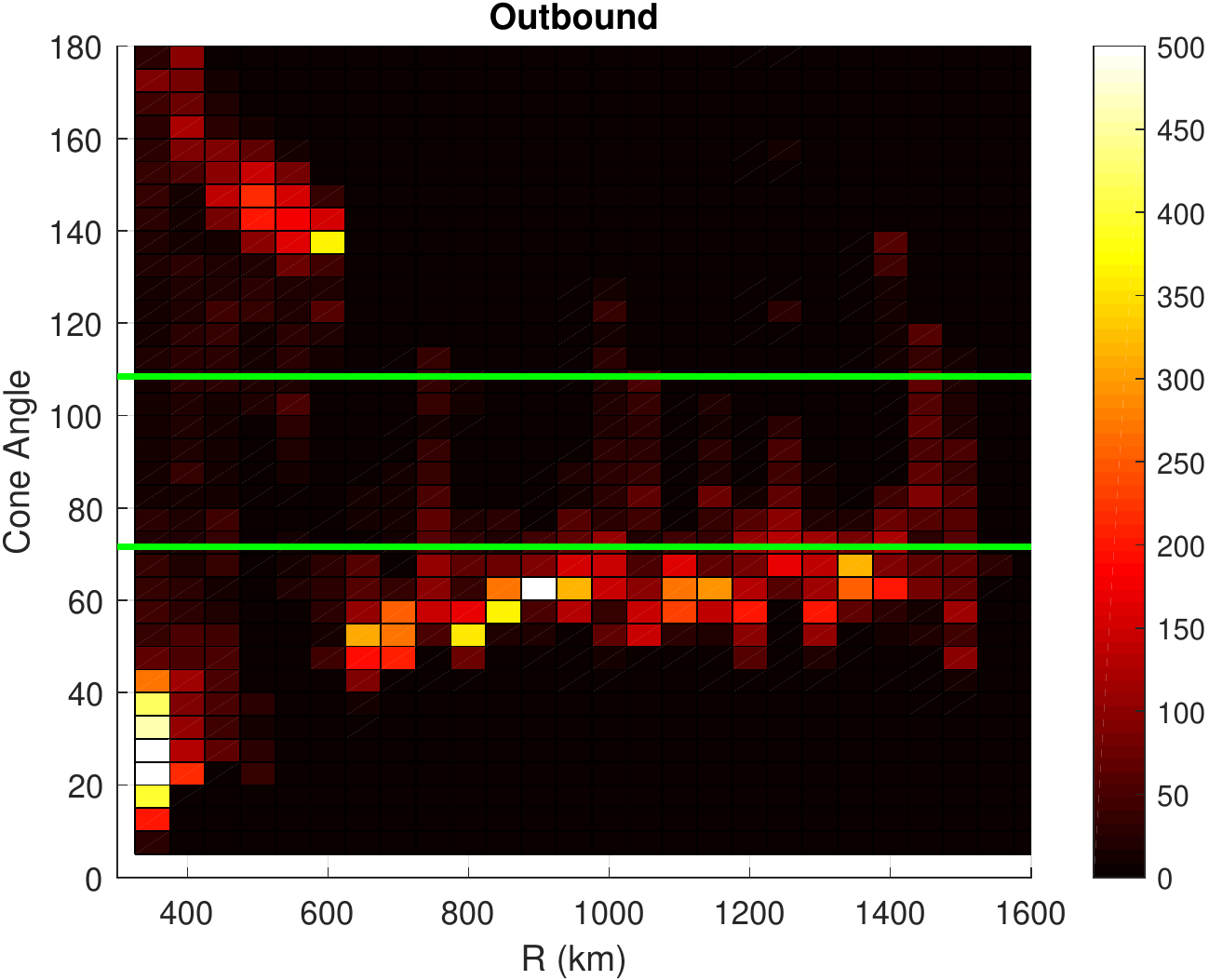} \includegraphics[width=0.4\textwidth]{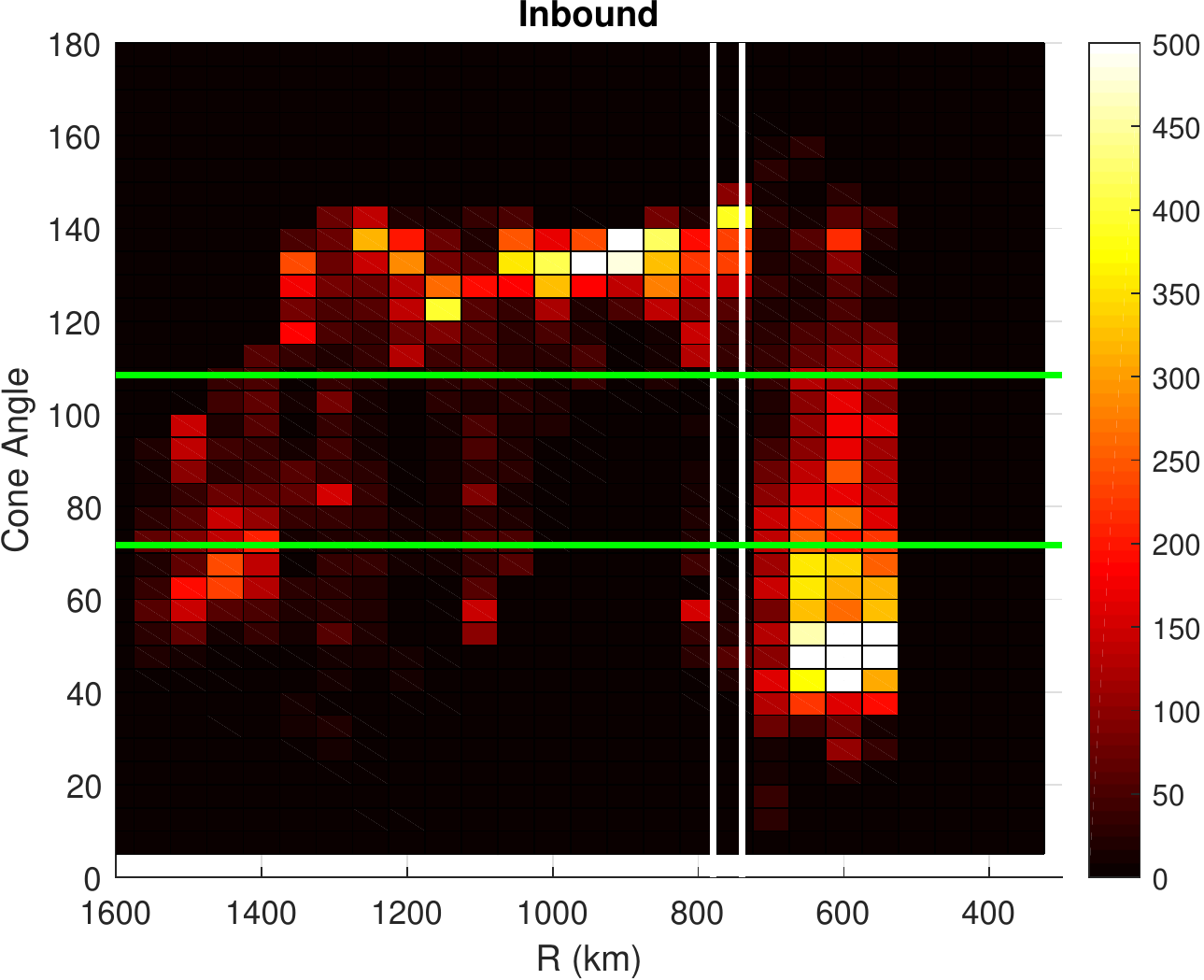}
    \caption{Two-dimensional histograms of the cone angle of the magnetic field along the \emph{Rosetta} orbit for the outbound (left) and inbound (right) leg. The green lines show the cone angles for the Parker spirals. The two white vertical lines show the interval during which the ICME interaction took place. From \citet{Volwerk2019}.}
    \label{fig:dyndrape}
\end{figure*}

The above described 1D model delivers the magnetic field strength in the subsolar cometary environment, but does not describe the full structure of the induced magnetosphere. Therefore it needs to be expanded with a second radial dimension, perpendicular to the comet-Sun line. This results in the magnetic field line being draped around the active nucleus as was first described by \citet{Alfven1957}. Through the finite conductivity, the magnetic field will diffuse through the plasma cloud around the nucleus, thereby avoiding another magnetic catastrophy, i.e.\ an infinite magnetic field strength. The IMF, however, varies in direction and the radial component of the magnetic field changes sign regularly. Thereby, differently directed magnetic field can be draped around the active nucleus. When the diffusion speed through the plasma is smaller than the flow velocity of the mass-loaded solar wind, a layering of differently directed magnetic field will be created upstream of the comet, which is referred to as ``nested draping.'' This phenomenon was first observed at comet 1P/Halley during the flybys of \emph{VEGA 1} \citep{Riedler1986} and \emph{Giotto} \citep{Raeder1987}. The boundaries between these different magnetic field directions will have to support current sheets. Indeed, during the early phase of \emph{Rosetta}, during the unbound pyramidal orbits, many rotations of the field were found with a minimum magnetic field strength and a maximum in electron density in the centre of the rotation \citep{Volwerk2017}. Again, this would argue for nested draping of the IMF if this were a ``snapshot'' flyby like \emph{Giotto}. However, \emph{Rosetta} moves at a speed of only a few meters per second, which needs a different interpretation. On the other hand, if the diffusion speed is fast enough, no such pile-up will occur, apart from a regular increase of the magnetic field strength created by the mass-loading and slowing down of the solar wind. 

Another situation occurred at comet 67P, during the dayside excursion by \emph{Rosetta}. The spacecraft took an upstream excursion of $\sim 20$ days up to a distance of $\sim 1500$ km from the nucleus. In order to find nested draping, \citet{Volwerk2019} determined the cone angle of the magnetic field, defined as:

\begin{align}
\theta_{\rm co} = \tan^{-1} \left\{ \frac{ \sqrt{ B_y^2 + B_z^2} }{B_x} \right\},
\end{align}

\noindent
where $\theta_{\rm co} = 0^{\circ}$ is purely sunward directed field and $\theta_{\rm co} = 180^{\circ}$ purely anti-sunward\footnote{$B_x$, $B_y$, and $B_z$ are defined in a CSEQ coordinate system, with the nucleus in the origin, $\Vec{e}_x$ points to the Sun, $\Vec{e}_z$ is parallel to the Solar rotational axis, and $\Vec{e}_y$ completes the right handed system. }. Indeed, regions of differently directed magnetic field were found, but when the spacecraft returned back after apo-apsis, a region was entered that did not exist during the outbound leg, see Fig.\ \ref{fig:dyndrape}. This indicates that the draping moves faster towards the nucleus than the spacecraft is moving, which is not difficult as \emph{Rosetta} moved at $\sim 1\,$m/s. The authors called this ``dynamic draping.'' 
The question of how dynamic and nested field line draping are related is still open, as there is either a snapshot observation by a fast moving flyby or there is only a local long-term observation by an orbiting spacecraft. 
A multi-spacecraft mission could allow us to get a better grip on this problem: different trajectories could be flown, like it is planned for the upcoming Comet Interceptor mission \citep{Snodgrass2019}.

Lately, \citet{delv14} re-investigated the Vega-1 magnetometer data taken during the flyby of comet 1P/Halley, to study the magnetic pile-up boundary (MPB) and the draping pattern of the magnetic field. The MPB is the boundary between the cometosheath (where there is no draping) and the pile-up region (where there is strong draping). In order to study the draping, an electromagnetic reference frame was calculated from the abberated cometocentric coordinate system, in which $x_{\rm IMF}$ is along ${\bf v}_{\rm SW}$ and the new $y_{\rm IMF}$-axis direction is given by $ - {\bf v}_{\rm SW} \times {\bf B}_{\rm IMF}$. The draping is then studied by the relationship between $B_{\rm x, IMF}$ and $B_{\rm rad}$ the field component pointing radially away from the cometary nucleus. There should be a strong negative correlation between the two variables if the spacecraft is inside the MPB.

One effect has not been taken into account in this discussion, and that is conservation of momentum related to the pick-up process through the acceleration of the heavy cometary ions. The newly-created ions are accelerated by the convective electric field of the solar wind. Through Newton's second law, this action will need an opposite reaction, and thus the solar wind ions will have to move in the direction opposite to the convective electric field (see also Sect. \ref{sec:cometosphere}). As the IMF is frozen into the solar wind plasma flow, this will result in the magnetic field draping in this direction. \citet{Broiles2015} first showed that indeed the solar wind and pick-up ions moved in opposite directions, and the magnetic field draping effect was shown for 28 March 2015 (low to medium activity), when \emph{Rosetta} had a closest approach of $\sim 15$ km \citep[see e.g. Fig. 5 in][]{Koenders2016}. This effect starts to become very important when the pick-up density starts to become similar to the solar wind density. 

The comet-solar wind interaction has influence on the structure of the induced cometary magnetosphere. This influence can be used to obtain information about the solar wind for missions, like \emph{Rosetta}, that do not have easy access to the undisturbed solar wind. Indeed, the interaction of a ICME and CIR with comet 67P, on 3 July 2015, led to the unprecedented high magnetic field strength of almost 300 nT measured by \emph{Rosetta's} magnetometer \citep{Goetz2019}. But not only such extreme field measurements are necessary, \citet{Timar2019} derived a proxy by using the magnetic pressure just outside of the boundary of the diamagnetic cavity. Assuming that the magnetic pressure in the induced magnetosphere needs to balance the dynamic pressure of the solar wind, they found a good correlation between the proxy and the propagated solar wind from the Earth to the comet.

\subsection{Ion/Plasma/Magnetotail}
The most notable structure of a comet is its tail, or rather tails: the dust tail curved along the comet's orbit; and the ion tail which points almost radially away from the Sun. Historically, the cometary tails were categorized by \citet{bredikhin1879} into three types: type I - a straight tail pointing away from the Sun (the ion/plasma tail) and types II and III - tails curved towards the orbit of the comet (the dust tails) with type II consisting of medium-heavy elements and type III consisting of heavy metals. With the available new observations, this classification is no longer valid. The following sections will focus on the plasma tail itself, but there is also some evidence that the interaction of a comet with the solar wind can affect the dust tail through Lorentz force acceleration of the charged dust particles \citep[e.g. ][]{Kramer2014,Price2019}. 

\subsubsection{Remote sensing of the plasma tail}
Remote sensing, for example with ground-based observations of comets, gives a large-scale view of the ion tail, which, because of its large extent cannot be studied by spacecraft missions. These observations give insight in the structure and the dynamics of the tail. However, as we know that the cometary tail is created through the interaction of the solar wind and IMF, observations of the tail can also provide information about the ambient plasma (i.e.\ the solar wind) conditions.

Photometric observations of comet 1P/Halley in 1986 showed many structures in the tail. \citet{bran82} already talked about rays, streams, kinks, knots, helices, condensations and disconnection events. \citet{sait86} added arcades in the collection of structures. \citet{mozh17} studied the large scale structure of more than 1500 photographs of comets to categorize two different plasma tail types: double structures -- two bright diverging rays from the coma, and outflow -- single large scale thin tail . The double structure is assumed to be created by a line-of-sight effect of a cylindrical distribution of plasma in the tail created by mass loading of the solar wind. The outflow tails are posited to be created by outflow of plasma from the coma driven by the dynamic pressure of the solar wind, similar to Martian ionospheric plasma outflow \citep{lund11}.

\subsubsection{Structure of the tail}
\citet{Biermann1951} and \citet{Alfven1957} laid the foundations for the generation of the ion tail, through mass loading and field line draping described above. On 11 September 1986, the first evidence for this magnetic field draping in the comet's magnetotail was established through observations by the International Cometary Explorer (ICE) spacecraft. It passed comet 21P/Giacobini-Zinner at the downstream side, crossing the induced magnetotail \citep{Smith1986, Slavin1986a}.

Assuming that the IMF is always perpendicular to the comet-Sun line is not realistic, although it makes the modeling easier. However, the IMF statistically follows the Parker spiral \citep{Parker1958} and thus arrives at the comet at an angle, making the draping pattern asymmetric \citep{Volwerk2014}, as also observed at Venus \citep{delv17}. 

Sometimes chance encouters with cometary tails by spacecraft happen, such as in the case of Ulysses traversing comet Hyakutake's (C/1996 B2) ion tail \citep{Jones2000}. It was detected at a distance of more than 3.8 au from its nucleus and had a diameter of at least 7 million km.

It took 30 years for a planned investigation of a cometary tail. In the period from 24 March until 10 April 2016 \emph{Rosetta} was sent on a tail excursion. In this case the spacecraft moved up to only $\sim 1000$ km from the nucleus so the very near-region was studied. \citet{Volwerk2018a} looked at the draping of the field, the cone-angle $\theta_{\rm co}$ was calculated, and the distribution peaked between $60^{\circ}$ and $80^{\circ}$. This means that the ``draped'' magnetic field is more cross tail than along the Sun-comet-line as one would expect. This direction of the magnetic field is most-likely caused by the deflected solar wind magnetic field \citep{Koenders2016} as discussed above, which is transported to the downstream side of the comet. Eventually the field stretches out to the more regular tail structure. 

There is a non-radial component to the orientation of the plasma tail, as it will be abberated through the motion of the comet itself. The plasma tail will make an angle $\epsilon$, with the radial direction to the Sun given by: 

\begin{align}
\epsilon \approx \tan^{-1} \left\{ \frac{|V_{\rm c} sin(\gamma)|}{|V_{\rm sw}|} \right\}
\end{align}

\noindent
where $V_{\rm c}$ is the cometary orbital velocity, $\gamma$ is the angle between $V_{\rm c}$ and the anti-solar radial direction and $V_{\rm sw}$ is the radial component of the solar wind \citep{mend07}, where typical values are at $\epsilon < 6^{\circ}$.

It was found that ion tails often show acceleration of condensations (i.e.\ bright blobs of plasma in the tail), which exceed the gravitational pull of the Sun by 1 to 2 orders of magnitude. \citet{opik64} used \citet{bobr30}'s observations of comet 1P/Halley from 1910, to show that the usage of a correct coordinate system leads to better esitmates of the acceleration, as sometimes values $> 1000$ were published. The general acceptance was that the acceleration was caused by radiation pressure \citep[see, e.g.][]{wurm43}.

\subsubsection{Plasma acceleration}
One of the open questions in (cometary) tail physics is how fast the plasma is accelerated. Knots are bright patches of plasma that are created by disconnection events in the tail. Studies by \citet{nied81}, \citet{sait87}, \citet{Rauer1993}, \citet{kino96} and \citet{buff08} gave a broad range of velocities of these knots varying from 20 to 100 km/s. Acceleration of the knots were measured as 21 cm/s$^2$ \citep{nied81} and 17 cm/s$^2$ \citep{sait87}. 

\citet{yagi15} used the Subaru Telescope on Mauna Kea to observe the tail of comet C/2013 R1 (Lovejoy) and study the short time behaviour through short exposures, and thereby were able to determine the initial velocity of the knots to be between 20 and 25 km/s. This is smaller than the earlier studies, but there the initial velocity was determined by interpolation and may be inaccurate. There are reasons why the initial velocity can differ, e.g.\ the radial distance from the Sun or the heliospheric latitude. Significant changes in the width of the tail were also found over a time scale of 7 minutes, which makes high-time resolution imaging of cometary tails a necessity to understand the influence of solar wind variations on the tail.

\subsubsection{Waves and oscillations}
An interesting effect was found at distances $\geq 500$ km from the nucleus of comet 67P/Churyumov-Gerasimenko. \citet{Volwerk2018a} calculated the clock angle, defined as:

\begin{align}
\phi_{\rm cl} = \tan^{-1} \left\{ \frac{B_{\rm z}}{B_{\rm y}} \right\}.
\end{align}

\noindent
The unwrapped clock angle showed a continuously increasing angle $\phi_{\rm cl}$, even when the spacecraft turns around at apoapsis and moves in the opposite direction. This was interpreted as a traveling rotational wave along the magnetic field. The angular frequency of this wave whilst \emph{Rosetta} moved outbound(inbound) was $\omega_{\rm h} \approx 5.65$($5.83)^{\circ}$/h, and the spacecraft speed was $v_{\rm ros} \approx 1.3(3)$ m/s. Assuming \emph{Rosetta} observed a Doppler shifted wave, a phase velocity for the wave was determined at  $v_{\rm h} \approx 136$ m/s. With an average magnetic field of $B_{\rm m} \approx 10$ nT, this velocity does not correspond any known specific velocity, nor does the frequency correspond to any gyro frequency.

\citet{nist18} studied oscillations of cometary tails to investigate the interaction of the solar wind with a cometary plasma tail, especially for Sun grazing comets. Thereby, insight can be obtained in the solar wind conditions. Using images of comets Encke and ISON by the STEREO-a/HI-2 telescope they found oscillations of the two cometary plasma tails, which they interpreted as so called ``vortex shedding'' \citep[see e.g.][]{john04}. Modeling the tail as a driven and damped harmonic oscillator, the authors obtained information about the local solar wind velocity. Indeed, \citet{rama14} makes a case that ion tails can be used to investigate the the inner structure of the whole heliosphere. This opens a new window in solar wind studies from historical and contemporary observations \citep[see e.g.][]{rama22}

Tail dynamics, such as waves, kinks, detachments and including cometary ion rays, need to be studied through well-planned campaigns, not in the least because of the enormous size of a cometary tail. What actually causes detachments, what accelerates the condensations in the tail, what drives the kinks? To answer these questions, detailed in-situ observations, combined with ground observation campaigns are necessary.

\subsubsection{Historical solar wind}
Historical observations of cometary tails can also be used to investigate the historic solar wind conditions. One interesting historic period is the so-called Maunder Minimum (1645 - 1715). \citet{guly15} studied drawings of cometary plasma tails (type I tails) made during the Maunder Minimum. More than 20 comets were observed during that time, including comet 1P/Halley. The presence of plasma tails is directly related to the presence of a solar wind. \citet{guly15} states that, in principle it should be possible to obtain the solar wind velocity by looking at shifts of individual features in consecutive drawings, however these are very hard to distinguish. 

\citet{zolo18} studied descriptions of comets from the eleventh to the eighteenth century, but found from colour descriptions that mainly dust tails were described. Using the determination of the aberration $\epsilon$ as determined by \citep{bredikhin1879} an average value of $\sim 10^{\circ}$ was found, also indicating that mainly dust tails were seen, because the plasma tail is much less intense and thus more difficult to observe with the unaided eye. \citet{zolo18} concludes that probably the first identification of a plasma tail was done for the great comet C/1769 P1. However, \citet{haya21} discuss three cases, earlier than 1769, in which historical sources talk about two tails at comets C/1577 V1, and 1P/Halley in both the years 760 and 837. Comparing these descriptions with simulations of the cometary tails showed that it was indeed plausible that the plasma tail was observed already centuries before 1769. Indeed, \citet{silv21} show that observations of the great comet of 1577 is clearly described as having two tails in various sources.

\subsubsection{Sounding the plasma tail}
\citet{iju15} used distant radio sources and scintillation of the signal to determine the electron density in the tail of comet C/2021 S1 ISON. This interplanetary scintillation method has been regularly used to obtain the electron density in the solar wind and its structures such as ICMEs \citep{hewi64, gapp82}. This effect was first seen in a cometary tail (C/1973 E1 Kohoutek) by \citet{anan75}, and \citet{lee76} used that signal to estimate the electron density fluctuations in the tail with a value of $\Delta N_e \approx 80$ cm$^{-3}$. 

Using five consecutive nights, 13 - 17 November 2013, \citet{iju15} obtained scintillation measurements of one radio source through the cometary tail, from which they could derive the local electron density. There were no ICMEs passing through the field of view on 13, 16 and 17 November, but the electron density was determined for all nights. Normally, the number density increases towards the centre of the cometary tail \citep[see e.g.][]{meye86}, however what was measured was that the electron density increased as the radio source moved towards the edge of the cometary tail. This increase is thought to be created by an outburst of the comet \citep{combi14} or by turbulence on the boundary between the tail and the solar wind. 

\begin{figure*}
	\begin{center}
		\includegraphics[width=0.8\textwidth]{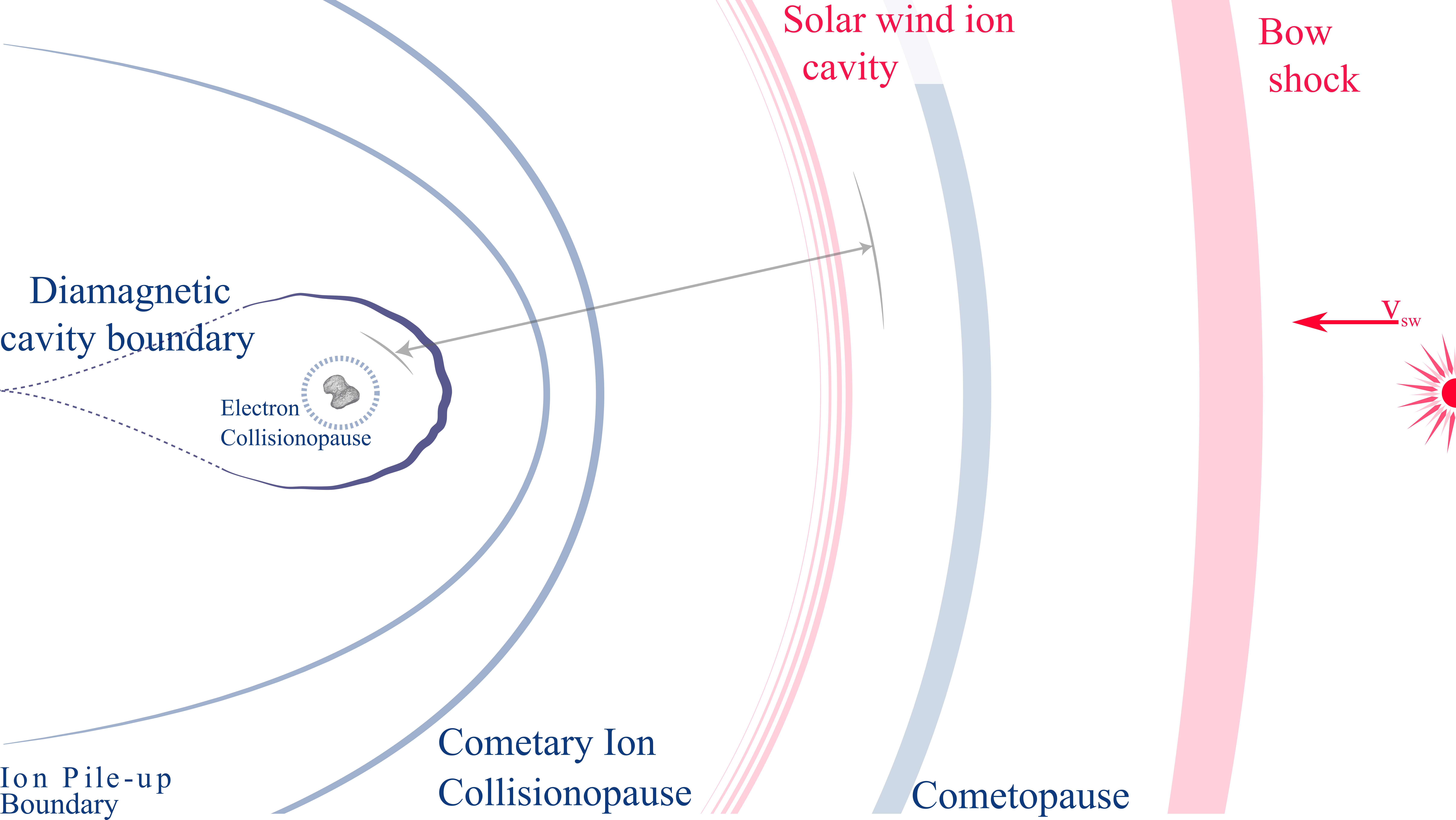}
		\caption{Overview of large scale boundaries observed by spacecraft at comets, not to scale.  The gray arrow indicates the boundaries covered by Rosetta. Credit: E. Behar, adapted from \cite{Goetz2021b}.} \label{bound_overview}
			\end{center}
\end{figure*}

\subsection{Induced Magnetospheric Activity}
A notable feature of comets are the near-linear regions of enhanced brightness emerging from the coma around the nucleaus, which bend backwards towards the tail \citep[e.g.,][]{wurm63}. These are so-called cometary or ion rays \citep{eddi10a, rahe68}. Attempts to understand the physics of ray formation did not start until  \citet{Alfven1957}'s paper on tail formation, which explains the bend-back of the rays, however, not the brightness variations. Observations of comet Morehouse, 1P/Halley and others showed that the rays are most likely created by ionization of \ch{H2O}, \ch{CO} and \ch{CO2} close to the nucleus, which happens in a discontinuous way \citep{Rahe1969}. Abrupt currents closing through the ``head'' of the comet, generated by tail reconnection (see Sect.\ \ref{sect:mrx}), can act as an extra source for ionization in the coma. \citet{wolf85} also considered cometary rays to be channeled outflow of cometary ions along magnetic flux tubes. However, flux tubes entering the cometary ionosphere through the Kelvin-Helmholtz instability on the ionopause \ref{sec:boundaries}. \citet{ip94} expanded on this model by positing that modification of the electron heat conduction is cause for variations along the flux tubes.

Estimates for the tail magnetic field strength were performed by \citet{Ip75}, who derived a magnetic field of  $\geq 100 \gamma$ (now nT) for comet C/1973 E1 Kohoutek, which is similar to the field measured at comet 21P/Giacobini-Zinner \citep{Slavin1986a} and around 67P \citep{Goetz2017}. This value was determined through observations of a helical structure that moved down the cometary tail \citep{hyde74}. Assuming that this wave moves at the Alfv\'en velocity, and that it was created by a kink-mode instability, \citet{hyde74} came up with a necessary tail current of $I \approx 2 \times 10^7$ A, which was later adjusted by \citet{Ip75} to $I \approx 2 \times 10^6$ A.

Similar helical waves were also observed in the tails of comets C/1973 E1 Kohoutek, C/1908 R1 Morehouse and C/1956 R1 Arend-Roland by \citet{ersh77}. These authors modeled the tail as a cylinder, which is separated from the solar wind by tangential discontinuities. They determined the dispersion relation for helical waves for an ideal compressible plasma, and for typical parameters the phase velocity of these waves was found to be close to the Alfv\'en velocity. The authors conclude that the origin of the waves lies in the Kelvin-Helmholtz instability.

Interest in tail currents was generated through the temporal variations in the observed brightness, e.g.\ in \ch{CO+} close to the nucleus \citep{wurm61} and the cometary rays. \citet{ip76} discussed how the folding of the cometary rays generates a strong tail field through accumulation of magnetic flux. Flux conservation leads to a field strength $B_{\rm t}$ of the tail:

\begin{equation}
B_{\rm t} \approx \frac{B_0 v_{\rm s} t_{\rm f}}{h_{\rm t}},
\label{eq:ipbtail}
\end{equation}

\noindent
where $B_0 \approx 5$ nT and $v_{\rm s} \approx 300$ km/s are the solar wind magnetic field and velocity and $h_{\rm t} \approx 5 \times 10^4$ km is the half-width of the tail. These assumptions lead to rather strong magnetic fields, $B_{\rm t} \approx 1500$ nT, but it was realised that this folding leads to, similar as in Earth's magnetotail, two regions of oppositely directed magnetic field, as in the model by \citet{Alfven1957}. Between these two lobes there needs to be a cross-tail current sheet, given by Amp\`ere's law:

\begin{align}
\nabla \times {\bf B} = \mu_0 {\bf J} + \frac{1}{c^2}\frac{\partial {\bf E}}{\partial t}.
\label{eq:ampere}
\end{align}

\noindent
\citet{ip79} assumed that (part) of this current can suddenly be channeled through the cometary atmosphere by a process to avoid a magnetic catastrophy. This could happen in a similar way as in Earth's magnetotail during a substorm \citep{bost74}. This current will then flow along the field lines towards the coma, and can be used to increase the ionization rate near the nucleus.

\citet{isra14} studied the magnetotails of Titan, Venus and comet 1P/Halley. Using the \emph{Giotto} flyby, they showed that the magnetic tension of the draped magnetic field lines, given by $T = ({\bf B}\cdot\nabla){\bf B}$ is always pointing tailward because of the specific geometry of the field. A mean tension of $T \ \approx 1600\ {\rm nT}^2/R_{\rm dc}$, with $R_{\rm dc} = 4000$ km, the radius of the diamagnetic cavity. With $B \approx 80$ nT they estimate a curvature radius of the field lines of $R_{\rm c} \approx 4 R_{\rm dc}$. This is important for tail detachment events discussed below.

The connection between the solar wind and the folding of cometary ion rays was clearly shown by \citet{degr08}, studying ion rays at comet C/2004 Q2 Machholz. They combinined telescopic U-band observations with SOHO solar wind measurements and suggested that the ion rays were formed and folded back through a sudden change of the solar wind from slow to fast.

\subsection{Magnetic Reconnection}
\label{sect:mrx}
The presence of oppositely directed magnetic field throughout the induced cometary magnetosphere, in the nested/dynamic draping upstream and the two lobes downstream of the nuclues gives rise to the assumption that reconnection events should be ubiquitous. 

As mentioned in Sect.\ \ref{sec:draping} changing directions of the IMF over time leads to a nested draping of magnetic field around the outgassing nucleus. The rotations of the magnetic field create current sheets, which can be observed when high-enough resolution data are available \citep[see, e.g.,][]{Volwerk2017}. These current sheets are usually the location where reconnection will occur.

There was much less nested draping observed by VEGA 1 \citep[see e.g.,][figure 3]{Volwerk2014}, and inside the coma the plasma instrument did not measure any solar wind ions (see Sect. \ref{ssec:sw_cavity}). But near closest approach the plasma instrument showed bursts of 100 - 1000 eV ions over a period of 5 minutes. \citet{veri87} showed the magnetic topology of the site at which the bursts were observed and proposed a reconnection region, which accelerates the ions.

\citet{kirs89} showed that during the magnetic field rotations there is a peak in field-aligned high-energy ions and a drop in the low-energy ions. They calculated the energy of the ions assuming that they are accelerated over 50000 km through reconnection and found a maximum energy of $E_{\rm i} \approx 22$ keV, but this is well below the observed energies of 97 - 145 keV. 

These are indirect indications that magnetic reconnection could be taking place in the nested draping region, and only a multi-probe mission could give a more precise answer. However, one should also realise the reconnection rate, as determined by \citet{swee56} and \citet{park57} is probaly very much reduced because of the high plasma densities in the coma. The rate is proportional to $1/\sqrt{\rho}$, where $\rho$ is the plasma density, for slow reconnection, and $\propto 1/\ln S$ for fast reconnection \citep{pets64}, where $S$ is the Lundquist (or magnetic Reynolds) number.

The magnetic field measurements by ICE showed a clear draping pattern of the magnetic field around the comet,  reminiscent of the Earth's magnetotail. The spacecraft passed through two regions of oppositely directed field, which were separated by a current sheet of $\sim 200$ km thickness \citep{Slavin1986b, Slavin1986a}. The current density was determined by \citet{mcco87a} and with that the ${\bf J}\times{\bf B}$-force was calculated, which was negative, re-accelerating the plasma tailward. Such a magnetotail, comparable with Earth's, will experience instabilities driven by external (e.g.\ by the solar wind) forces or internally through e.g.\ flux accumulation. These instabilities can give rise to magnetic reconnection and in the most drastic version a tail disconnection event can take place \citep[e.g.][]{Niedner1978} 

Solar wind sector boundaries interact with a comet's induced magnetosphere, as well as other structures such as ICMEs \citep[see e.g.,][]{moes14}, will influence the magnetospheric activity. \citet{Vourlidas2007} studied comet 2P/Encke interacting with an ICME using STEREO-A's SECCHI HI-1 instrument. 
As an ICME passes over the comet, part of the tail is disconnected and transported away by the solar wind. After the tail disconnetion, the magnetotail slowly grows again to ``full length.''

To further study tail detachment events one has to turn to numerical simulations to model the interaction of a cometary tail and the solar wind and its structures. \citet{jia09} modeled the tail detachment event of comet 2P/Encke. This interaction was modeled with an MHD code, which showed that a sudden $180^{\circ}$ rotation of the IMF caused a reconnection event at the upstream side of the comet. As the model was based on only a single species of ions, the speed of the evolution was slower than observed. A change to multiple ion species, cometary and solar wind ions, might improve the agreement with the observations. Also, this model only looks at one type of ICME-tail interaction. 

Comet 153/Ikeya-Zhang experienced multiple interactions with ICMEs \citep{jone04}. These interactions showed evidence of possible shocks travelling along the length of the tail, which created a disruption and a disconnection of part of the tail.

\citet{Edberg2016CIR} found that with the CIR interaction (see Sect. \ref{sec:cometosphere_external}) there was a sudden reduction in the piled-up magnetic field, combined with an increase in energetic electrons. This could be evidence a tail disconnection event the comet's ionospheric response. 

Similarly, there was an interaction with an ICME (see Sect. \ref{sec:cometosphere_external}), which compressed the magnetophere \citep{Edberg2016}. Large spikes on the magnetic field data were observed and these were assumed to be created through reconnection of the magnetic field lines of the ICME with the draped field lines around the nucleus. Due to the lack of plasma tail observations for 67P, it is not clear whether this interaction led to a tail disconnection event, and it need not have done.

Even though IMF reversals, through whatever means (sector boundaries, ICMEs, CIRs), or shocks can generate tail disconnection events, there is not a one-to-one correlation between the two. \citet{delv91} used the data from the VEGA 1/2 for the period of 1 December 1985 to 1 May 1986 (which includes the flybys at 6 and 9 March 1986 of comet 1P/Halley) and disconnection events observed by ground-based observatories. The authors correlated the crossing of sector boundaries of the solar wind with tail disconnection events. If changes in the solar wind \citep{jock85} generate day-side reconnection \citep{Niedner1978}, then one should find a good correlation. In all, \citet{delv91} determined that 50\% of the events showed a correlation between sector boundary crossing and tail disconnection. The authors also state that density enhancements in the solar wind are connected to events in the tail.

The first direct observation of an ICME interacting with a comet (but not the first one published in the literature) was done with the Solar Mass Ejection Imager (SMEI) looking at comet C/2001 Q4 (NEAT) in 2004 \citep{kuch08}. The ICME first created a kink in the tail, which moved anti-sunward, which developed into knots. This seems to be a general behaviour, as seen in a handful of other events. The generation of the knots, evidence for the disruption of the tail, can result from various effects like shocks or polarity reversals of the magnetic field in the ICME.

Comparison of events at comet C/2001 Q4 (NEAT) with those of comet C/2002 T7 (LINEAR), and tracing them back along the Parker spiral shows that these occur preferably when the comet is near the heliospheric current sheet \citep{kuch08}. \citet{bran99} and \citet{bran00} also concluded this effect using observations of comets 1P/Halley, 122P/de Vico, C/1996 B2 Hyakutake, and C/1995 O1 Hale-Bopp. 

\citet{li18} performed laboratory experiments to study tail detachment events. They used a laser-driven plasma which collides with a cylindrical obstacle. Behind the obstacle detached tails are measured. At the same time numerical simulations of this interaction were performed. However, it should be noted that this was an unmagnetized interaction. The experiment showed that an electrostatic field is created in the plasma when the density is high, because of a temperature difference between the ions and the electrons. This field leads the plasma ions to converge in a tail and move away form the obstacle, giving the impression of a detached tail. The authors argue that this is a viable process to explain a detached tail, because the physical sizes related to cometary effects are shorter than the gyro radius of cometary ions in the solar wind, invalidating an MHD approach.

\section{Boundaries/Regions}
\label{sec:boundaries}

\subsection{Overview}
\label{ssec:overview}

Comet interactions with the solar wind begin at great distances from the comet because the high neutral outgassing rate leads to mass-loading of the solar wind \citep{Galeev1988}. Closer to the nucleus, this interaction leads to the formation of large scale structures in the form of boundaries that separate regions characterized by plasmas with differing parameters. Prior to spacecraft flybys of comets, computer simulations of the solar wind interaction with the comet predicted the existence of two permanent boundaries: a bow shock and a contact surface \citep{SchmidtWegmann1982,IpAxford1982}.
These boundaries established three regions of comet-solar wind interaction: an upstream region outside of the bow shock, a region between the bow shock and contact surface termed the cometosheath, and a diamagnetic cavity located between the contact surface and the nucleus. After spacecraft made in situ observations during several flybys and the escort of comet 67P by \emph{Rosetta}, several other boundaries were also observed: the cometopause, an ion pileup boundary, and ion-neutral and electron-neutral collisionopauses. Plasma interaction boundaries observed at a comet can be permanent features, solar wind and interplanetary magnetic field boundaries, or small-scale transient features created by waves or instabilities \citep{Cravens1989}.
Permanent features include shocks and collisionopause boundaries. Shocks form when the relative velocity of the plasma and an obstacle, in this case the cometary plasma, exceed the characteristic speed of waves in the plasmas. As discussed in Sect. \ref{ssec:in_collisionopause}, a collisionopause forms when collisions between neutrals and ions or electrons change characteristics of the plasma such as composition or velocity \citep{Mendis1986,Cravens1989,Cravens1991}.
\cite{Cravens1991} outlined various types of collisionopause for both ions and electrons depending on the collision processes and the reactions that may occur, including ion-neutral charge transfer and ion-neutral chemistry. Fig. \ref{bound_overview} illustrates the boundaries that are discussed in this section. We review here observations of five boundaries that were determined to be permanent features \citep{Gringauz1991}.

\subsection{Bow Shock}
\label{ssec:bow_shock}

\begin{figure}
	\centering
		\includegraphics[width=\columnwidth]{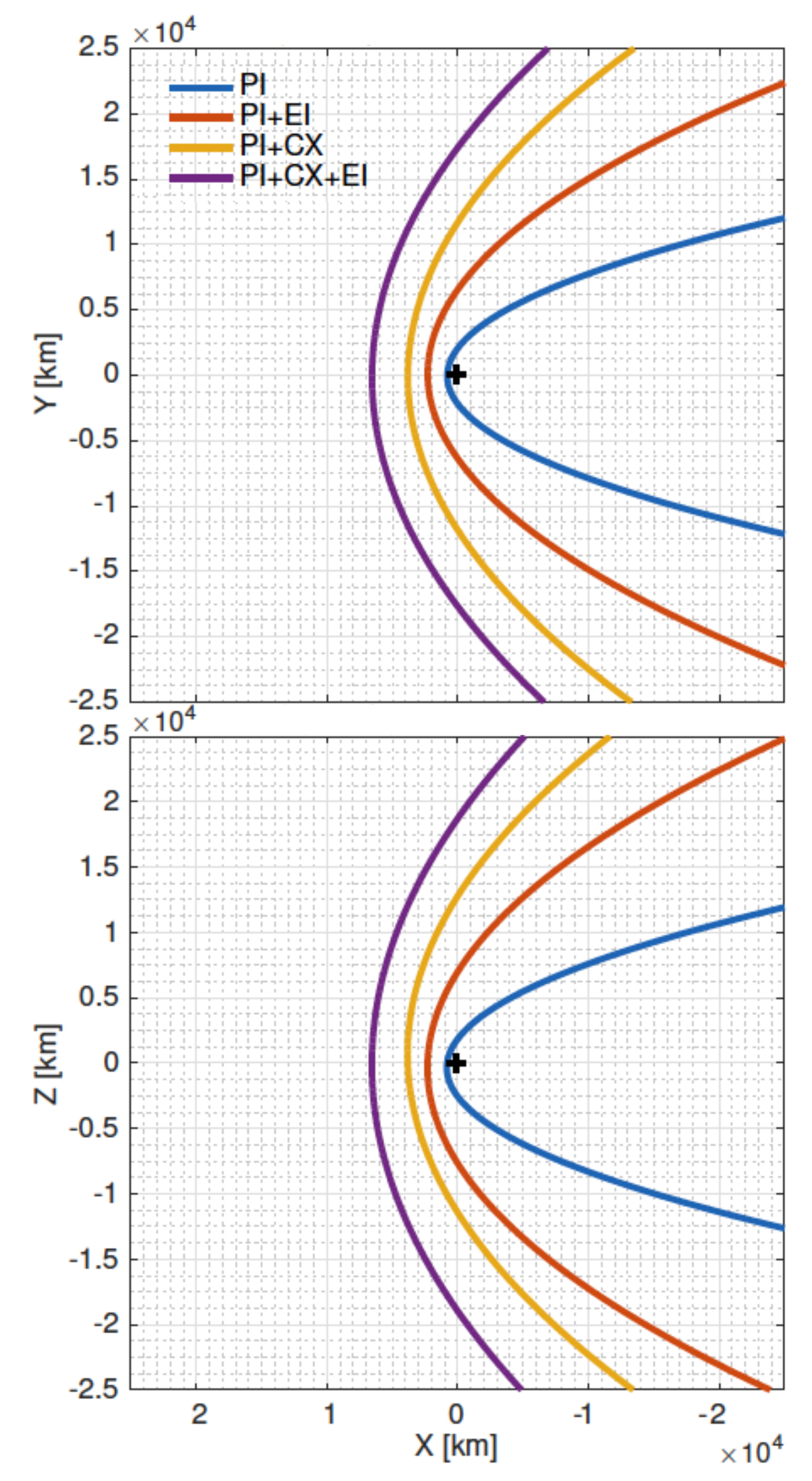}
		\caption{Illustration of how the simulated distance of the bow shock from the nucleus increases compared to simulations with only photo-ionization (PI) when the influence of electron-impact ionization (EI) and charge-exchange processes (CX) are included (Adapted from Fig. 11 in \citet{SimonWedlund2017}).} 
		\label{ionization}
\end{figure}

The bow shock is the location where the solar wind flow transitions from supersonic to subsonic as a result of mass loading. Simulations predicted, and spacecraft observations confirmed that this boundary would be broad and weak compared to planetary bow shocks because the slowing of the solar wind begins well upstream of the shock (see also Sect. \ref{ssec:magnetic_structure}). Spacecraft observations have noted sudden changes in the plasma density and velocity across the bow shock and an enhancement of the magnetic field \citep{Galeev1988}. The detailed physics of the cometary bow shock formation and early modeling efforts are described in Section 3 of \cite{Ip2004} while spacecraft observations prior to \emph{Rosetta} are outlined in Section 4.1 of \cite{Ip2004}. \emph{Rosetta}'s exploration of the plasma environment around 67P has significantly advanced our understanding of cometary bow shocks for weakly outgassing comets. Hybrid modeling of the comet found that including electron impact ionisation and charge-exchange processes in addition to photo-ionization increased the standoff distance of the bow shock along the Sun-comet line by more than a factor of six, as illustrated in Fig. \ref{ionization}. Furthermore, simulations also showed that asymmetric outgassing and illumination driven outgassing can extend the standoff distance even further \citep{Huang2016,alho_hybrid_2020}. The \emph{Rosetta} spacecraft did not travel far enough from the comet to cross the bow shock during outgassing rates relevant to these simulations.

\begin{table*}[!t]
\centering
\begin{tabular}{||c|c|lclclclcl||}
\hline
Number & Type & \multicolumn{9}{c|}{Example}\\
\hline
R1 & Ion-neutral charge transfer & H$_2$O$_\text{fast}^+$ &+& H$_2$O &$\rightarrow$& H$_2$O$_\text{fast}$ &+& H$_2$O$^+$&&\\
R2 & Ion-neutral chemistry & H$_2$O$^+$ &+& H$_2$O& $\rightarrow$ &H$_3$O$^+$& +& OH&&\\
R3 & Ionization & H$_\text{fast}^+$& +& H$_2$O &$\rightarrow$& H$_\text{fast}$& +& H$_2$O$^+$& +& e$^-$\\
R4 & Electron removal & O &+& H$_2$O &$\rightarrow$& O$^+$& +& H$_2$O& + &e$^-$\\
R5 & Ion-electron recombination & H$_3$O$^+$& +& e$^-$ &$\rightarrow$& OH &+& 2H&&\\
R6 & Ion-ion coulomb interaction & H$_2$O$^+$& +& H$_2$O$^+$& $\rightarrow$& H$_2$O$^+$& +& H$_2$O$^+$&&\\
\hline
\end{tabular}
\caption{Summary table of collisional processes that can take place in the cometary plasma.}
\label{tab: CollisionRxns}
\end{table*}

The greatest advantage of \emph{Rosetta} in advancing our understanding of cometary bow shocks was thanks to the extended amount of time spent escorting the comet around the Sun. This allowed detection of the bow shock as it began to form, or an infant bow shock \citep{Gunell2018a}. Statistical analyses of the full \emph{Rosetta} dataset showed that the spacecraft encountered the infant bow shock multiple times between 3.0 and 1.7 AU inbound to the Sun and then again outbound starting at around 1.8 AU \citep{Goetz2021}. Additionally, a method for detecting the existence of a bow shock from inside the shock was discovered. The energy of ions at the spacecraft location that were accelerated by a constant electric field upstream of the spacecraft is proportional to the distance from the place where the ions were initially created. This was used to determine that a bow shock had formed at a distance of $\sim 4000\,$km from the nucleus when the comet was 1.4 AU from the Sun \citep{Nilsson2018}. This was confirmed with hybrid simulations \citep{Alho2019} that were later used to determine the spectral features in ion energy that could indicate the presence of a bow shock upstream of the spacecraft \citep{alho_hybrid_2020}. 
Similarly, a bimodal ion distribution found in the environment of comet 1P/Halley indicated that ions produced before and after the bow shock are observed as two distinct populations, indicating that the acceleration history of the ions is preserved \citep{Thomsen1987}.

Although we have learned a great deal about cometary bow shocks, we are lacking observations that cover sufficient space and production rate to fully understand the transition between an infant bow shock and a full shock as well as the implications of asymmetric outgassing on the location and shape of the bow shock. Future observations should include simultaneous measurements at multiple points in space \citep{Snodgrass2019} and multi-spacecraft missions over long time periods \citep{Goetz2019wp}, which should be combined with simulations that treat electrons and ions kinetically resolving all relevant scales \citep{Balogh2013}.

\subsection{Cometopause}
\label{ssec:cometopause}
The concept of the cometopause is introduced in Section 4.3 of \cite{Ip2004}, which discusses ion properties inside of the bow shock over a wide distance range. We have learned a great deal more from \emph{Rosetta} about the interaction between the solar wind and the coma inside of the cometopause. We outline here what new understanding we have gained about the cometopause and introduce new regions and boundaries that were discovered by \emph{Rosetta}.  

The cometopause is described as the boundary where the ion composition changes from predominantly solar wind ions to predominantly cometary ions \citep{Gringauz1986,Mendis1989,Coates1997}. Although some dispute the existence of this boundary \citep{Reme1994}, it was determined by several researchers to be a permanent feature \citep{Gringauz1991,Sauer1995}. Several authors have proposed that the cometopause is the location where collisions causing charge exchange between solar wind protons and cometary neutrals become dominant \citep{Gringauz1986,Gombosi1987,Cravens1989,Ip1989}, a form of collisionopause boundary ((see Table \ref{tab: CollisionRxns}) and \ref{ssec:in_collisionopause}). However, this is not an appropriate explanation for the cometopause as explained in \ref{ssec:in_collisionopause}. Instead, as theoretical simulations suggest, the cometopause is best explained by deflection of solar wind protons \citep{Sauer1995}. \emph{Giotto} observed plasma flux deflection that indicated plasma flow forced around a comet \citep{Perez1989}, suggesting that the cometopause could be the beginning of the solar wind ion cavity (see Sect. \ref{ssec:sw_cavity}). Previous work stated that \emph{Rosetta's} instruments did not observe the cometopause \citep{Mandt2016}, but if this is the boundary of the solar wind ion cavity then it may have been observed several times as it was forming \citep{Behar2017}. Further work is needed to better understand the physics of this region of the cometosphere.

\subsection{Solar wind ion cavity}
\label{ssec:sw_cavity}
\begin{figure}
	\begin{center}
		\includegraphics[width=\columnwidth]{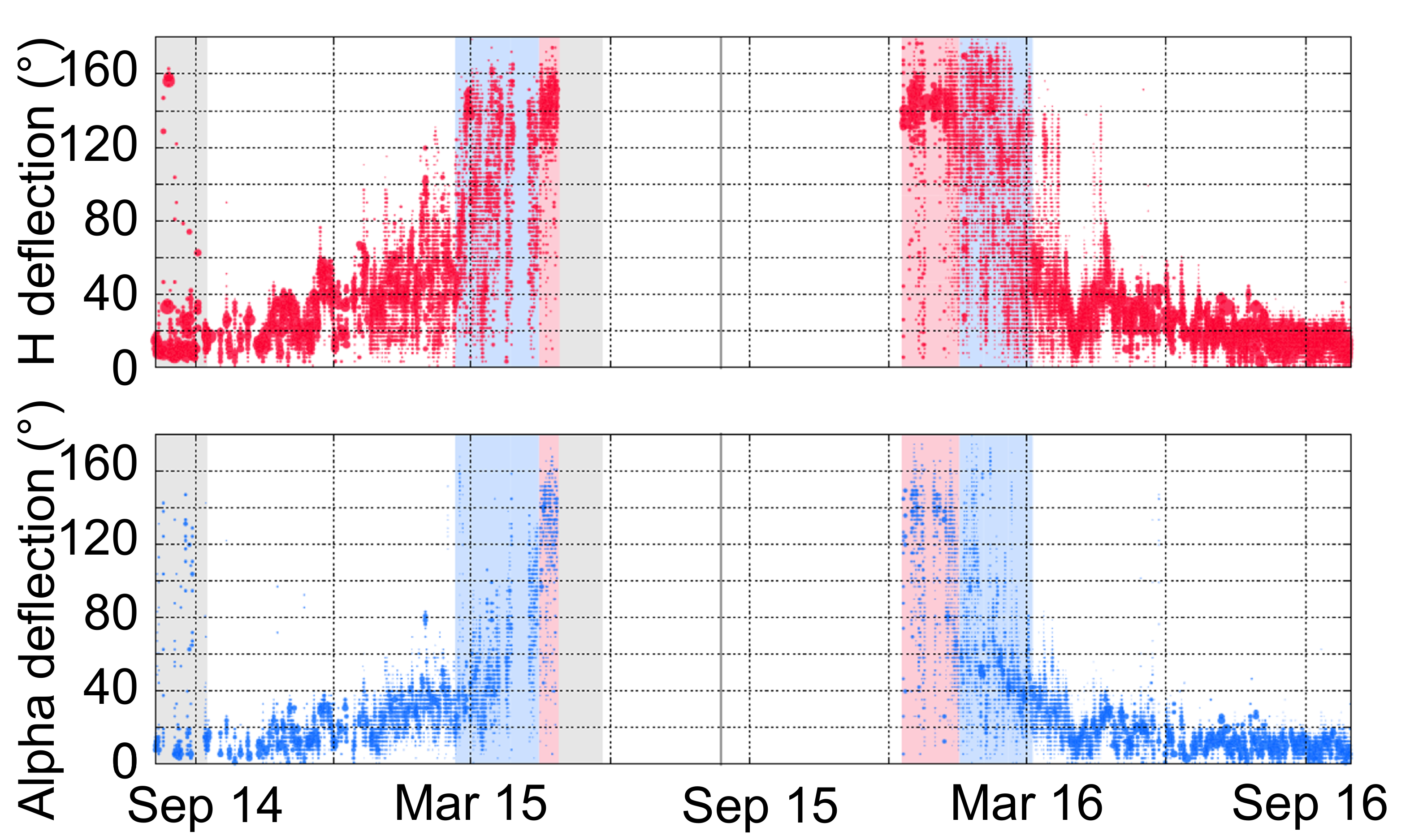}
		\caption{Observations of the increasing deflection of solar wind protons and alpha particles as comet 67P approached the Sun from September 2014 to May 2015, the formation of the solar wind ion cavity between May and December 2015, and the decreasing deflection of solar wind ions as the comet moved away from the Sun after December 2015 (adapted from Fig. 1 of \citet{Behar2017}).}
		\label{deflection}
	\end{center}
\end{figure}

From May to December 2015 \emph{Rosetta} was located in a region that was mostly free from solar wind ions, a region termed the solar wind ion cavity \citep{Nilsson2017,Behar2017}. This cavity was determined to have formed as a result of the deflection of the solar wind ions by magnetic pileup that results from mass loading of the solar wind. Prior to the disappearance of the solar wind ions, the ion instruments observed deflection of the solar wind that increased over time as shown in Fig. \ref{deflection}. When the ion deflections were greater than 90 degrees, ion observations became less frequent and had lower densities until they disappeared. Deflections were observed to be very large near the boundary of the cavity, possibly in the region of the cometopause. A simple analytical model that estimates the global dynamics of solar wind protons for a given heliocentric distance provided strong agreement with observed deflection angles during \emph{Rosetta's} nightside excursion, validating understanding of how the solar wind ions gyrate withing the coma and how they are repelled from the inner-most region to form an ion cavity \citep{behar2018aa_ns}. Although the solar wind ions are deflected away from the solar wind ion cavity, cometary ions picked up by the solar wind and the solar wind magnetic field are still present even though the pickup ions are deflected in a similar manner to the solar wind ions \citep{Nilsson2020}. The boundary also does not constitute a discontinuity in the momentum budget, but just a change in the composition of the plasma flow \citep{Williamson2020}. \cite{Edberg2016} showed that the impact of an ICME can compress the solar wind ion cavity, indicating that the solar wind dynamic pressure regulates the size of this region.

\subsection{Ion-neutral collisionopause}
\label{ssec:in_collisionopause}
Collisions within an atmosphere or coma influence the dynamics and chemistry of neutrals and ions. The term collisionopause is used to define a boundary inside of which collisions play a dominant role. We describe in this section ion-neutral collisionopause boundaries, and cover the electron-neutral collisionopause in Sect. \ref{ssec:en_collisionopause}. 

The location of an ion-neutral collisionopause is calculated in a manner similar to determine the location of the exobase in aeronomy, or the study of neutral atmospheres. In the case of a steady state atmosphere, the exobase is the location is where the Knudsen number, or the ratio of the mean free path, $\lambda$ to the scale height, is one.

\begin{align}
\lambda = \frac{1}{n_n\sigma}.
\end{align}

\noindent
where $n_n$ is the neutral density and $\sigma$ is the collision cross section. In aeronomy, when determining the location of the neutral exobase the cross-section is the species-specific neutral collision cross-section. This means that each species has its own exobase. For an ion-neutral collisionopause, the cross-section would be the collision cross-section for the relevant ion species with water. In the coma, the neutral density, $n_n$, as a function of distance from the comet can be approximated as

\begin{align}
n_n(r)=n_\text{s/c} \left( \frac{r_\text{s/c}}{r} \right)^2
\end{align}

\noindent
where $r$ is the distance from the nucleus, $n_{s/c}$ is the density at the spacecraft, and $r_{s/c}$ is the distance of the spacecraft from the nucleus. 

The scale height is the distance over which the density of a neutral species or the plasma reduces by a factor of $e$. \emph{Rosetta} observations suggested that the plasma scale height at the location of \emph{Rosetta} when escorting 67P could be approximated by the distance of the spacecraft from the comet, $r$ \citep{Edberg2015}. Therefore the location of both ion-neutral and electron-neutral collisionopause boundaries could be found by setting the mean free path equal to the distance from the comet

\begin{align}
\lambda = r = n_\text{s/c} \sigma r_\text{s/c}^2
\label{eq:in_dist}
\end{align}

In the case of \emph{Rosetta}, the neutral density was measured at the spacecraft location, making this a reasonable equation to use for determining where a collisionopause would be located based on local measurements. A collisionopause distance can also be calculated by substituting for $n_{s/c}$ with the outgassing rate, $Q$, which is a function of $n_{s/c}$ and the neutral outflow velocity, $v_n$

\begin{align}
Q = 4 \pi r^2n_\text{s/c}v_n
\end{align}

\cite{Cravens1991} outlined various types of collisionopause for both ions and electrons depending on the collision processes involved, which are outlined in Table \ref{tab: CollisionRxns}. Reactions of type R1 are charge transfer reactions where the charge is transferred from a fast ($< 300$ km/s) ion in the mass loaded solar wind flow (e.g. H$^+$ or H$_2$O$^+$) to a neutral travelling away from the comet at velocities of $1\,$km/s or less. As a result of this reaction, the ion becomes an energetic neutral, and the bulk velocity of the ions is reduced. Chemical reactions like R2 not only reduce the bulk velocity of the ions, but also alter the relative composition of the ion population. For example, in reaction R2, a proton is transferred from H$_2$O$^+$ to H$_2$O, creating an H$_3$O$^+$ ion, increasing the bulk mass 19/18 ratio of the ion population.  If H$_2$O$^+$ is fast it will end as an energetic OH neutral while H$_3$O$^+$ will have the same energy as the neutrals. Another chemical reaction that should be noted is the proton transfer from H$_3$O$^+$ to NH$_3$ producing NH$_4^+$ in the dense coma \citep{Beth2016}. This reaction depends on the volume mixing ratio of NH$_3$ in the coma and would reduce the mass 19/18 ratio, countering the effect of producing H$_3$O$^+$. 

\begin{figure}
	\begin{center}
		\includegraphics[width=\columnwidth]{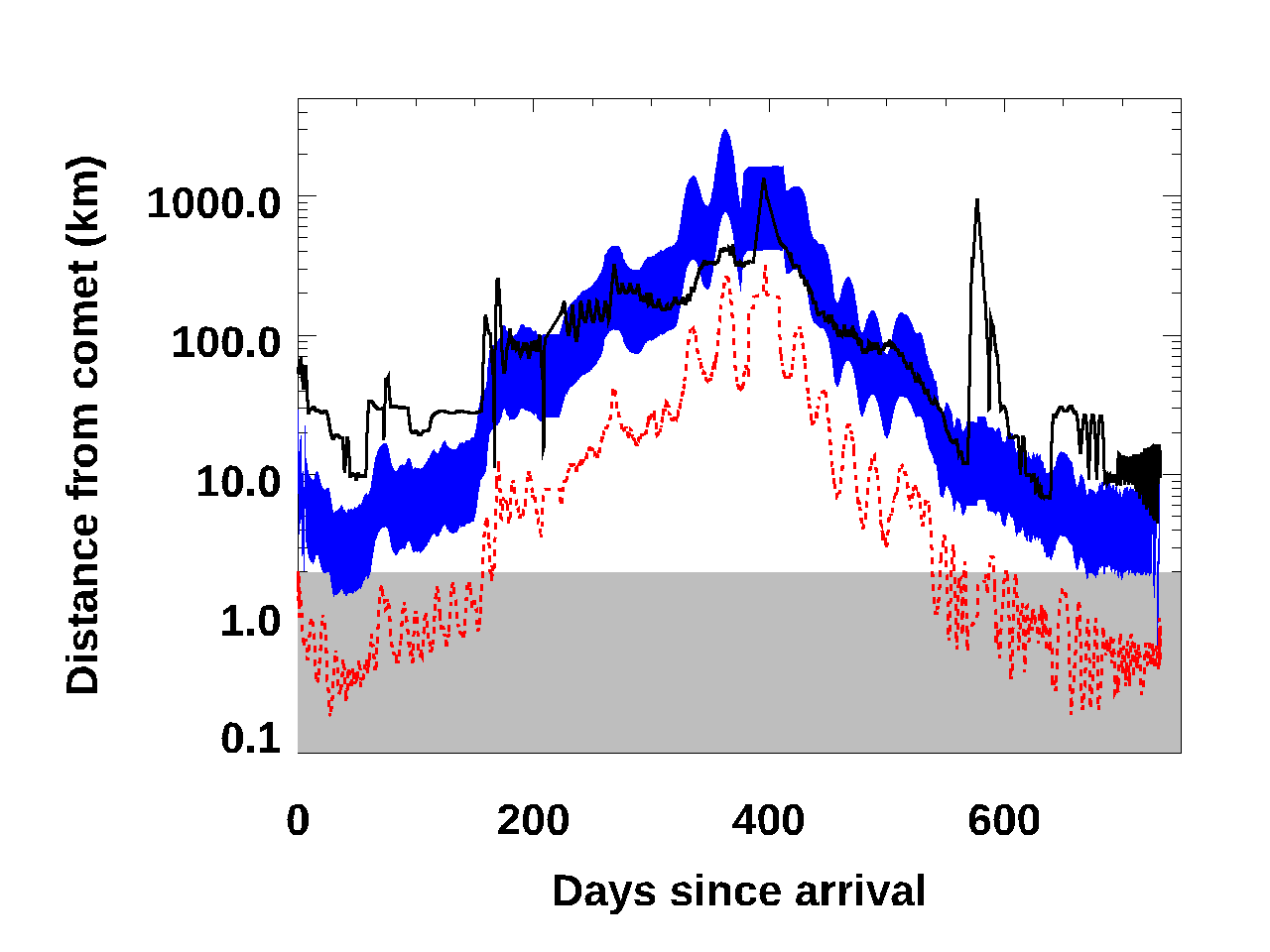}
		\caption{Calculated location of the cometary ion collisionopause (blue shaded region) and the electron exobase (red dashed line) compared to the distance of \emph{Rosetta} (black line) from the comet nucleus (gray shaded region).}
		\label{collisionopause}
			\end{center}
\end{figure}

As outlined in \cite{Mandt2019}, the collision cross-section is the greatest source of uncertainty for calculating ion-neutral collisionopause distances. In early studies, the ion-neutral cross-section was estimated to be $2 \times 10^{-15} \text{cm}^2$ for solar wind ions and $8 \times 10^{-15} \text{cm}^2$ for mass loaded solar wind with a bulk composition of $H_2O^+$ \citep{Mendis1986}. We illustrate in the shaded blue region of Fig. \ref{collisionopause} the calculated location of the collisionopause throughout the \emph{Rosetta} mission based on these cross-sections. Note that because the solar wind ion cross-section is smaller than the cross-section for water ions, the collisionopause location calculated with Eq. \eqref{eq:in_dist} for charge transfer of solar wind ions will be closer to the nucleus than the collisionopause for picked up cometary ions. \emph{Rosetta} made several observations of a boundary where the bulk ion velocity transitioned between values greater than 10 km/s outside of the boundary to velocities too low to measure directly because the ions were seen at the value of the spacecraft potential that were interpreted to represent a cometary ion collisionopause \citep{Mandt2016}. However, most of these observations were within the solar wind ion cavity \citep{Mandt2016}, meaning that the spacecraft was inside of the cometopause. This demonstrates that the cometopause is not a collisionopause and is instead a boundary formed as a result of deflection of the solar wind ions.

Furthermore, although the boundary was observed within the range of predicted locations illustrated in the shaded blue region of Fig. \ref{collisionopause}, it appeared to vary in distance depending on solar wind dynamic pressure \citep{Mandt2016} similar to the cometopause \citep{Edberg2016}. Laboratory measurements show that the cross-sections for reactions (R1) and (R2) depend on the energy of the ions \citep{Lishawa1990,Fleshman2012}. A comparison of these cross-sections as a function of energy with the \emph{Rosetta} observations of cometary ion collisionopause crossings \citep{Mandt2016}, observations made by \emph{Giotto} at 1P/Halley \citep{Schwenn1988,Altwegg1993}, and observations made by Deep Space 1 at 19P/Borelly \citep{Young2004} found that the  collisionopause had been observed at all three comets and appeared to form as a result of a combination of reactions (R1) and (R2) \citep{Mandt2019}. 

Many questions remain about this region of the cometosphere and the role of collisions in influencing the dynamics of cometary ions that have been picked up by the solar wind. A greater understanding of cross sections is needed and more observations of ion and neutral composition would be of high value. Additionally, measurements of the electron temperatures are needed to constrain the electron recombination rates in this region and inside of the collisionopause.

\subsection{Electron-neutral collisionopause}
\label{ssec:en_collisionopause}
Collisions between electrons and neutrals in the coma will cool the electrons to temperatures similar to the neutral gas temperature \citep{Eberhardt1995}. The electron-neutral collisionopause, also called the electron exobase, represents the boundary outside of which collisions do not efficiently influence the electron temperatures. The location of the electron exobase calculated using Eq. \eqref{eq:in_dist} and a cross-section of $5 \times 10^{-16} \text{cm}^2$ is illustrated in Fig. \ref{collisionopause}. As shown here, \emph{Rosetta} was not expected to cross inside of the electron exobase. However, additional processes can enhance the electron collision rate, and thus electron cooling can extend the location where electrons thermalize to the same temperature as the neutrals beyond the calculated exobase location \citep{Henri2017}. These processes include rotational and vibrational cooling for electrons orginating close to the nucleus where electrons pass through the densest gas \citep{Engelhardt2018}, and possibly through an ambipolar electric field either trapping electrons in the dense neutral region \citep{Madanian_etal_2016,Vigren2019} or increasing the electron collision cross-section by slowing their outward movement \citep{Engelhardt2018}. 

At several points during the mission \emph{Rosetta's} instruments observed a population of cold electrons that at times made up as much as 90\% of the electron population \citep{Engelhardt2018,Gilet2020a}. \cite{Henri2017} found a relationship between the location of the electron exobase and the diamagnetic cavity boundary (DCB, Sect. \ref{ssec:cavity}).  These observations confirmed that electron cooling is more efficient than suggested by the simple approximation shown in Eq. \eqref{eq:in_dist} and that a more comprehensive cooling model is required to estimate the location of the electron exobase (see also Sect. \ref{sec:cometosphere}). Furthermore, because \emph{Rosetta} did not explore the coma at a close enough distance to the nucleus, in situ measurements of the electron neutral collisionopause are lacking. Future exploration of this region, measuring the ion and neutral composition and the electron densities and temperatures would be of high value.

\subsection{Diamagnetic cavity}
\label{ssec:cavity}
The mass-loading of the solar wind by cometary ions was predicted to decelerate the solar wind until the velocity reached zero upstream of the comet under high outgassing rates \citep{Biermann1967}. Because the solar wind magnetic field is tied to the solar wind plasma, the magnetic field was predicted to come to a stop creating a magnetic field free region around the nucleus \citep{SchmidtWegmann1982,IpAxford1982}. The first confirmation that a diamagnetic cavity could form in the conditions provided by a coma was during the Active Magnetospheric Particle Tracer Experiment (AMPTE). The magnetic field briefly dropped to zero after release of a Barium cloud, confirming the formation of a field-free region \citep{Luehr1988}. The boundary of this region was found to be where the electron thermal pressure equalled the magnetic field pressure.

The \emph{Giotto} spacecraft flew close enough to comet 1P/Halley to pass within the predicted region where the diamagnetic cavity was expected to exist. The magnetic field measurements showed that the field magnitude dropped to and remained at close to $0\,$nT around $5000\,$km from the comet nucleus. The observations suggested that the boundary of the diamagnetic cavity was not symmetric about the nucleus, and may have ripples across its surface \citep{Neubauer1988}. The position of the boundary was expected to be determined by a balance of outward ion-neutral drag and inward magnetic gradient forces \citep{Gringauz1991,Coates1997}. 
Besides Giotto at 1P/Halley, no other spacecraft directly observed the diamagnetic cavity previous to the Rosetta mission. However, a decrease in HCO$^+$ emissions in the innermost coma of comet Hale-Bopp was speculated to be associated to the diamagnetic cavity. As Hale-Bopp is much more active than 1P/Halley, the diamagnetic cavity was predicted to be around $500000\,$km in size, which coincided with a drop in emissions \citep{Womack1997}.
It should be noted that in early works the solar wind ion cavity boundary and the diamagnetic cavity boundary were identified to be in the same place. However, Rosetta observations have clearly shown that the solar wind ion cavity is most of the time much larger than the diamagnetic cavity. This is because the solar wind ions are deflected upstream and the pick up ions take over the role of the solar wind flow and carry the magnetic field into the inner coma where it is eventually stopped at the diamagnetic cavity boundary.

Prior to the Rosetta mission, several modelling studies for 67P were conducted to project the location of the diamagnetic cavity. Single fluid MHD simulations predicted that the cavity would form around 50 km from the nucleus within two months of perihelion, while multifluid simulations showed that the boundary could be extended on the sunward side by an asymmetric outgassing profile \citep{Huang2016}. Hybrid simulations agreed with these results. However, clear signatures of a magnetic field free region were observed at distances as great as 400 km from the nucleus much earlier in the mission when the comet was within 2 AU of the sun inbound and continued until the comet had reached 2.4 AU from the Sun outbound \citep{Goetz2016a,Goetz2019}. In fact, the diamagnetic cavity was crossed into by the spacecraft over 700 times, indicating a highly variable boundary. The extension of the diamagnetic cavity was shown to be dependent on the gas production rate, with higher gas production rates leading to larger sizes of the diamagnetic cavity. 

The ions inside the diamagnetic cavity are usually found to be quite constant in density and velocity, except for sporadic enhancements (see below). The ion velocity of the constant low velocity population is around $5 - 10\,$km/s \citep{Bergman2021}. This is significantly higher than the neutral gas velocity \citep[$\leq 1\,$km/s,][]{Hansen2016}, which indicates that there must be an acceleration mechanism for those ions. It is speculated that an ambipolar field caused by the electron pressure gradient of the expanding plasma accelerates these ions \citep{Vigren2019}. At the same time, the ions cannot collide frequently with the neutral gas, as that would cause ion cooling and a reduction in velocity. 

This is quite interesting since for comet 1P/Halley, the collisions between neutral gas and ions were identified as the process that prevents the magnetic field from diffusing or convecting into the diamagnetic cavity \citep{Cravens1987}. Therefore another process must be at play at comet 67P, but it is still an open question which one it is. 

A series of asymmetric, steepened waves were visible in the magnetic field and plasma density outside of the boundary \citep{Goetz2016a,StenbergWieser2017,Hajra2018,Ostaszewski2020} and in the shape of the boundary itself \citep{Goetz2016}. The unmagnetised plasma density inside of the cavity scales well with the neutral density \citep{Henri2017}. Observations of dense plasma events when the spacecraft was inside of the boundary \citep{Hajra2018,Masunaga2019} provided indications that plasma from outside of the diamagnetic cavity boundary could possibly penetrate into the cavity. 

The electron environment changes at the diamagnetic cavity boundary: electrons in an energy range of $100-200\,$eV are ubiquitous outside of the diamagnetic cavity, but depleted inside the diamagnetic cavity \citep{Nemeth2016}. These electrons probably originate from the solar wind (strahl electrons) and are tied to the magnetic field. As the magnetic field decreases into the diamagnetic cavity, the electrons are adiabatically transported and become field aligned. Therefore, without a perpendicular component, they cannot cross field lines and enter the diamagnetic cavity \citep{Madanian2020}. 

The relationship between observations of the electron exobase and the diamagnetic cavity boundary \citep{Henri2017} suggest that the mechanism determining the distance of the boundary from the nucleus is related to electron neutral collisions and that the location of the boundary could change quite rapidly as a result of instabilities. \citet{Huang2018} showed that the introduction of a Hall term in an MHD multi-fluid model could also lead to an extension of the diamagnetic cavity as well as the formation of filaments extending away from the main diamagnetic cavity boundary, similar to an instability as described above.
In general, the \emph{Rosetta} observations were unexpected and many questions remain about the mechanisms involved in forming and determining the location of the dimagnetic cavity boundary. In particular, the pressure balance at the boundary is not well understood, nor the role of the changing plasma environment in the movement of the boundary.
Why is there evidence for ion acceleration at 67P within the cavity but not at 1P/Halley?
The processes that cool and/or accelerate electrons and ions in and near the cavity are poorly known, as well as the process for transmitting plasma enhancements through the boundary and inside the cavity. Finally, we do not know the origin of the asymmetry in the in- and outbound crossing of the boundary. Answering these questions would likely require multi-point measurements of the diamagnetic cavity, its boundary, and the upstream plasma conditions \citep{Goetz2019}.

\section{Waves}
\label{sec:waves}

\begin{figure*}
    \centering
    \includegraphics[width=\textwidth]{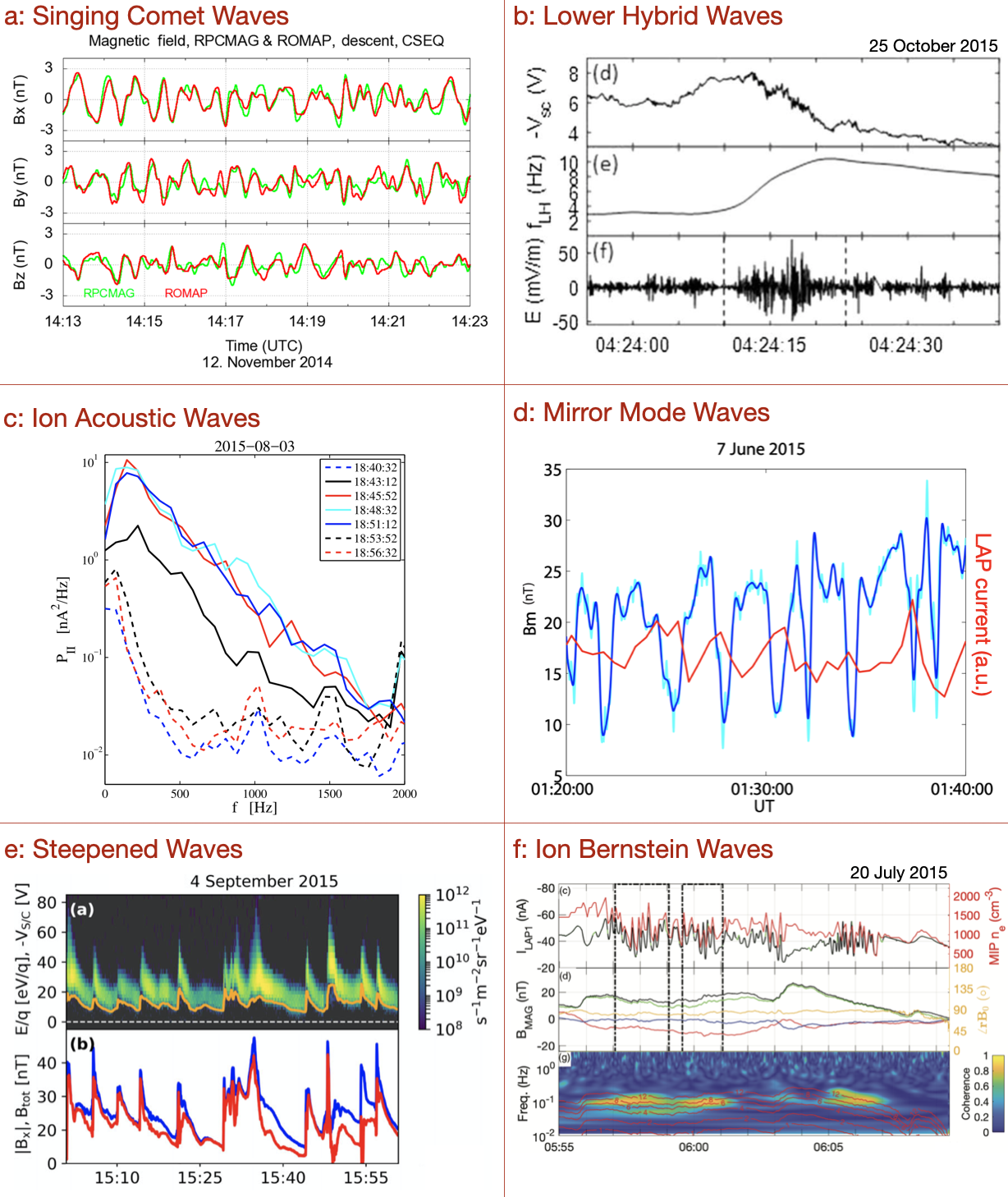}
    \caption{Examples of wave observations at comet 67P. Note that all panels show timeseries, except for panel c) which displays a frequency spectrum for better visibility. For descriptions see text. Credits: a) \cite{Richter2016}, Annales Geophysicae; b) \cite{Karlsson2017}, Geophysical Research Letters; c) \cite{Gunell2017a}, MNRAS; d) \cite{Volwerk2016}, Annales Geophysicae; e) \cite{StenbergWieser2017}, MNRAS; f) \cite{Odelstad2020}, Journal of Geophysical Research}
    \label{fig:waves}
\end{figure*}

In general, waves in plasmas are oscillations in the properties of a coupled system. Since any oscillations have to adhere to the plasma equations, wave modes are discrete. Different approximations of the plasma will result in different wave dispersion relations, thus it is important to always check the underlying assumption of any approximation and make sure it is applicable to the situation.

Usually, a plasma, which is initially in equilibrium, becomes unstable because a source of free energy is added, at a comet, this source of free energy is the presence of newly ionised cometary ions in the solar wind flow. \cite{Goetz2017} found that, in general, the wave activity, or overall power spectral density of the magnetic field at 67P, is modulated by the neutral gas production rate, demonstrating that the addition of more ions leads to more free energy that needs to be distributed in the plasma in order to reach equilibrium again.
This addition of free energy will induce wave-like disturbances at a multitude of frequencies. Most of them are damped quickly and only if the disturbances are at a frequency near a wave mode (e.g. ion cyclotron mode) can a wave actually develop and propagate. Due to various processes, the wave is dispersed and dissipated until the energy contained in the original instability is evenly distributed.
For a wave to be detectable by instruments, its amplitude needs to be larger than the underlying thermal fluctuations. In addition, plasma wave excitation and propagation depends on the direction of the magnetic field.

Plasma waves in the cometary environment contribute to the heating and cooling of the plasma and couple fields and particles as well as different particle populations. Therefore the study of these waves is important in understanding the energy and momentum transfer as well as the behaviour of particles in the environment. 
Hereunder, the reader may find a list of waves that have been detected at comets, with an emphasis on the new results from the \emph{Rosetta} mission to comet 67P.

\paragraph{Pick-up induced waves}
If a cometary neutral is ionised in the solar wind, the resulting ion is moving at a velocity of the negative of the solar wind velocity ($- v_\text{sw}$) in the solar wind frame of reference. It will therefore be subject to the solar wind convective electric field 
\begin{equation}
    E_\text{conv} = -\vec{v}_\text{sw} \times \vec{B}_\text{IMF},
\end{equation}
where $\vec{B}_\text{IMF}$ is the interplanetary magnetic field.
Along with the ion's gyromotion in the magnetic field, this will lead to an $E\times B$-drift. In velocity space, this motion describes a circle with the solar wind ion velocity at its center. If ions are produced over a region greater than the ion gyroradius the cometary ion distribution function will form a full ring distribution, if the ions are produced over a region smaller than the ion gyroradius, the ring distribution is only partial \citep[see e.~g.][]{thesis_behar}. If the interplanetary magnetic field is parallel to the solar wind velocity, the convective electric field is zero, and the cometary ions are not accelerated. Then, they form a beam distribution in the solar wind frame of reference. Both the beam and ring distribution coexist with the solar wind beam distribution.
Therefore, upstream of a comet, two ion distributions can be used to approximate the situation: a ring-beam and a beam-beam distribution.
Both these distributions are unstable and give rise to wave activity \citep[e.g.][]{Coates1993}.

These waves were detected at active comets and usually at high cometocentric distances ($r_c > 1000\,$km). In this regime, the large interaction region allows for the ring distribution of the cometary ions to fully develop. This was not the case for most of the \emph{Rosetta} mission, where the ion gyroradius was larger or of the same order of magnitude as the size of the interaction region. 
In the high activity case during the \emph{Rosetta} mission, the plasma environment was larger than the ion gyroradius, but it was also inhomogeneous at those length scales which prevents the ring distribution from developing. Therefore a full, classical ring-beam distribution does not develop \citep{Nicolaou2017}.
As a result none of the pick-up induced waves were observed at 67P and most of the results pertaining to ring/ring-beam instabilities stem from earlier works. For more information on those results, the reader is referred to \cite{Ip2004}.

\paragraph{Singing Comet Waves}
These ultra low frequency (ULF) waves were first (and only) detected at comet 67P at low to medium gas production rates \citep{Richter2015,Goetz2020}.
The singing comet waves are characterized by large amplitude magnetic field magnitude fluctuations in the frequency range of $10-100\,$mHz and can be detected ubiquitously in the plasma environment of comet 67P for gas production rates $Q<5\times 10^{26}\,$s$^{-1}$. Figure \ref{fig:waves}a shows an example of the magnetic field measurements from the two magnetometers of \emph{Rosetta} and Philae. Their frequency is not correlated with the magnetic field magnitude and it was therefore concluded that it was not due to an ion-cyclotron resonance (for which the frequency correlates with the magnetic field). A new mechanism for the generation of these particular waves was found: an unstable cross-field current in the comet's reference frame. At low gas production rates, the cometary pickup ions have gyroradii much larger than the scale length of the cometary environment ($10000$s$\,$km vs $100$s$\,$km) and therefore the ions cannot complete a full ring distribution as it was seen at more active comets with much larger cometary environments. Instead, they are accelerated linearly along the electric field. Since all ions are moving in the same direction, this constitutes a current that is parallel to the convective electric field, but perpendicular to the magnetic field. Such a configuration is unstable to the ion-Weibel instability and will produce waves. This mechanism was investigated in hybrid simulations \citep{Koenders2016waves} and in an analytical model \citep{Meier2016}. Although the hybrid simulation suggests that the waves should be more ubiquitous in the hemisphere that has a positive convective electric field, this cannot be seen from data, where the waves are detected everywhere, without a preferential location \citep{Goetz2020}. This is not necessarily a contradiction, because the waves could be generated in a region around the nucleus that is larger than the distances covered by the measurements. 
There are some indications that this is the case, e.g. from two-point measurements which constrain the generation region to between $100\,$km and more than $800\,$km in size, which is larger or of the same order than the measurement range \citep[up to $260\,$km in the interval covered by the study by][]{Volwerk2018a}. But the exact extent of the generation region is as of yet unknown and requires more investigation.

Two point measurements constrained the wavelength to $100$s$\,$km and found that the length scale over which the wave trains are coherent is larger than $\sim 50\,$km (the separation of the two measurement points). This is easily seen in the measurements (Figure \ref{fig:waves}a) where both magnetometers show the same wave form, with only marginal deviation. 
The waves are compressional and, in isolated occasions, also observable in the plasma density \citep{Breuillard2019}. While a case study has found an anti-correlation of the wave frequency with the plasma density, a statistical study covering several months of observations could not confirm this correlation. Therefore the exact relationship between the plasma density and the magnetic field remains an open question.

\paragraph{Lower hybrid waves}
While singing comet waves are mostly detected in the magnetic field observations, electric field measurements also exhibit wave activity. Most prominent among these signatures are waves in the lower hybrid frequency range (a couple of Hz). This particular type of wave can transfer energy between ions and electrons and is typically found in plasmas where the ions are not magnetized, but the electrons are. If there are density gradients present in such a plasma, a lower hybrid drift instability (LHDI) can occur and cause lower hybrid waves (LHW) to grow.

The lower hybrid frequency $ f_\text{LH} $ is defined as:
\begin{equation}
    f_\text{LH} = \frac{1}{2 \pi} \sqrt{\frac{\omega_\text{gi}\,\omega_\text{ge}}{1+ \frac{\omega_\text{ge}^2}{\omega_\text{pe}^2} }} \approx \frac{1}{2 \pi} \sqrt{\omega_\text{ge} \,\omega_\text{gi}},
\end{equation}
where $\omega_\text{gi}$ and $\omega_\text{ge}$ are the ion gyrofrequency and the electron gyrofrequency respectively and $\omega_\text{pe}$ is the electron plasma frequency. At comet 67P, the approximation is usually satisfied as the electron plasma frequency is typically much larger than the electron gyrofrequency. Values for $f_\text{LH}$ are in the $1-20\,$Hz range at 67P. Interaction of electrons with LH waves have been suggested as a possible heating mechanism for the electrons, but no studies have attempted to verify this.

Observations show that wave packets in the electric field are often observed at plasma density gradients and with frequencies near those associated with the lower hybrid wave \citep{Karlsson2017,Andre2017}.
The upper panel in Figure \ref{fig:waves}b shows the spacecraft potential, (a proxy for the plasma density), during a plasma density gradient. The middle panel shows the derived lower hybrid frequency, and the lower panel shows the electric field measurements with the LH wave activity clearly visible. The amplitudes are largest during the density gradient.
Estimates of the LHDI criterion and model results show that these observations are consistent with LH waves generated by a LHDI and that the growth rate can be quite large so that the wave packets can grow to significant amplitudes within a couple of seconds.
The LHW can also influence the plasma as a whole, by e.g. forcing the diamagnetic boundary to oscillate slightly. This in turn can lead to a mode conversion, where LHW that are generated at density gradients outside the diamagnetic cavity can be converted to ion acoustic waves (IAW) that propagate in the unmagnetized plasma of the diamagnetic cavity \citep{Madsen2018}. 

It should be noted that collisions can significantly inhibit wave growth as they cool and slow down the electrons. Therefore, for dense plasmas such as those encountered at comet 1P/Halley and at comet 67P close to perihelion and/or very close to the nucleus are not favorable for LHW growth. 

\paragraph{Ion acoustic waves}
Ion acoustic waves (IAWs) are compressional waves in an unmagnetized plasma, or in a plasma where the gyrofrequencies are lower than the wave frequency and the gyroradii are larger than the wavelength. They have been observed at comet 1P/Halley's foreshock \citep{Oya1986} and at the artificial comet AMPTE \citep{Gurnett1985grl}. The presence of IAW was reported at comet 67P at multiple times during the \emph{Rosetta} mission, all in the plasma in the innermost coma of 67P. 
IAWs can be observed in a range of frequencies, from 100s Hz up to the kHz range.

In order to verify that the observed waves are indeed IAW, calculations of the dispersion relation using the observed ion and electron distribution function as well as the measured plasma density can be used. 
Notably, all observations were made in the inner coma, where a significant cold ion population exists due to ion-neutral collisions. IAW grow if $T_e \gg T_i$, so that any wave activity should be damped quickly in regions where ion cooling is insignificant \citep{Gunell2017a}. 
In the case that there are accelerated water ions present, this population constitutes a beam-like part of the ion population which makes the situation unstable and can lead to IAW growth.
In the absence of such a beam, a current driven instability can add to the growth rates of the IAW. \emph{Rosetta}'s close flyby of comet 67P made it possible to observe the large scale current that is associated with the magnetic field draping near the nucleus \citep{Koenders2016} and the wave signatures in that region. It was found that in the presence of this current, IAW are produced, while outside of the region containing the current, the waves are propagating away from the current and eventually damped \citep{Gunell2021}.

IAW were also detected inside the diamagnetic cavity, close to the DCB, but not outside of it. Figure \ref{fig:waves}c shows the power spectral density of the current (a proxy for the density) for short intervals inside (solid lines), outside (dashed lines) and in the boundary (black line) of the diamagnetic cavity. Clearly, the power spectral density is about two orders of magnitude higher inside the diamagnetic cavity than it is outside.
Here, again the dominance of the cold ion population leads to the growth of the waves. However, an additional current was speculated to be in place. Considering that the diamagnetic cavity boundary is wavy, there might be a current closing through the protruding parts of the diamagnetic cavity that drives the wave generation \citep{Gunell2017b}. Further studies and observations, ideally by multiple spacecraft, are necessary to confirm this generation mechanism.
Through a combination of data analysis and modeling of dispersion relations, the wave observations can be used to constrain the plasma parameters in the generation region of the waves.

\paragraph{Mirror modes and magnetic holes}
The pick-up process at the comet leads to the generation of a ring or ring-beam distribution of the heavy ions. This distribution is unstable, and can cause the generation of mirror-modes in a high-$\beta$ plasma. Mirror modes are compressional, pressure equilibrium structures. They have large amplitudes and do not propagate in the plasma, instead they are convected with the plasma flow.
Mirror modes have been observed at comet 1P/Halley and comet 21P/Giacobini-Zinner. \cite{Volwerk2014} found that at 1P/Halley that changes in the dynamic pressure of the solar wind influenced the generation of mirror-mode waves in the cometosheath, as well as outgassing rate changes. Increased solar wind dynamic pressure compresses the magnetosheath and inhibits the growth of mirror-modes, but increased outgassing will enhance ion pick up and thereby assist the growth of mirror modes. \cite{Schmid2014} showed evidence for the Bohm-type diffusion of mirror-modes as they move from the source region further into the magnetosheath. The mirror-modes grow in size over time as the high-frequency parts of the structures diffuse faster than the low-frequencies.

Mirror modes have been observed in the pile-up region at comet 67P \citep{Volwerk2016} with timescales of $100\,$s to $150\,$s which corresponds to sizes of 10s of water ion gyroradii. Figure \ref{fig:waves}d shows an example of the magnetic field measurements during a mirror mode wave train, the LAP current (a density proxy) is added to show that density and magnetic field are out of phase. 
Outside of the pile-up region, the timescale of the mirror modes is smaller ($\sim 10$s) and the scale size is just a few water ion gyroradii. These mirror mode signatures are more asymmetric with either the decrease or increase of the field being steeper than the other side.
The presence of mirror modes indicates that there is a full ring/ring-beam distribution present in the coma, i.e. that the water ions have had enough time to go through an entire gyration before they reach the spacecraft in the inner coma. The larger mirror mode structures are thought to be caused by the diffusion of smaller scale mirror modes as they are convected through the coma, whereas the asymmetry of the structure could be caused by the interaction of different mirror modes.

Magnetic holes are thought to be a further development stage of these mirror modes \citep{Winterhalter2000}.
They are omnipresent in the solar wind \citep{Volwerk2020,Volwerk2021} and therefore should impact the pick-up and pile-up processes in the coma. Magnetic holes were detected at 67P, inside and outside of the solar wind ion cavity \citep{Plaschke2018}, which indicates that the magnetic field structures are moving into the coma along with the electron fluid, while the solar wind ions are substituted by accelerated cometary ions.

\paragraph{Steepened magnetosonic waves}
Steepened magnetic field structures were first detected near the diamagnetic cavity of comet 67P \citep{Goetz2016a}. The diamagnetic cavity entry and exit is also asymmetric, with the former being usually longer than the latter.

High time resolution observations of the low energy ion environment in the inner coma showed periodic enhancements in the ion energy with a sharp increased followed by a longer relaxation time, as illustrated in the upper panel of Figure \ref{fig:waves}e. Some of this observed increase was due to the spacecraft potential increase, but taking the spacecraft potential into account still allows identification of the asymmetric structures in the measured ion energy and flux \citep{StenbergWieser2017}. The occurrence rate of these structures was highest near the diamagnetic cavity.

The magnetic field observations (see lower panel of Figure \ref{fig:waves}e cover a larger time span of the mission time, and therefore a larger study could be performed. Using machine learning, over 70000 individual steepened wave structures were detected in the magnetic field \citep{Ostaszewski2021}.
More steepened waves occur when there is more mass-loading of the plasma, which means that the peak in wave activity is around perihelion. During the dayside excursion, the only time that \emph{Rosetta} left the innermost coma, the number of wave detections decreased. 
There is no evidence that the occurance of steepened waves is correlated to the solar wind parameters.

At high activity levels, the waves are steeper but have lower amplitudes, this indicates wave evolution based on the interaction region properties.
Using a 1D MHD model it is possible to model the steepening of a wave packet in a fluid with non-negligible viscosity and resistivity. It shows that the plasma environment is large enough for the wave packet to steepen to the values of skewness observed in the magnetic field observations. Comparison of the model parameters and the measured wave properties allows to infer the viscosity of the plasma \citep{Ostaszewski2021}. 

Some of these structures can also be detected inside the diamagnetic cavity \citep{Masunaga2019,Hajra2017}. There, the magnetic field of the structure remains close to zero, but the density and ion flux are similar to the steepened waves upstream of the boundary. This indicates that while the magnetic field remains zero, the diamagnetic cavity boundary is permeable to the heavy ions observed in the inner coma. The exact mechanism of the transmission of those wave packets through the boundary is not yet clear and requires further modelling and analysis. 

\paragraph{Ion Bernstein Waves}
A closer inspection of the steepened wave plasma density observations reveals a substructure of wave activity in the descending, longer part of the steepened wave. A corresponding signature in the magnetic field was found to be of lower amplitude and phase shifted by $90^\circ$.  These were tentatively attributed to ion bernstein waves, which is an electrostatic wave mode that can be excited by a ring/ring-beam instability \citep{Odelstad2020}.
In Figure \ref{fig:waves}f the current (plasma density), magnetic field and coherence of these two parameters is shown. There are clear signatures in the coherence at around $100\,$Hz for several minutes at a time.

\section{Summary and Outlook}

In this chapter we have shown the richness of the processes that arise when the cometary ion cloud interacts with the charged solar wind. 
The cometary plasma environment is not only highly variable in time, but also in spatial dimension. Depending on the properties of the nucleus and its distance to the Sun, the extension of the coma can vary by four orders of magnitude. It therefore is an ideal laboratory to explore processes at different scale sizes and cross-scale interactions in a multi-ion plasma. 

Simply put, the environment is created by the ions and electrons of cometary origin and the interaction with the solar wind distributes energy and momentum in this plasma to achieve a stable state, where the solar wind plasma and the cometary plasma are fully mixed. This fundamental process is often referred to as mass-loading and manifests itself in different ways, depending on the scale sizes of the interaction region and the particles in the plasma.
Most importantly, the ion gyroradius and collision length scales determine how the solar wind particles and the cometary particles are interacting with each other and amongst each other.
At a high activity comet, a more fluid-like behavior is common, while a kinetic approach is preferred at low activity comets and in the inner coma of any comet.

For all comets except the most active, the large gyroradius means that the solar wind ions are not just slowed down, but also deflected. Eventually the cometary ions are picked up and substitute the solar wind ions in the flow. Often, two distinct cometary ion populations are detected: accelerated pick-up ions and slow, newborn ions.

On small scales, the electrons and ions decouple and electrons are accelerated into the inner coma by an ambipolar field. There they can be trapped by collisions and cooled. Therefore three electron populations are present: cold, warm, and hot. The interplay of electric fields and collisions changes the electron energy.

Transient solar wind events like ICMEs and CIRs increase not only the solar wind pressures but also the compression factor of the cometosphere and the electron impact ionisation rate. This leads to an increase in the cometary ion density that is greater than the changes of the solar wind parameters itself. This also increases the magnetic field to unprecedented values.

The interplanetary magnetic field is piled up and draped around the inner coma, creating an induced magnetosphere. Nested draping creates current sheets, and the small gyroradius can lead to draping in a different direction due to ions being deflected. Thus, while the ions are not directly tied to the magnetic field, they still have an influence on it via the electrons.
The plasma tail is structured, with cometary rays and density enhancements being visible from Earth remotely. Magnetic reconnection at current sheets could potentially cause tail disconnection events.

There are multiple boundaries that form depending on the gas production rate. The first is a bow shock that is broad and weak due to the mass-loaded nature of the plasma. Here, small scale processes, like electron impact ionisation and charge exchange, can affect the bow shock standoff distance, demonstrating the importance of cross scale coupling.
While a cometopause seems to exist at most comets, a solar wind ion cavity also appears and is the more obvious boundary. Collisionopauses are broad regions where different collisional processes dominate.

The diamagnetic cavity at comets seems highly unstable, asymmetric and the boundary is often dominated by surface waves. While a diamagnetic cavity exists at both comet 1P/Halley and at the lower activity comet 67P, the mechanism that forms this region seems entirely different.

The free energy that is added to the plasma by the creation of heavy water ions modulates the overall wave activity that is observable in the environment as waves are a way to distribute energy. Often, the existence of certain wave modes allows us to learn more about the plasma in which they were generated. Steepening of waves can be used to diagnose the resistivity and viscosity of the plasma and how waves evolve when they travel through it.

All of the phenomena described here can be investigated in their own right, but the coupling between processes of different temporal and spatial scales necessitates that a more rounded approach is taken.
While the comet nucleus is quite small, the plasma environment can extend up to millions of km, with the plasma tail spanning multiple AU at times. Therefore a large parameter space is covered and a multitude of processes can be observed to have effects on the plasma environment. 

In the corresponding chapter of Comets II \citep{Ip2004}, high hopes were put on the results from the Rosetta mission. However, in hindsight, it is very difficult to compare Rosetta results with those from previous flyby missions to more active comets. It turned out that the plasma environment at 67P was very different in terms of collisionality and gyro radius effects. Instead of answering the questions posed in Comets II, Rosetta has provided a whole new set of results that expand our knowledge of the plasma environment of low to medium activity comets. 

Many open questions remain with regards to this topic and future missions to gather a more complete data set are necessary. First and foremost among these should be a multi-spacecraft mission that will be able to provide spatial and temporal coverage at the same time and allow us to disentangle the influence of the different contributions (solar wind processes and internal processes) to the plasma. Only then can we take full advantage of this intriguing plasma laboratory that presents itself to us any time a comet is explored.

\bibliographystyle{sss-three.bst}
\bibliography{ALLTHEREFERENCES.bib}

\begin{thebibliography}{233}
\providecommand{\natexlab}[1]{#1}
\parskip=0pt \itemsep=0pt \small \baselineskip=11pt

\bibitem[{\emph{{Alfv\'en}}(1957)}]{Alfven1957}
{Alfv\'en} H. (1957) \emph{{On the theory of comet tails}}, \emph{Tellus},
  \emph{9}.

\bibitem[{\emph{Alho et~al.}(2020)\emph{Alho, Jarvinen, Simon~Wedlund, Nilsson,
  Kallio, and Pulkkinen}}]{alho_hybrid_2020}
Alho M., Jarvinen R., Simon~Wedlund C. et~al. (2020) \emph{Remote sensing of
  cometary bow shocks: Modelled asymmetric outgassing and pickup ion
  observations}, \emph{Submitted to MNRAS}, \emph{-}, 1--14.

\bibitem[{\emph{{Alho} et~al.}(2019)\emph{{Alho}, {Simon Wedlund}, {Nilsson},
  {Kallio}, {Jarvinen}, and {Pulkkinen}}}]{Alho2019}
{Alho} M., {Simon Wedlund} C., {Nilsson} H. et~al. (2019) \emph{{Hybrid
  modelling of cometary plasma environments. II. Remote sensing of a cometary
  bow shock}}, \emph{\aap}, \emph{1}.

\bibitem[{\emph{{Altwegg} et~al.}(1993)\emph{{Altwegg}, {Balsiger}, {Geiss},
  {Goldstein}, {Ip}, {Meier}, {Neugebauer}, {Rosenbauer}, and
  {Shelley}}}]{Altwegg1993}
{Altwegg} K., {Balsiger} H., {Geiss} J. et~al. (1993) \emph{{The ion population
  between 1300 KM and 230000 KM in the coma of comet P/Halley}}, \emph{\aap},
  \emph{279}, 260--266.

\bibitem[{\emph{Ananthakrishnan et~al.}(1975)\emph{Ananthakrishnan, Bhandai,
  and Pramesh}}]{anan75}
Ananthakrishnan S., Bhandai S.~M., and Pramesh R.~A. (1975) \emph{Occultation
  of radio source {PKS} 2025-15 by comet {Kohoutek} (1973f)}, \emph{Astrophys.
  Space Sci.}, \emph{37}, 275 -- 282.

\bibitem[{\emph{{Andr{\'e}} et~al.}(2017)\emph{{Andr{\'e}}, {Odelstad},
  {Graham}, {Eriksson}, {Karlsson}, {Stenberg Wieser}, {Vigren}, {Norgren},
  {Johansson}, {Henri}, {Rubin}, and {Richter}}}]{Andre2017}
{Andr{\'e}} M., {Odelstad} E., {Graham} D.~B. et~al. (2017) \emph{{Lower hybrid
  waves at comet 67P/Churyumov-Gerasimenko}}, \emph{Monthly Notices of the
  Royal Astronomical Society}, \emph{469}, S29--S38.

\bibitem[{\emph{{Balogh} et~al.}(1999)\emph{{Balogh}, {Bothmer}, {Crooker},
  {Forsyth}, {Gloeckler}, {Hewish}, {Hilchenbach}, {Kallenbach}, {Klecker},
  {Linker}, {Lucek}, {Mann}, {Marsch}, {Posner}, {Richardson}, {Schmidt},
  {Scholer}, {Wang}, {Wimmer-Schweingruber}, {Aellig}, {Bochsler}, {Hefti}, and
  {Miki{\'c}}}}]{Balogh_etal_SSR_1999}
{Balogh} A., {Bothmer} V., {Crooker} N.~U. et~al. (1999) \emph{{The Solar
  Origin of Corotating Interaction Regions and Their Formation in the Inner
  Heliosphere}}, \emph{\ssr}, \emph{89}, 141--178.

\bibitem[{\emph{{Balogh} and {Treumann}}(2013)}]{Balogh2013}
{Balogh} A. and {Treumann} R.~A. (2013) \emph{Physics of Collisionless Shocks},
  vol.~12 of \emph{ISSI Scientific Report Series}, Springer, New York, NY.

\bibitem[{\emph{Behar}(2018)}]{thesis_behar}
Behar E. (2018) \emph{Solar Wind Dynamics within The Atmosphere of Comet
  67P/Churyumov-Gerasimenko}, Ph.D. thesis, Lule\aa University of Technology.

\bibitem[{\emph{{Behar} et~al.}(2016{\natexlab{a}})\emph{{Behar}, {Lindkvist},
  {Nilsson}, {Holmstr{\"o}m}, {Stenberg-Wieser}, {Ramstad}, and
  {G{\"o}tz}}}]{Behar2016aa}
{Behar} E., {Lindkvist} J., {Nilsson} H. et~al. (2016{\natexlab{a}})
  \emph{{Mass-loading of the solar wind at 67P/Churyumov-Gerasimenko.
  Observations and modelling}}, \emph{\aap}, \emph{596}, A42.

\bibitem[{\emph{{Behar} et~al.}(2017)\emph{{Behar}, {Nilsson}, {Alho}, {Goetz},
  and {Tsurutani}}}]{Behar2017}
{Behar} E., {Nilsson} H., {Alho} M. et~al. (2017) \emph{{The birth and growth
  of a solar wind cavity around a comet - Rosetta observations}},
  \emph{\mnras}, \emph{469}, S396--S403.

\bibitem[{\emph{Behar et~al.}(2018)\emph{Behar, Nilsson, Henri, Bercic,
  Nicolaou, Stenberg~Wieser, Wieser, Tabone, Saillenfest, and
  Goetz}}]{behar2018aa_ns}
Behar E., Nilsson H., Henri P. et~al. (2018) \emph{The root of a comet tail --
  rosetta ion observations at comet 67p/churyumov-gerasimenko}, \emph{Astronomy
  \& Astrophysics}.

\bibitem[{\emph{{Behar} et~al.}(2016{\natexlab{b}})\emph{{Behar}, {Nilsson},
  {Wieser}, {Nemeth}, {Broiles}, and {Richter}}}]{Behar2016}
{Behar} E., {Nilsson} H., {Wieser} G.~S. et~al. (2016{\natexlab{b}})
  \emph{{Mass loading at 67P/Churyumov-Gerasimenko: A case study}},
  \emph{Geophysical Research Letters}, \emph{43}, 1411--1418.

\bibitem[{\emph{Ber{\v c}i{\v c} et~al.}(2018)\emph{Ber{\v c}i{\v c}, Behar,
  Nilsson, Nicolaou, Stenberg{ }Wieser, Wieser, and Goetz}}]{Bercic2018}
Ber{\v c}i{\v c} L., Behar E., Nilsson H. et~al. (2018) \emph{Cometary ion
  dynamics observed in the close vicinity of comet
  {67P/C}huryumov-{G}erasimenko during the intermediate activity period},
  \emph{Astronomy \& Astrophysics}, \emph{613}, A57.

\bibitem[{\emph{{Bergman} et~al.}(2021)\emph{{Bergman}, {Stenberg Wieser},
  {Wieser}, {Johansson}, {Vigren}, {Nilsson}, {Nemeth}, {Eriksson}, and
  {Williamson}}}]{Bergman2021}
{Bergman} S., {Stenberg Wieser} G., {Wieser} M. et~al. (2021) \emph{{Ion bulk
  speeds and temperatures in the diamagnetic cavity of comet 67P from RPC-ICA
  measurements}}, \emph{\mnras}, \emph{503}, 2733--2745.

\bibitem[{\emph{{Beth} et~al.}(2016)\emph{{Beth}, {Altwegg}, {Balsiger},
  {Berthelier}, {Calmonte}, {Combi}, {De Keyser}, {Dhooghe}, {Fiethe},
  {Fuselier}, {Galand}, {Gasc}, {Gombosi}, {Hansen}, {Hässig}, {Heritier},
  {Kopp}, {Le Roy}, {Mandt}, {Peroy}, {Rubin}, {Sémon}, {Tzou}, and
  {Vigren}}}]{Beth2016}
{Beth} A., {Altwegg} K., {Balsiger} H. et~al. (2016) \emph{First in situ
  detection of the cometary ammonium ion {NH}$_4^{+}$ (protonated ammonia
  {NH}$_3$) in the coma of {67P/C-G} near perihelion}, \emph{Monthly Notices of
  the Royal Astronomical Society}, \emph{462}, S562--S572.

\bibitem[{\emph{{Bieler} et~al.}(2016)\emph{{Bieler}, {Altwegg}, {Balsiger},
  {Berthelier}, {Calmonte}, {Combi}, {De Keyser}, {Fiethe}, {Fuselier}, {Gasc},
  {Gombosi}, {Hansen}, {H{\"a}ssig}, {Korth}, {Le Roy}, {Mall}, {R{\`e}me},
  {Rubin}, {S{\'e}mon}, {Tenishev}, {Tzou}, {Waite}, and
  {Wurz}}}]{Bieler_etal_SPIE_2016}
{Bieler} A., {Altwegg} K., {Balsiger} H. et~al. (2016) in \emph{Systems
  Contamination: Prediction, Control, and Performance 2016} (J.~{Egges}, C.~E.
  {Soares}, and E.~M. {Wooldridge}, eds.), vol. 9952 of \emph{Society of
  Photo-Optical Instrumentation Engineers (SPIE) Conference Series}, p. 99520E.

\bibitem[{\emph{{Biermann}}(1951)}]{Biermann1951}
{Biermann} L. (1951) \emph{{Kometenschweife und solare Korpuskularstrahlung}},
  \emph{Zeitschrift f{\"u}r Astrophysik}, \emph{29}, 274.

\bibitem[{\emph{{Biermann} et~al.}(1967)\emph{{Biermann}, {Brosowski}, and
  {Schmidt}}}]{Biermann1967}
{Biermann} L., {Brosowski} B., and {Schmidt} H.~U. (1967) \emph{{The
  interactions of the solar wind with a comet}}, \emph{Solar Physics},
  \emph{1}, 254--284.

\bibitem[{\emph{Bobrovnikoff}(1930)}]{bobr30}
Bobrovnikoff N.~T. (1930) \emph{Halley’s comet in its apparition of
  1909—1911}, \emph{Lick Obs. Public.}, \emph{17}, 305.

\bibitem[{\emph{Bodewits et~al.}(2004)\emph{Bodewits, Juhasz, Hoekstra, and
  Tielens}}]{Bodewits2004}
Bodewits D., Juhasz Z., Hoekstra R. et~al. (2004) \emph{{Catching Some Sun:
  Probing the Solar Wind with Cometary X-Ray and Far-Ultraviolet Emission}},
  \emph{Astrophysical Journal}, \emph{606}, L81--L84.

\bibitem[{\emph{Bodewits et~al.}(2016)\emph{Bodewits, Lara, A’hearn,
  La~Forgia, Gicquel, Kovacs, Knollenberg, Lazzarin, Lin, Shi
  et~al.}}]{bodewits2016changes}
Bodewits D., Lara L.~M., A’hearn M.~F. et~al. (2016) \emph{Changes in the
  physical environment of the inner coma of 67p/churyumov--gerasimenko with
  decreasing heliocentric distance}, \emph{The astronomical journal},
  \emph{152}, 130.

\bibitem[{\emph{{Bostr\"om}}(1974)}]{bost74}
{Bostr\"om} R. (1974) in \emph{Magnetospheric Physics} (B.~M. {McCormac}, ed.),
  pp. 45 -- 59, D. Reidel Publishing Co., Dordrecht.

\bibitem[{\emph{Brandt}(1982)}]{bran82}
Brandt J.~C. (1982) in \emph{Comets} (L.~L. Wilkening, ed.), pp. 519 -- 537,
  Arizona University Press.

\bibitem[{\emph{Brandt et~al.}(1999)\emph{Brandt, Caputo, Hoeksema, Niedner,
  Yi, and Snow}}]{bran99}
Brandt J.~C., Caputo F.~M., Hoeksema J.~T. et~al. (1999) \emph{Disconnection
  events {(DEs)} in {Halley's} comet 1985-1986:. the correlation with crossings
  of the heliospheric current sheet {(HCS)}}, \emph{Icarus}, \emph{137}, 69 --
  83.

\bibitem[{\emph{Brandt and Snow}(2000)}]{bran00}
Brandt J.~C. and Snow M. (2000) \emph{Heliospheric latitude variations of
  properties of cometary plasma tails: A test of the {Ulysses} comet watch
  paradigm}, \emph{Icarus}, \emph{148}, 52 -- 64.

\bibitem[{\emph{Bredikhin}(1879)}]{bredikhin1879}
Bredikhin T. (1879) \emph{Annales de l'obervatioire de {Moscou}}, Imprimerie F.
  {Neub\"urger}, Moscow, Russia.

\bibitem[{\emph{Breuillard et~al.}(2019)\emph{Breuillard, Henri, Bucciantini,
  Volwerk, Karlsson, Eriksson, Johansson, Odelstad, Richter, Goetz, Vallières,
  and Hajra}}]{Breuillard2019}
Breuillard H., Henri P., Bucciantini L. et~al. (2019) \emph{The properties of
  the singing comet waves in the {67P/C}huryumov–{G}erasimenko plasma
  environment as observed by the {R}osetta mission}, \emph{\aap}.

\bibitem[{\emph{{Broiles} et~al.}(2016)\emph{{Broiles}, {Burch}, {Chae},
  {Clark}, {Cravens}, {Eriksson}, {Fuselier}, {Frahm}, {Gasc}, {Goldstein},
  {Henri}, {Koenders}, {Livadiotis}, {Mandt}, {Mokashi}, {Nemeth}, {Odelstad},
  {Rubin}, and {Samara}}}]{Broiles2016}
{Broiles} T.~W., {Burch} J.~L., {Chae} K. et~al. (2016) \emph{{Statistical
  analysis of suprathermal electron drivers at 67P/Churyumov- Gerasimenko}},
  \emph{\mnras}, \emph{462}, S312--S322.

\bibitem[{\emph{{Broiles} et~al.}(2015)\emph{{Broiles}, {Burch}, {Clark},
  {Koenders}, {Behar}, {Goldstein}, {Fuselier}, {Mandt}, {Mokashi}, and
  {Samara}}}]{Broiles2015}
{Broiles} T.~W., {Burch} J.~L., {Clark} G. et~al. (2015) \emph{{Rosetta
  observations of solar wind interaction with the comet
  67P/Churyumov-Gerasimenko}}, \emph{\aap}, \emph{583}, A21.

\bibitem[{\emph{Buffington et~al.}(2008)\emph{Buffington, Bisi, Clover, Hick,
  Jackson, and Kuchar}}]{buff08}
Buffington A., Bisi M.~M., Clover J.~M. et~al. (2008) \emph{Analysis of
  plasma-tail motions for comets {C/2001 Q4 (NEAT) and C/2002 T7 (LINEAR)}
  using observations from {SMEI}}, \emph{Astrophys. J.}, \emph{677}, 798 --
  807.

\bibitem[{\emph{Clark et~al.}(2015)\emph{Clark, {Broiles, T. W.}, {Burch, J.
  L.}, {Collinson, G. A.}, {Cravens, T.}, {Frahm, R. A.}, {Goldstein, J.},
  {Goldstein, R.}, {Mandt, K.}, {Mokashi, P.}, {Samara, M.}, and {Pollock, C.
  J.}}}]{Clark2015a}
Clark G., {Broiles, T. W.}, {Burch, J. L.} et~al. (2015) \emph{Suprathermal
  electron environment of comet {67P/Churyumov-Gerasimenko}: Observations from
  the {Rosetta} ion and electron sensor}, \emph{Astronomy \& Astrophysics}.

\bibitem[{\emph{Coates}(1997)}]{Coates1997}
Coates A. (1997) \emph{Ionospheres and magnetospheres of comets},
  \emph{Advances in Space Research}, \emph{20}, 255--266.

\bibitem[{\emph{{Coates} et~al.}(2015)\emph{{Coates}, {Burch}, {Goldstein},
  {Nilsson}, {Stenberg Wieser}, {Behar}, and {the RPC Team}}}]{Coates2015}
{Coates} A.~J., {Burch} J.~L., {Goldstein} R. et~al. (2015) in \emph{Journal of
  Physics Conference Series}, vol. 642 of \emph{Journal of Physics Conference
  Series}, p. 012005.

\bibitem[{\emph{{Coates} et~al.}(1993)\emph{{Coates}, {Johnstone}, {Wilken},
  and {Neubauer}}}]{Coates1993}
{Coates} A.~J., {Johnstone} A.~D., {Wilken} B. et~al. (1993) \emph{{Velocity
  space diffusion and nongyrotropy of pickup water group ions at comet
  Grigg-Skjellerup}}, \emph{Journal of Geophysical Research}, \emph{98},
  20985--20994.

\bibitem[{\emph{{Combi} and {Feldman}}(1993)}]{Combi_Feldman_Icarus_1993}
{Combi} M.~R. and {Feldman} P.~D. (1993) \emph{{Water Production Rates in Comet
  P/Halley from IUE Observations of HI Lyman-{\ensuremath{\beta}}}},
  \emph{\icarus}, \emph{105}, 557--567.

\bibitem[{\emph{Combi et~al.}(2014)\emph{Combi, Fougere, {M\"akinen}, Bertaux,
  {Qu\'emerais}, and Ferron}}]{combi14}
Combi M.~R., Fougere N., {M\"akinen} J. T.~T. et~al. (2014) \emph{Unusual water
  production activity of comet {C/2012 S1 (ISON)}: Outbursts and continuous
  fragmentation}, \emph{Astrophys. J. Lett.}, \emph{788}, L7.

\bibitem[{\emph{{Combi} et~al.}(2004)\emph{{Combi}, {Harris}, and
  {Smyth}}}]{Combi2004}
{Combi} M.~R., {Harris} W.~M., and {Smyth} W.~H. (2004) in \emph{Comets II}
  (M.~C. {Festou}, H.~U. {Keller}, and H.~A. {Weaver}, eds.), p. 523,
  University of Arizona Press.

\bibitem[{\emph{Cravens}(1991)}]{Cravens1991}
Cravens T. (1991) \emph{Collisional processes in cometary plasmas},
  \emph{Cometary Plasma Processes}, \emph{61}, 27--35.

\bibitem[{\emph{{Cravens}}(1987)}]{Cravens1987}
{Cravens} T.~E. (1987) \emph{{Theory and observations of cometary
  ionospheres}}, \emph{Advances in Space Research}, \emph{7}, 147--158.

\bibitem[{\emph{Cravens}(1989)}]{Cravens1989}
Cravens T.~E. (1989) \emph{{Galactic cosmic rays and cell-hit frequencies
  outside the magnetosphere}}, \emph{Advances in Space Research}, \emph{9},
  293--298.

\bibitem[{\emph{{Deca} et~al.}(2017)\emph{{Deca}, {Divin}, {Henri}, {Eriksson},
  {Markidis}, {Olshevsky}, and {Hor{\'a}nyi}}}]{Deca2017}
{Deca} J., {Divin} A., {Henri} P. et~al. (2017) \emph{{Electron and Ion
  Dynamics of the Solar Wind Interaction with a Weakly Outgassing Comet}},
  \emph{\prl}, \emph{118}, 205101.

\bibitem[{\emph{Deca et~al.}(2019)\emph{Deca, Henri, Divin, Eriksson, Galand,
  Beth, Ostaszewski, and Hor{\'a}nyi}}]{Deca2019}
Deca J., Henri P., Divin A. et~al. (2019) \emph{Building a weakly outgassing
  comet from a generalized {O}hm's law}, \emph{\prl}, in press.

\bibitem[{\emph{Degroote et~al.}(2008)\emph{Degroote, Bodewits, and
  Reyniers}}]{degr08}
Degroote P., Bodewits D., and Reyniers M. (2008) \emph{Folding ion rays in
  comet {C/2004 Q2 (Machholz)} and the connection with the solar wind},
  \emph{Astron. Astrophys.}, \emph{477}, L41 -- L44.

\bibitem[{\emph{Delva et~al.}(2014)\emph{Delva, Bertucci, Schwingenschuh,
  Volwerk, and Romanelli}}]{delv14}
Delva M., Bertucci C., Schwingenschuh K. et~al. (2014) \emph{Magnetic pileup
  boundary and field draping at comet halley}, \emph{Planet. Space Sci.},
  \emph{96}, 125 -- 131.

\bibitem[{\emph{Delva et~al.}(1991)\emph{Delva, Schwingenschuh, Niedner, and
  Gringauz}}]{delv91}
Delva M., Schwingenschuh K., Niedner M.~B. et~al. (1991) \emph{Comet {Halley}
  remote plasma tail observations and in situ solar wind properties - {Vega-1/2
  IMF/plasma} observations and ground-based optical observations from 1
  {December} 1985 to 1 {May} 1986}, \emph{Planet. Space Sci.}, \emph{39}, 697
  -- 708.

\bibitem[{\emph{Delva et~al.}(2017)\emph{Delva, Volwerk, Jarvinen, and
  Bertucci}}]{delv17}
Delva M., Volwerk M., Jarvinen R. et~al. (2017) \emph{Asymmetries in the
  magnetosheath field draping on {Venus'} nightside}, \emph{J. Geophys. Res.},
  \emph{122}, 10396 -- 10407.

\bibitem[{\emph{Divin et~al.}(2020)\emph{Divin, Deca, Eriksson, Henri, Lapenta,
  Olshevsky, and Markidis}}]{Divin2020}
Divin A., Deca J., Eriksson A. et~al. (2020) \emph{A fully kinetic perspective
  of electron acceleration around a weakly outgassing comet}, \emph{The
  Astrophysical Journal}, \emph{889}, L33.

\bibitem[{\emph{Eberhardt and Krankowsky}(1995)}]{Eberhardt1995}
Eberhardt P. and Krankowsky D. (1995) \emph{The electron temperature in the
  inner coma of comet p/halley.}, \emph{Astronomy and Astrophysics},
  \emph{295}, 795.

\bibitem[{\emph{{Edberg} et~al.}(2016{\natexlab{a}})\emph{{Edberg}, {Alho},
  {Andr{\'e}}, {Andrews}, {Behar}, {Burch}, {Carr}, {Cupido}, {Engelhardt},
  {Eriksson}, {Glassmeier}, {Goetz}, {Goldstein}, {Henri}, {Johansson},
  {Koenders}, {Mandt}, {M{\"o}stl}, {Nilsson}, {Odelstad}, {Richter}, {Simon
  Wedlund}, {Stenberg Wieser}, {Szego}, {Vigren}, and {Volwerk}}}]{Edberg2016}
{Edberg} N.~J.~T., {Alho} M., {Andr{\'e}} M. et~al. (2016{\natexlab{a}})
  \emph{{CME impact on comet 67P/Churyumov-Gerasimenko}}, \emph{\mnras},
  \emph{462}, S45--S56.

\bibitem[{\emph{{Edberg} et~al.}(2015)\emph{{Edberg}, {Eriksson}, {Odelstad},
  {Henri}, {Lebreton}, {Gasc}, {Rubin}, {Andr{\'e}}, {Gill}, {Johansson},
  {Johansson}, {Vigren}, {Wahlund}, {Carr}, {Cupido}, {Glassmeier},
  {Goldstein}, {Koenders}, {Mandt}, {Nemeth}, {Nilsson}, {Richter}, {Wieser},
  {Szego}, and {Volwerk}}}]{Edberg2015}
{Edberg} N.~J.~T., {Eriksson} A.~I., {Odelstad} E. et~al. (2015) \emph{{Spatial
  distribution of low-energy plasma around comet 67P/CG from Rosetta
  measurements}}, \emph{\grl}, \emph{42}, 4263--4269.

\bibitem[{\emph{{Edberg} et~al.}(2016{\natexlab{b}})\emph{{Edberg}, {Eriksson},
  {Odelstad}, {Vigren}, {Andrews}, {Johansson}, {Burch}, {Carr}, {Cupido},
  {Glassmeier}, {Goldstein}, {Halekas}, {Henri}, {Koenders}, {Mandt},
  {Mokashi}, {Nemeth}, {Nilsson}, {Ramstad}, {Richter}, and
  {Wieser}}}]{Edberg2016CIR}
{Edberg} N.~J.~T., {Eriksson} A.~I., {Odelstad} E. et~al. (2016{\natexlab{b}})
  \emph{{Solar wind interaction with comet 67P: Impacts of corotating
  interaction regions}}, \emph{Journal of Geophysical Research (Space
  Physics)}, \emph{121}, 949--965.

\bibitem[{\emph{{Edberg} et~al.}(2019)\emph{{Edberg}, {Johansson}, {Eriksson},
  {Andrews}, {Hajra}, {Henri}, {Simon Wedlund}, {Alho}, and
  {Thiemann}}}]{Edberg2019flare}
{Edberg} N.~J.~T., {Johansson} F.~L., {Eriksson} A.~I. et~al. (2019)
  \emph{{Solar flares observed by Rosetta at comet 67P/Churyumov-Gerasimenko}},
  \emph{\aap}, \emph{630}, A49.

\bibitem[{\emph{Eddington}(1910)}]{eddi10a}
Eddington A.~S. (1910) \emph{The envelopes of comet {Morehouse} (1098 c)},
  \emph{Mon. Not. Roy. Astron. Soc.}, \emph{70}, 442 -- 458.

\bibitem[{\emph{{Engelhardt} et~al.}(2018{\natexlab{a}})\emph{{Engelhardt},
  {Eriksson}, {Vigren}, {Valii{\`e}res}, {Rubin}, {Gilet}, and
  {Henri}}}]{Engelhardt2018b}
{Engelhardt} I.~A.~D., {Eriksson} A.~I., {Vigren} E. et~al.
  (2018{\natexlab{a}}) \emph{{Cold electrons at comet
  67P/\linebreak[0]{}Chury\-umov-Gerasimenko}}, \emph{ArXiv e-prints}.

\bibitem[{\emph{{Engelhardt} et~al.}(2018{\natexlab{b}})\emph{{Engelhardt},
  {Eriksson}, {Vigren}, {Valli{\'e}res}, {Rubin}, {Gilet}, and
  {Henri}}}]{Engelhardt2018}
{Engelhardt} I.~A.~D., {Eriksson} A.~I., {Vigren} E. et~al.
  (2018{\natexlab{b}}) \emph{{Cold electrons at comet
  67P/Churyumov-Gerasimenko}}, \emph{\aap}, \emph{616}, A51.

\bibitem[{\emph{{Eriksson} et~al.}(2017)\emph{{Eriksson}, {Engelhardt},
  {Andr{\'e}}, {Bostr{\"o}m}, {Edberg}, {Johansson}, {Odelstad}, {Vigren},
  {Wahlund}, {Henri}, {Lebreton}, {Miloch}, {Paulsson}, {Simon Wedlund},
  {Yang}, {Karlsson}, {Jarvinen}, {Broiles}, {Mandt}, {Carr}, {Galand},
  {Nilsson}, and {Norberg}}}]{Eriksson2017}
{Eriksson} A.~I., {Engelhardt} I.~A.~D., {Andr{\'e}} M. et~al. (2017)
  \emph{{Cold and warm electrons at comet 67P/Churyumov-Gerasimenko}},
  \emph{\aap}, \emph{605}, A15.

\bibitem[{\emph{Ershkovich and Heller}(1977)}]{ersh77}
Ershkovich A.~I. and Heller A.~B. (1977) \emph{Helical waves in type-1 comet
  tails}, \emph{Astrophys. Space Sci.}, \emph{48}, 365 -- 377.

\bibitem[{\emph{{Feldman} et~al.}(1975)\emph{{Feldman}, {Asbridge}, {Bame},
  {Montgomery}, and {Gary}}}]{Feldman_etal_JGR_1975}
{Feldman} W.~C., {Asbridge} J.~R., {Bame} S.~J. et~al. (1975) \emph{{Solar wind
  electrons}}, \emph{\jgr}, \emph{80}, 4181.

\bibitem[{\emph{{Fleshman} et~al.}(2012)\emph{{Fleshman}, {Delamere},
  {Bagenal}, and {Cassidy}}}]{Fleshman2012}
{Fleshman} B.~L., {Delamere} P.~A., {Bagenal} F. et~al. (2012) \emph{{The roles
  of charge exchange and dissociation in spreading Saturn's neutral clouds}},
  \emph{Journal of Geophysical Research (Planets)}, \emph{117}, E05007.

\bibitem[{\emph{{Fougere} et~al.}(2016)\emph{{Fougere}, {Altwegg},
  {Berthelier}, {Bieler}, {Bockel{\'e}e-Morvan}, {Calmonte}, {Capaccioni},
  {Combi}, {De Keyser}, {Debout}, {Erard}, {Fiethe}, {Filacchione}, {Fink},
  {Fuselier}, {Gombosi}, {Hansen}, {H{\"a}ssig}, {Huang}, {Le Roy}, {Leyrat},
  {Migliorini}, {Piccioni}, {Rinaldi}, {Rubin}, {Shou}, {Tenishev}, {Toth}, and
  {Tzou}}}]{Fougere2016}
{Fougere} N., {Altwegg} K., {Berthelier} J.-J. et~al. (2016)
  \emph{{Three-dimensional direct simulation Monte-Carlo modeling of the coma
  of comet 67P/Churyumov-Gerasimenko observed by the VIRTIS and ROSINA
  instruments on board Rosetta}}, \emph{\aap}, \emph{588}, A134.

\bibitem[{\emph{Fougere et~al.}(2016)\emph{Fougere, Altwegg, Berthelier,
  Bieler, Bockel\'ee-Morvan, Calmonte, Capaccioni, Combi, De~Keyser, Debout,
  Erard, Fiethe, Filacchione, Fink, Fuselier, Gombosi, Hansen, H\"assig, Huang,
  Le~Roy, Leyrat, Migliorini, Piccioni, Rinaldi, Rubin, Shou, Tenishev, Toth,
  Tzou, the, and the}}]{Fougere2016MNRAS}
Fougere N., Altwegg K., Berthelier J.~J. et~al. (2016) \emph{Direct simulation
  monte carlo modelling of the major species in the coma of comet
  67{P}/{C}huryumov-{G}erasimenko}, \emph{Monthly Notices of the Royal
  Astronomical Society}, \emph{462}, S156--S169.

\bibitem[{\emph{Fuselier et~al.}(1991)\emph{Fuselier, Shelley, Goldstein,
  Goldstein, Neugebauer, Ip, Balsiger, and Reme}}]{Fuselier1991}
Fuselier S.~A., Shelley E.~G., Goldstein B.~E. et~al. (1991)
  \emph{{Observations of solar wind ion charge exchange in the Comet Halley
  coma}}, \emph{Astrophysical Journal}, \emph{379}, 734.

\bibitem[{\emph{Galand et~al.}(2020)\emph{Galand, Feldman, Bockel\'ee-Morvan,
  Biver, Cheng, Rinaldi, Rubin, and et~al.}}]{Galand2020}
Galand M., Feldman P.~D., Bockel\'ee-Morvan D. et~al. (2020) \emph{Far
  ultraviolet aurora identified at comet 67p/churyumov-gerasimenko},
  \emph{Nature Astronomy}, \emph{00}, 00--00.

\bibitem[{\emph{{Galand} et~al.}(2016)\emph{{Galand}, {H{\'e}ritier},
  {Odelstad}, {Henri}, {Broiles}, {Allen}, {Altwegg}, {Beth}, {Burch}, {Carr},
  {Cupido}, {Eriksson}, {Glassmeier}, {Johansson}, {Lebreton}, {Mandt},
  {Nilsson}, {Richter}, {Rubin}, {Sagni{\`e}res}, {Schwartz}, {S{\'e}mon},
  {Tzou}, {Valli{\`e}res}, {Vigren}, and {Wurz}}}]{Galand2016}
{Galand} M., {H{\'e}ritier} K.~L., {Odelstad} E. et~al. (2016)
  \emph{{Ionospheric plasma of comet 67P probed by Rosetta at 3 au from the
  Sun}}, \emph{\mnras}, \emph{462}, S331--S351.

\bibitem[{\emph{{Galeev} et~al.}(1985)\emph{{Galeev}, {Cravens}, and
  {Gombosi}}}]{Galeev1985}
{Galeev} A.~A., {Cravens} T.~E., and {Gombosi} T.~I. (1985) \emph{{Solar wind
  stagnation near comets}}, \emph{\apj}, \emph{289}, 807--819.

\bibitem[{\emph{{Galeev} and {Sagdeev}}(1988)}]{Galeev1988}
{Galeev} A.~A. and {Sagdeev} R.~Z. (1988) \emph{{Alfv{\'e}n waves in a space
  plasma and its role in the solar wind interaction with comets}},
  \emph{\apss}, \emph{144}, 427--438.

\bibitem[{\emph{{Gan} and {Cravens}}(1990)}]{GanCravens1990}
{Gan} L. and {Cravens} T.~E. (1990) \emph{{Electron energetics in the inner
  coma of Comet Halley}}, \emph{Journal of Geophysical Research}, \emph{95},
  6285--6303.

\bibitem[{\emph{Gapper et~al.}(1982)\emph{Gapper, Hewish, Purvis, and
  {Duffett-Smith}}}]{gapp82}
Gapper G.~R., Hewish A., Purvis A. et~al. (1982) \emph{Observing interplanetary
  disturbances from the ground}, \emph{Nature}, \emph{296}, 633 -- 636.

\bibitem[{\emph{{Gilet} et~al.}(2020)\emph{{Gilet}, {Henri}, {Wattieaux},
  {Traor{\'e}}, {Eriksson}, {Valli{\`e}res}, {Mor{\'e}}, {Rand riamboarison},
  {Odelstad}, {Johansson}, and {Rubin}}}]{Gilet2020a}
{Gilet} N., {Henri} P., {Wattieaux} G. et~al. (2020) \emph{{Observations of a
  mix of cold and warm electrons by RPC-MIP at 67P/Churyumov-Gerasimenko}},
  \emph{\aap}, \emph{640}, A110.

\bibitem[{\emph{Goetz et~al.}(2021{\natexlab{a}})\emph{Goetz, Behar, Beth,
  Bodewits, bRomley, Burch, Deca, Divin, Erkisson, Feldman, Galand, Gunell,
  Henri, Heritier, Jones, Mandt, Nisson, Noonan, Odelstad, Parker, Rubin,
  {Simon Wedlund}, Stephenson, Taylor, Vigren, Vines, and
  Volwerk}}]{Goetz2021b}
Goetz C., Behar E., Beth A. et~al. (2021{\natexlab{a}}) \emph{The plasma
  environment of comet {67P/Dhuryumov-Gerasimenko}}, \emph{Space Sci. Rev.},
  \emph{-}, Under Review.

\bibitem[{\emph{Goetz et~al.}(2021{\natexlab{b}})\emph{Goetz, Gunell,
  Johansson, LLera, Nilsson, Glassmeier, and Taylor}}]{Goetz2021}
Goetz C., Gunell H., Johansson F. et~al. (2021{\natexlab{b}}) \emph{Warm
  protons at comet 67p/churyumov--gerasimenko -- implications for the infant
  bow shock}, \emph{Annales Geophysicae}, \emph{39}, 379--396.

\bibitem[{\emph{{Goetz} et~al.}(2016{\natexlab{a}})\emph{{Goetz}, {Koenders},
  {Hansen}, {Burch}, {Carr}, {Eriksson}, {Fr{\"u}hauff}, {G{\"u}ttler},
  {Henri}, {Nilsson}, {Richter}, {Rubin}, {Sierks}, {Tsurutani}, {Volwerk}, and
  {Glassmeier}}}]{Goetz2016a}
{Goetz} C., {Koenders} C., {Hansen} K.~C. et~al. (2016{\natexlab{a}})
  \emph{{Structure and evolution of the diamagnetic cavity at comet
  67P/Churyumov-Gerasimenko}}, \emph{\mnras}, \emph{462}, S459--S467.

\bibitem[{\emph{{Goetz} et~al.}(2016{\natexlab{b}})\emph{{Goetz}, {Koenders},
  {Richter}, {Altwegg}, {Burch}, {Carr}, {Cupido}, {Eriksson}, {G{\"u}ttler},
  {Henri}, {Mokashi}, {Nemeth}, {Nilsson}, {Rubin}, {Sierks}, {Tsurutani},
  {Vallat}, {Volwerk}, and {Glassmeier}}}]{Goetz2016}
{Goetz} C., {Koenders} C., {Richter} I. et~al. (2016{\natexlab{b}})
  \emph{{First detection of a diamagnetic cavity at comet
  67P/Churyumov-Gerasimenko}}, \emph{\aap}, \emph{588}, A24.

\bibitem[{\emph{Goetz et~al.}(2020)\emph{Goetz, Plaschke, and
  Taylor}}]{Goetz2020}
Goetz C., Plaschke F., and Taylor M. G. G.~T. (2020) \emph{Singing comet waves
  in a solar wind convective electric field frame}, \emph{Geophysical Research
  Letters}, \emph{47}, e2020GL087418.

\bibitem[{\emph{{Goetz} et~al.}(2018)\emph{{Goetz}, {Tsurutani, B. T.}, {Henri,
  P.}, {Volwerk, M.}, {Behar, E.}, {Edberg, N. J. T.}, {Eriksson, A.},
  {Goldstein, R.}, {Mokashi, P.}, {Nilsson, H.}, {Richter, I.}, {Wellbrock,
  A.}, and {Glassmeier, K. H.}}}]{Goetz2019}
{Goetz} C., {Tsurutani, B. T.}, {Henri, P.} et~al. (2018) \emph{{Unusually high
  magnetic fields in the coma of 67P/Churyumov-Gerasimenko during its
  high-activity phase}}, \emph{\aap}.

\bibitem[{\emph{{Goetz} et~al.}(2017)\emph{{Goetz}, {Volwerk}, {Richter}, and
  {Glassmeier}}}]{Goetz2017}
{Goetz} C., {Volwerk} M., {Richter} I. et~al. (2017) \emph{{Evolution of the
  magnetic field at comet 67P/Churyumov-Gerasimenko}}, \emph{\mnras},
  \emph{469}, S268--S275.

\bibitem[{\emph{{Goldstein} et~al.}(2019)\emph{{Goldstein}, {Burch}, {Llera},
  {Mokashi}, {Nilsson}, {Dokgo}, {Eriksson}, {Odelstad}, and
  {Richter}}}]{Goldstein2019a}
{Goldstein} R., {Burch} J.~L., {Llera} K. et~al. (2019) \emph{{Electron
  acceleration at comet 67P/Churyumov-Gerasimenko}}, \emph{\aap}, \emph{630},
  A40.

\bibitem[{\emph{{Gombosi}}(1987)}]{Gombosi1987}
{Gombosi} T.~I. (1987) \emph{{Charge exchange avalanche at the cometopause}},
  \emph{\grl}, \emph{14}, 1174--1177.

\bibitem[{\emph{Gombosi}(2015)}]{Gombosi2015}
Gombosi T.~I. (2015) \emph{Physics of Cometary Magnetospheres}, chap.~10, pp.
  169--188, American Geophysical Union (AGU).

\bibitem[{\emph{{G{\"o}tz} et~al.}(2019)\emph{{G{\"o}tz}, {Gunell}, {Volwerk},
  {Beth}, {Eriksson}, {Galand}, {Henri}, {Nilsson}, {Simon Wedlund}, {Alho},
  {Andersson}, {Andre}, {De Keyser}, {Deca}, {Ge}, {Gla{\ss}meier}, {Hajra},
  {Karlsson}, {Kasahara}, {Kolmasova}, {LLera}, {Madanian}, {Mann}, {Mazelle},
  {Odelstad}, {Plaschke}, {Rubin}, {Sanchez-Cano}, {Snodgrass}, and
  {Vigren}}}]{Goetz2019wp}
{G{\"o}tz} C., {Gunell} H., {Volwerk} M. et~al. (2019) \emph{{Cometary Plasma
  Science -- A White Paper in response to the Voyage 2050 Call by the European
  Space Agency}}, \emph{arXiv e-prints}, arXiv:1908.00377.

\bibitem[{\emph{Gringauz and Verigin}(1991)}]{Gringauz1991}
Gringauz K. and Verigin M. (1991) \emph{Permanent and nonstationary plasma
  phenomena in comet halley's head}, \emph{Cometary Plasma Processes},
  \emph{61}, 107--116.

\bibitem[{\emph{{Gringauz} et~al.}(1986)\emph{{Gringauz}, {Gombosi},
  {T{\'a}trallyay}, {Verigin}, {Remizov}, {Richter}, {Ap{\'a}thy}, {Szemerey},
  {Dyachkov}, {Balakina}, and {Nagy}}}]{Gringauz1986}
{Gringauz} K.~I., {Gombosi} T.~I., {T{\'a}trallyay} M. et~al. (1986)
  \emph{{Detection of a new {\textquotedblleft}chemical{\textquotedblright}
  boundary at comet Halley}}, \emph{Geophysical Research Letters}, \emph{13},
  613--616.

\bibitem[{\emph{{Gr{\"u}n} et~al.}(2016)\emph{{Gr{\"u}n}, {Agarwal},
  {Altobelli}, {Altwegg}, {Bentley}, {Biver}, {Della Corte}, {Edberg},
  {Feldman}, {Galand}, {Geiger}, {G{\"o}tz}, {Grieger}, {G{\"u}ttler}, {Henri},
  {Hofstadter}, {Horanyi}, {Jehin}, {Kr{\"u}ger}, {Lee}, {Mannel}, {Morales},
  {Mousis}, {M{\"u}ller}, {Opitom}, {Rotundi}, {Schmied}, {Schmidt}, {Sierks},
  {Snodgrass}, {Soja}, {Sommer}, {Srama}, {Tzou}, {Vincent},
  {Yanamandra-Fisher}, {A'Hearn}, {Erikson}, {Barbieri}, {Barucci}, {Bertaux},
  {Bertini}, {Burch}, {Colangeli}, {Cremonese}, {Da Deppo}, {Davidsson},
  {Debei}, {De Cecco}, {Deller}, {Feaga}, {Ferrari}, {Fornasier}, {Fulle},
  {Gicquel}, {Gillon}, {Green}, {Groussin}, {Guti{\'e}rrez}, {Hofmann},
  {Hviid}, {Ip}, {Ivanovski}, {Jorda}, {Keller}, {Knight}, {Knollenberg},
  {Koschny}, {Kramm}, {K{\"u}hrt}, {K{\"u}ppers}, {Lamy}, {Lara}, {Lazzarin},
  {L{\`o }pez-Moreno}, {Manfroid}, {Epifani}, {Marzari}, {Naletto}, {Oklay},
  {Palumbo}, {Parker}, {Rickman}, {Rodrigo}, {Rodr{\`\i}guez}, {Schindhelm},
  {Shi}, {Sordini}, {Steffl}, {Stern}, {Thomas}, {Tubiana}, {Weaver},
  {Weissman}, {Zakharov}, and {Taylor}}}]{Gruen2016}
{Gr{\"u}n} E., {Agarwal} J., {Altobelli} N. et~al. (2016) \emph{{The 2016 Feb
  19 outburst of comet 67P/CG: an ESA Rosetta multi- instrument study}},
  \emph{\mnras}, \emph{462}, S220--S234.

\bibitem[{\emph{Gulyaev}(2015)}]{guly15}
Gulyaev R.~A. (2015) \emph{Type {I} cometary tails and the solar wind at the
  epoch of the {Maunder} minimum}, \emph{Astron. Rep.}, \emph{59}, 791 -- 794.

\bibitem[{\emph{Gunell et~al.}(2017{\natexlab{a}})\emph{Gunell, Goetz,
  Eriksson, {Nilsson}, {Simon Wedlund}, {Henri}, {Maggiolo}, {Hamrin}, {De
  Keyser}, {Rubin}, {Stenberg Wieser}, {Cessateur}, {Dhooghe}, and
  {Gibbons}}}]{Gunell2017b}
Gunell H., Goetz C., Eriksson A. et~al. (2017{\natexlab{a}}) \emph{{Plasma
  waves confined to the diamagnetic cavity of comet 67P/Churyumov-
  Gerasimenko}}, \emph{\mnras}, \emph{469}, S84--S92.

\bibitem[{\emph{Gunell et~al.}(2021)\emph{Gunell, Goetz, Odelstad, Beth,
  Hamrin, Henri, Johansson, Nilsson, and Stenberg{ }Wieser}}]{Gunell2021}
Gunell H., Goetz C., Odelstad E. et~al. (2021) \emph{Ion acoustic waves near a
  comet nucleus: {R}osetta observations at comet
  {67P/C}huryumov-{G}erasimenko}, \emph{Annales Geophysicae}, \emph{39},
  53--68.

\bibitem[{\emph{{Gunell} et~al.}(2018)\emph{{Gunell}, {Goetz}, {Simon Wedlund},
  {Lindkvist}, {Hamrin}, {Nilsson}, {Llera}, {Eriksson}, and
  {Holmstr{\"o}m}}}]{Gunell2018a}
{Gunell} H., {Goetz} C., {Simon Wedlund} C. et~al. (2018) \emph{{The infant bow
  shock: a new frontier at a weak activity comet}}, \emph{\aap}, \emph{619},
  L2.

\bibitem[{\emph{{Gunell} et~al.}(2019)\emph{{Gunell}, {Lindkvist}, {Goetz},
  {Nilsson}, and {Hamrin}}}]{Gunell2019}
{Gunell} H., {Lindkvist} J., {Goetz} C. et~al. (2019) \emph{{Polarisation of a
  small-scale cometary plasma environment. Particle-in-cell modelling of comet
  67P/Churyumov-Gerasimenko}}, \emph{\aap}, \emph{631}, A174.

\bibitem[{\emph{Gunell et~al.}(2017{\natexlab{b}})\emph{Gunell, Nilsson,
  Hamrin, Eriksson, Odelstad, Maggiolo, Henri, Valli{\`e}res, Altwegg, Tzou,
  Rubin, Glassmeier, Stenberg{ }Wieser, {Simon Wedlund}, {De Keyser}, Dhooghe,
  Cessateur, and Gibbons}}]{Gunell2017a}
Gunell H., Nilsson H., Hamrin M. et~al. (2017{\natexlab{b}}) \emph{Ion acoustic
  waves at comet {67P/C}huryumov-{G}erasimenko -- {O}bservations and
  computations}, \emph{Astronomy and Astrophysics}, \emph{600}, A3.

\bibitem[{\emph{Gurnett et~al.}(1985)\emph{Gurnett, Anderson, H\"ausler,
  Haerendel, Bauer, Treumann, Koons, Holzworth, and Lühr}}]{Gurnett1985grl}
Gurnett D.~A., Anderson R.~R., H\"ausler B. et~al. (1985) \emph{Plasma waves
  associated with the ampte artificial comet}, \emph{Geophysical Research
  Letters}, \emph{12}, 851--854.

\bibitem[{\emph{{Hajra} et~al.}(2018)\emph{{Hajra}, {Henri}, {Myllys},
  {H{\'e}ritier}, {Galand}, {Simon Wedlund}, {Breuillard}, {Behar}, {Edberg},
  {Goetz}, {Nilsson}, {Eriksson}, {Goldstein}, {Tsurutani}, {Mor{\'e}},
  {Valli{\`e}res}, and {Wattieaux}}}]{Hajra2018}
{Hajra} R., {Henri} P., {Myllys} M. et~al. (2018) \emph{{Cometary plasma
  response to interplanetary corotating interaction regions during 2016
  June-September: a quantitative study by the Rosetta Plasma Consortium}},
  \emph{\mnras}, \emph{480}, 4544--4556.

\bibitem[{\emph{{Hajra} et~al.}(2017)\emph{{Hajra}, {Henri}, {Valli{\`e}res},
  {Galand}, {H{\'e}ritier}, {Eriksson}, {Odelstad}, {Edberg}, {Burch},
  {Broiles}, {Goldstein}, {Glassmeier}, {Richter}, {Goetz}, {Tsurutani},
  {Nilsson}, {Altwegg}, and {Rubin}}}]{Hajra2017}
{Hajra} R., {Henri} P., {Valli{\`e}res} X. et~al. (2017) \emph{{Impact of a
  cometary outburst on its ionosphere. Rosetta Plasma Consortium observations
  of the outburst exhibited by comet 67P /Churyumov-Gerasimenko on 19 February
  2016}}, \emph{\aap}, \emph{607}, A34.

\bibitem[{\emph{{Hansen} et~al.}(2016)\emph{{Hansen}, {Altwegg}, {Berthelier},
  {Bieler}, {Biver}, {Bockel{\'e}e-Morvan}, {Calmonte}, {Capaccioni}, {Combi},
  {De Keyser}, {Fiethe}, {Fougere}, {Fuselier}, {Gasc}, {Gombosi}, {Huang}, {Le
  Roy}, {Lee}, {Nilsson}, {Rubin}, {Shou}, {Snodgrass}, {Tenishev}, {Toth},
  {Tzou}, {Simon Wedlund}, and {Rosina Team}}}]{Hansen2016}
{Hansen} K.~C., {Altwegg} K., {Berthelier} J.-J. et~al. (2016) \emph{{Evolution
  of water production of 67P/Churyumov-Gerasimenko: An empirical model and a
  multi-instrument study}}, \emph{\mnras}, \emph{462}, S491--S506.

\bibitem[{\emph{{Haser}}(1957)}]{Haser1957}
{Haser} L. (1957) \emph{{Distribution d'intensit{\'e} dans la t{\^e}te d'une
  com{\`e}te}}, \emph{Bulletin de la Societe Royale des Sciences de Liege},
  \emph{43}, 740--750.

\bibitem[{\emph{Hayakawa et~al.}(2021)\emph{Hayakawa, Fujii, Murata, Mitsuma,
  Cheng, Nogami, K., Sano, Tsumura, Kawamoto, and Nishino}}]{haya21}
Hayakawa H., Fujii Y.~I., Murata K. et~al. (2021) \emph{Three case reports on
  the cometary plasma tail in the historical documents}, \emph{J. Space Weather
  Space Clim.}, \emph{11}, 21.

\bibitem[{\emph{{Henri} et~al.}(2017)\emph{{Henri}, {Valli{\`e}res}, {Hajra},
  {Goetz}, {Richter}, {Glassmeier}, {Galand}, {Rubin}, {Eriksson}, {Nemeth},
  {Vigren}, {Beth}, {Burch}, {Carr}, {Nilsson}, {Tsurutani}, and
  {Wattieaux}}}]{Henri2017}
{Henri} P., {Valli{\`e}res} X., {Hajra} R. et~al. (2017) \emph{{Diamagnetic
  region(s): structure of the unmagnetized plasma around Comet 67P/CG}},
  \emph{\mnras}, \emph{469}, S372--S379.

\bibitem[{\emph{Heritier et~al.}(2018)\emph{Heritier, {Galand, M.}, {Henri,
  P.}, {Johansson, F. L.}, {Beth, A.}, {Eriksson, A. I.}, {Valli\`eres, X.},
  {Altwegg, K.}, {Burch, J. L.}, {Carr, C.}, {Ducrot, E.}, {Hajra, R.}, and
  {Rubin, M.}}}]{Heritier2018}
Heritier K.~L., {Galand, M.}, {Henri, P.} et~al. (2018) \emph{Plasma source and
  loss at comet 67{P} during the {R}osetta mission}, \emph{A\&A}, \emph{618},
  A77.

\bibitem[{\emph{Hewish et~al.}(1964)\emph{Hewish, Scott, and Wills}}]{hewi64}
Hewish A., Scott P.~F., and Wills D. (1964) \emph{Interplanetary scintillation
  of small diameter radio sources}, \emph{Nature}, \emph{203}, 1214 -- 1217.

\bibitem[{\emph{{Huang} et~al.}(2018)\emph{{Huang}, {T{\'o}th}, {Gombosi},
  {Jia}, {Combi}, {Hansen}, {Fougere}, {Shou}, {Tenishev}, {Altwegg}, and
  {Rubin}}}]{Huang2018}
{Huang} Z., {T{\'o}th} G., {Gombosi} T.~I. et~al. (2018) \emph{{Hall effect in
  the coma of 67P/Churyumov-Gerasimenko}}, \emph{\mnras}, \emph{475},
  2835--2841.

\bibitem[{\emph{{Huang} et~al.}(2016)\emph{{Huang}, {T{\'o}th}, {Gombosi},
  {Jia}, {Rubin}, {Fougere}, {Tenishev}, {Combi}, {Bieler}, {Hansen}, {Shou},
  and {Altwegg}}}]{Huang2016}
{Huang} Z., {T{\'o}th} G., {Gombosi} T.~I. et~al. (2016) \emph{{Four-fluid MHD
  simulations of the plasma and neutral gas environment of comet
  67P/Churyumov-Gerasimenko near perihelion}}, \emph{Journal of Geophysical
  Research (Space Physics)}, \emph{121}, 4247--4268.

\bibitem[{\emph{{Huddleston} et~al.}(1993)\emph{{Huddleston}, {Coates},
  {Johnstone}, and {Neubauer}}}]{Huddleston1993}
{Huddleston} D.~E., {Coates} A.~J., {Johnstone} A.~D. et~al. (1993) \emph{{Mass
  loading and velocity diffusion models for heavy pickup ions at comet
  Grigg-Skjellerup}}, \emph{\jgr}, \emph{98}, 20995--21002.

\bibitem[{\emph{Hyder et~al.}(1974)\emph{Hyder, Brandt, and Roosen}}]{hyde74}
Hyder C.~L., Brandt J.~C., and Roosen R.~G. (1974) \emph{Tail structures far
  from the head of comet {Kohoutek}. {I}}, \emph{Icarus}, \emph{23}, 601 --
  610.

\bibitem[{\emph{Iju et~al.}(2015)\emph{Iju, Abe, Tokumaru, and Fujiki}}]{iju15}
Iju T., Abe S., Tokumaru M. et~al. (2015) \emph{Plasma distribution of {Comet
  ISON (C/2021 S1)} observed using the radio scintillation method},
  \emph{Icarus}, \emph{252}, 301 -- 310.

\bibitem[{\emph{Ip}(1979)}]{ip79}
Ip W.-H. (1979) \emph{Currents in the cometary atmosphere}, \emph{Planet. Space
  Sci.}, \emph{27}, 121 -- 125.

\bibitem[{\emph{Ip}(1989)}]{Ip1989}
Ip W.-H. (1989) \emph{On charge exchange effect in the vicinity of the
  cometopause of comet {H}alley}, \emph{Astrophysical Journal}, \emph{343},
  946.

\bibitem[{\emph{Ip}(1994)}]{ip94}
Ip W.-H. (1994) \emph{On a thermodynamical origin of the cometary ion rays},
  \emph{Astrophys. J.}, \emph{432}, L143 -- L145.

\bibitem[{\emph{{Ip}}(2004)}]{Ip2004}
{Ip} W.~H. (2004) in \emph{Comets II} (M.~C. {Festou}, H.~U. {Keller}, and
  H.~A. {Weaver}, eds.), p. 605.

\bibitem[{\emph{{Ip} and {Axford}}(1982)}]{IpAxford1982}
{Ip} W.~H. and {Axford} W.~I. (1982) in \emph{IAU Colloq. 61: Comet
  Discoveries, Statistics, and Observational Selection} (L.~L. {Wilkening},
  ed.), pp. 588--634.

\bibitem[{\emph{Ip and Mendis}(1975)}]{Ip75}
Ip W.-H. and Mendis D.~A. (1975) \emph{The cometary magnetic field and its
  associated electric currents}, \emph{Icarus}, \emph{26}, 457 -- 461.

\bibitem[{\emph{Ip and Mendis}(1976)}]{ip76}
Ip W.-H. and Mendis D.~A. (1976) \emph{The generation of magnetic fields and
  electric currents in cometary plasma tails}, \emph{Icarus}, \emph{29}, 147 --
  151.

\bibitem[{\emph{Israelevich and Ershkovich}(2014)}]{isra14}
Israelevich P. and Ershkovich A. (2014) \emph{Magnetic tension in the tails of
  {Titan, Venus and comet Halley}}, \emph{Planet. Space Sci.}, \emph{103}, 339
  -- 346.

\bibitem[{\emph{{Itikawa} and {Mason}}(2005)}]{ItikawaMason2005}
{Itikawa} Y. and {Mason} N. (2005) \emph{{Cross Sections for Electron
  Collisions with Water Molecules}}, \emph{Journal of Physical and Chemical
  Reference Data}, \emph{34}, 1--22.

\bibitem[{\emph{Jia et~al.}(2009)\emph{Jia, Russell, Jian, Manchester, Cohen,
  Vourlidas, Hansen, Combi, and Gombosi}}]{jia09}
Jia Y.~D., Russell C.~T., Jian L.~K. et~al. (2009) \emph{Study of the 2007
  {April} 20 {CME-comet} interaction event with an {MHD} model},
  \emph{Astrophys. J.}, \emph{696}, L56 -- L60.

\bibitem[{\emph{Jockers}(1985)}]{jock85}
Jockers K. (1985) \emph{The ion tail of {Comet Kohoutek 1973 XII} during 17
  days of solar wind gusts}, \emph{Astron. Astrophys. Suppl. Ser.}, \emph{62},
  791 -- 838.

\bibitem[{\emph{Johnson et~al.}(2004)\emph{Johnson, Thompson, and
  Hourigan}}]{john04}
Johnson S.~A., Thompson M.~C., and Hourigan K. (2004) \emph{Predicted low
  frequency structures in the wake of elliptical cylinders}, \emph{Eur. J.
  Mech. B.}, \emph{23}, 229 -- 239.

\bibitem[{\emph{{Jones} et~al.}(2000)\emph{{Jones}, {Balogh}, and
  {Horbury}}}]{Jones2000}
{Jones} G.~H., {Balogh} A., and {Horbury} T.~S. (2000) \emph{Identification of
  comet {H}yakutake's extremely long ion tail from magnetic field signatures},
  \emph{\nat}, \emph{404}, 574--576.

\bibitem[{\emph{Jones and Brandt}(2004)}]{jone04}
Jones G.~H. and Brandt J.~C. (2004) \emph{The interaction of comet
  {153P/Ikeya-Zhang} with interplanetary coronal mass ejections: Identification
  of fast {ICME} signatures}, \emph{Geophys. Res. Lett.}, \emph{31}, L20805.

\bibitem[{\emph{{Karlsson} et~al.}(2017)\emph{{Karlsson}, {Eriksson},
  {Odelstad}, {Andr{\'e}}, {Dickeli}, {Kullen}, {Lindqvist}, {Nilsson}, and
  {Richter}}}]{Karlsson2017}
{Karlsson} T., {Eriksson} A.~I., {Odelstad} E. et~al. (2017) \emph{{R}osetta
  measurements of lower hybrid frequency range electric field oscillations in
  the plasma environment of comet {67P}}, \emph{Geophysical Research Letters},
  \emph{44}, 1641--1651.

\bibitem[{\emph{Kinoshita et~al.}(1996)\emph{Kinoshita, Fukushima, Watanabe,
  and Yamamoto}}]{kino96}
Kinoshita D., Fukushima H., Watanabe J.-I. et~al. (1996) \emph{Ion tail
  disturbance of comet {C/Hyakutake 1996B2} observed around the closest
  approach to the {Earth}}, \emph{Publ. Aston. Soc. Japan}, \emph{48}, L83 --
  L86.

\bibitem[{\emph{Kirsch et~al.}(1989)\emph{Kirsch, {McKenna-Lawlor}, Daly,
  Korth, Neubauer, {O'Sullivan}, Thompson, and Wenzel}}]{kirs89}
Kirsch E., {McKenna-Lawlor} S., Daly P. et~al. (1989) \emph{Evidence for the
  field line reconnection process in the particle and magnetic field
  measurements obtained during the {Giotto-Halley} encounter}, \emph{Ann.
  Geophys.}, \emph{7}, 107 -- 113.

\bibitem[{\emph{{Koenders} et~al.}(2013)\emph{{Koenders}, {Glassmeier},
  {Richter}, {Motschmann}, and {Rubin}}}]{Koenders2013}
{Koenders} C., {Glassmeier} K.-H., {Richter} I. et~al. (2013) \emph{{Revisiting
  cometary bow shock positions}}, \emph{Planetary and Space Science},
  \emph{87}, 85--95.

\bibitem[{\emph{{Koenders} et~al.}(2015)\emph{{Koenders}, {Glassmeier},
  {Richter}, {Ranocha}, and {Motschmann}}}]{Koenders2015}
{Koenders} C., {Glassmeier} K.-H., {Richter} I. et~al. (2015) \emph{{Dynamical
  features and spatial structures of the plasma interaction region of
  67P/Churyumov-Gerasimenko and the solar wind}}, \emph{\planss}, \emph{105},
  101--116.

\bibitem[{\emph{{Koenders} et~al.}(2016{\natexlab{a}})\emph{{Koenders},
  {Goetz}, {Richter}, {Motschmann}, and {Glassmeier}}}]{Koenders2016}
{Koenders} C., {Goetz} C., {Richter} I. et~al. (2016{\natexlab{a}})
  \emph{{Magnetic field pile-up and draping at intermediately active comets:
  results from comet 67P/Churyumov-Gerasimenko at 2.0 AU}}, \emph{\mnras},
  \emph{462}, S235--S241.

\bibitem[{\emph{{Koenders} et~al.}(2016{\natexlab{b}})\emph{{Koenders},
  {Perschke}, {Goetz}, {Richter}, {Motschmann}, and
  {Glassmeier}}}]{Koenders2016waves}
{Koenders} C., {Perschke} C., {Goetz} C. et~al. (2016{\natexlab{b}})
  \emph{{Low-frequency waves at comet 67P/Churyumov-Gerasimenko. Observations
  compared to numerical simulations}}, \emph{\aap}, \emph{594}, A66.

\bibitem[{\emph{{Kramer} et~al.}(2014)\emph{{Kramer}, {Fernandez}, {Lisse},
  {Kelley}, and {Woodney}}}]{Kramer2014}
{Kramer} E.~A., {Fernandez} Y.~R., {Lisse} C.~M. et~al. (2014) \emph{{A
  dynamical analysis of the dust tail of Comet C/1995 O1 (Hale-Bopp) at high
  heliocentric distances}}, \emph{\icarus}, \emph{236}, 136--145.

\bibitem[{\emph{Kuchar et~al.}(2008)\emph{Kuchar, Buffington, Arge, Hick,
  Howard, Jackson, Johnston, Mizuno, Tappin, and Webb}}]{kuch08}
Kuchar T.~A., Buffington A., Arge C.~N. et~al. (2008) \emph{Observations of a
  comet tail disruption by the passage of a {CME}}, \emph{J. Geophys. Res.},
  \emph{113}, A04101.

\bibitem[{\emph{{Lavorenti} et~al.}(2021)\emph{{Lavorenti}, {Henri},
  {Califano}, {Aizawa}, and {Andr{\'e}}}}]{Lavorenti_etal_2021}
{Lavorenti} F., {Henri} P., {Califano} F. et~al. (2021) \emph{{Electron
  acceleration driven by the lower-hybrid-drift instability: an extended
  quasilinear model}}, \emph{arXiv e-prints}, arXiv:2104.05011.

\bibitem[{\emph{Lee}(1976)}]{lee76}
Lee L.~C. (1976) \emph{Plasma irregularities in the comet's tails},
  \emph{Astrophys. J.}, \emph{210}, 254 -- 257.

\bibitem[{\emph{Li et~al.}(2018)\emph{Li, Li, Wang, Yuan, Zhu, y.~Zhong, Wei,
  Li, Han, Zhang, Pei, Zhang, Zhao, Lui, Liao, Fang, ans X.-G.~Wang, Sakawa,
  ans X.~Lu, Hua, Zhu, Morita, Kuramitsu, Huang, Fu, Zhu, Zhao, and
  Zhang}}]{li18}
Li Y.-F., Li Y.-T., Wang W.-M. et~al. (2018) \emph{Laboratory study on
  disconnection events in comets}, \emph{Sci. Reports}, \emph{8}, 463.

\bibitem[{\emph{{Lishawa} et~al.}(1990)\emph{{Lishawa}, {Dressler}, {Gardner},
  {Salter}, and {Murad}}}]{Lishawa1990}
{Lishawa} C.~R., {Dressler} R.~A., {Gardner} J.~A. et~al. (1990) \emph{{Cross
  sections and product kinetic energy analysis of H$_{2}$O$^{ + }$-H$_{2}$O
  collisions at suprathermal energies}}, \emph{\jcp}, \emph{93}, 3196--3206.

\bibitem[{\emph{{Luehr} et~al.}(1988)\emph{{Luehr}, {Kloecker}, and
  {Acu{\~n}a}}}]{Luehr1988}
{Luehr} H., {Kloecker} N., and {Acu{\~n}a} M.~H. (1988) \emph{{The diamagnetic
  effect during AMPTE's tail releases - Initial results}}, \emph{Advances in
  Space Research}, \emph{8}, 11--14.

\bibitem[{\emph{Lundin}(2011)}]{lund11}
Lundin R. (2011) \emph{Ion acceleration and outflow from {Mars} and {Venus}: an
  overview}, \emph{Space Sci. Rev.}, \emph{162}, 309 -- 334.

\bibitem[{\emph{Madanian et~al.}(2020)\emph{Madanian, Burch, Eriksson, Cravens,
  Galand, Vigren, Goldstein, Nemeth, Mokashi, Richter, and
  Rubin}}]{Madanian2020}
Madanian H., Burch J., Eriksson A. et~al. (2020) \emph{Electron dynamics near
  diamagnetic regions of comet 67p/churyumov- gerasimenko}, \emph{Planetary and
  Space Science}, \emph{187}, 104924.

\bibitem[{\emph{{Madanian} et~al.}(2016)\emph{{Madanian}, {Cravens}, {Rahmati},
  {Goldstein}, {Burch}, {Eriksson}, {Edberg}, {Henri}, {Mandt}, {Clark},
  {Rubin}, {Broiles}, and {Reedy}}}]{Madanian_etal_2016}
{Madanian} H., {Cravens} T.~E., {Rahmati} A. et~al. (2016) \emph{{Suprathermal
  electrons near the nucleus of comet 67P/Churyumov-Gerasimenko at 3 AU: Model
  comparisons with Rosetta data}}, \emph{Journal of Geophysical Research (Space
  Physics)}, \emph{121}, 5815--5836.

\bibitem[{\emph{{Madsen} et~al.}(2018)\emph{{Madsen}, {Simon Wedlund},
  {Eriksson}, {Goetz}, {Karlsson}, {Gunell}, {Spicher}, {Henri},
  {Valli{\`e}res}, and {Miloch}}}]{Madsen2018}
{Madsen} B., {Simon Wedlund} C., {Eriksson} A. et~al. (2018) \emph{Extremely
  low-frequency waves inside the diamagnetic cavity of comet
  {67P/C}huryumov-{G}erasimenko}, \emph{Geophysical Research Letters},
  \emph{45}, 3854--3864.

\bibitem[{\emph{Mandt et~al.}(2019)\emph{Mandt, Eriksson, Beth, Galand, and
  Vigren}}]{Mandt2019}
Mandt K., Eriksson A., Beth A. et~al. (2019) \emph{Influence of collisions on
  ion dynamics in the inner comae of four comets}, \emph{Astronomy \&
  Astrophysics}, \emph{630}, A48.

\bibitem[{\emph{{Mandt} et~al.}(2016)\emph{{Mandt}, {Eriksson}, {Edberg},
  {Koenders}, {Broiles}, {Fuselier}, {Henri}, {Nemeth}, {Alho}, {Biver},
  {Beth}, {Burch}, {Carr}, {Chae}, {Coates}, {Cupido}, {Galand}, {Glassmeier},
  {Goetz}, {Goldstein}, {Hansen}, {Haiducek}, {Kallio}, {Lebreton},
  {Luspay-Kuti}, {Mokashi}, {Nilsson}, {Opitz}, {Richter}, {Samara}, {Szego},
  {Tzou}, {Volwerk}, {Simon Wedlund}, and {Stenberg Wieser}}}]{Mandt2016}
{Mandt} K.~E., {Eriksson} A., {Edberg} N.~J.~T. et~al. (2016) \emph{{RPC
  observation of the development and evolution of plasma interaction boundaries
  at 67P/Churyumov-Gerasimenko}}, \emph{\mnras}, \emph{462}, S9--S22.

\bibitem[{\emph{{Marshall} et~al.}(2019)\emph{{Marshall}, {Rezac}, {Hartogh},
  {Zhao}, and {Attree}}}]{Marshall2019}
{Marshall} D., {Rezac} L., {Hartogh} P. et~al. (2019) \emph{{Interpretation of
  heliocentric water production rates of comets}}, \emph{\aap}, \emph{623},
  A120.

\bibitem[{\emph{{Masunaga} et~al.}(2019)\emph{{Masunaga}, {Nilsson}, {Behar},
  {Stenberg Wieser}, {Wieser}, and {Goetz}}}]{Masunaga2019}
{Masunaga} K., {Nilsson} H., {Behar} E. et~al. (2019) \emph{{Flow pattern of
  accelerated cometary ions inside and outside the diamagnetic cavity of comet
  67P/Churyumov-Gerasimenko }}, \emph{\aap}.

\bibitem[{\emph{{McComas} et~al.}(1987)\emph{{McComas}, Gosling, Bame, Slavin,
  Smith, and Steinberg}}]{mcco87a}
{McComas} D.~J., Gosling J.~T., Bame S.~J. et~al. (1987) \emph{The
  {Giacobini-Zinner} magnetotail: Tail configuration and current sheet},
  \emph{J. Geophys. Res.}, \emph{92}, 1139 -- 1152.

\bibitem[{\emph{Meier et~al.}(2016)\emph{Meier, Glassmeier, and
  Motschmann}}]{Meier2016}
Meier P., Glassmeier K.-H., and Motschmann U. (2016) \emph{Modified
  ion-{W}eibel instability as a possible source of wave activity at {C}omet
  {67P/C}huryumov-{G}erasimenko}, \emph{Annales Geophysicae}, \emph{34},
  691--707.

\bibitem[{\emph{Mendis et~al.}(1989)\emph{Mendis, Flammer, Reme, Sauvaud,
  d'Uston, Cotin, Cros, Anderson, Carlson, Curtis et~al.}}]{Mendis1989}
Mendis D., Flammer K., Reme H. et~al. (1989) in \emph{Annales Geophysicae},
  vol.~7, pp. 99--106.

\bibitem[{\emph{Mendis et~al.}(1986)\emph{Mendis, Smith, Tsurutani, Slavin,
  Jones, and Siscoe}}]{Mendis1986}
Mendis D., Smith E., Tsurutani B. et~al. (1986) \emph{Comet-solar wind
  interaction: Dynamical length scales and models}, \emph{Geophysical research
  letters}, \emph{13}, 239--242.

\bibitem[{\emph{Mendis}(2007)}]{mend07}
Mendis D.~A. (2007) in \emph{Handbook of the Solar-Terrestrial Environment}
  (Y.~Kamide and A.~C.-L. Chian, eds.), p. 494, Springer, Berlin, DE.

\bibitem[{\emph{{Meyer-Vernet} et~al.}(1986)\emph{{Meyer-Vernet}, Couturier,
  Hoang, Perche, and Steinberg}}]{meye86}
{Meyer-Vernet} N., Couturier P., Hoang S. et~al. (1986) \emph{Physical
  parameters fro hot and cold electron populations in comet {Giacobini-Zinner}
  with the {ICE} radio experiment}, \emph{Geophys. Res. Lett}, \emph{13}, 179
  -- 181.

\bibitem[{\emph{{M\"ostl} et~al.}(2014)\emph{{M\"ostl}, Amla, Hall, Liewer, {De
  Jong}, Coaninno, Veronig, Rollett, Temmer, Peinhart, Davies, Lugaz, Liu,
  Farrugia, Luhmann, {Vr$\check{\rm s}$nak}, Harrison, and Galvin}}]{moes14}
{M\"ostl} C., Amla K., Hall J.~R. et~al. (2014) \emph{Connecting speeds,
  directions and arrival times of 22 coronal mass ejections from the {Sun} to 1
  {AU}}, \emph{Astrophys. J.}, \emph{787}, 119.

\bibitem[{\emph{Mozhenkov and Vaisberg}(2017)}]{mozh17}
Mozhenkov R.~T. and Vaisberg O.~L. (2017) \emph{On the classification of comet
  plasma tails}, \emph{Sol. Sys. Res.}, \emph{51}, 258 -- 270.

\bibitem[{\emph{Myllys et~al.}(2019)\emph{Myllys, Henri, Galand, {Heritier},
  {Gilet}, {Goldstein}, {Eriksson}, {Johansson}, and {Deca}}}]{Myllys2019}
Myllys M., Henri P., Galand M. et~al. (2019) \emph{Plasma properties of
  suprathermal electrons near comet 67{P/C}huryumov-{G}erasimenko with
  {R}osetta}, \emph{Astronomy \& Astrophysics}, \emph{630}, A42.

\bibitem[{\emph{{Myllys} et~al.}(2021)\emph{{Myllys}, {Henri}, {Valli{\`e}res},
  {Gilet}, {Nilsson}, {Palmerio}, {Turc}, {Wellbrock}, {Goldstein}, and
  {Witasse}}}]{Myllys2021}
{Myllys} M., {Henri} P., {Valli{\`e}res} X. et~al. (2021) \emph{{Electric field
  measurements at the plasma frequency around comet 67P by RPC-MIP on board
  Rosetta}}, \emph{\aap}, \emph{652}, A73.

\bibitem[{\emph{{Nemeth} et~al.}(2016)\emph{{Nemeth}, {Burch}, {Goetz},
  {Goldstein}, {Henri}, {Koenders}, {Madanian}, {Mandt}, {Mokashi}, {Richter},
  {Timar}, and {Szego}}}]{Nemeth2016}
{Nemeth} Z., {Burch} J., {Goetz} C. et~al. (2016) \emph{{Charged particle
  signatures of the diamagnetic cavity of comet 67P/Churyumov-Gerasimenko}},
  \emph{\mnras}, \emph{462}, S415--S421.

\bibitem[{\emph{{Neubauer}}(1988)}]{Neubauer1988}
{Neubauer} F.~M. (1988) \emph{{The ionopause transition and boundary layers at
  Comet Halley from Giotto magnetic field observations}}, \emph{Journal of
  Geophysical Research}, \emph{93}, 7272--7281.

\bibitem[{\emph{{Neubauer} et~al.}(1986)\emph{{Neubauer}, {Glassmeier}, {Pohl},
  {Raeder}, {Acu{\~n}a}, {Burlaga}, {Ness}, {Musmann}, {Mariani}, {Wallis},
  {Ungstrup}, and {Schmidt}}}]{Neubauer1986}
{Neubauer} F.~M., {Glassmeier} K.~H., {Pohl} M. et~al. (1986) \emph{{First
  results from the Giotto magnetometer experiment at comet Halley}},
  \emph{Nature}, \emph{321}, 352--355.

\bibitem[{\emph{{Nicolaou} et~al.}(2017)\emph{{Nicolaou}, {Behar}, {Nilsson},
  {Wieser}, {Yamauchi}, {Ber{\v{c}}i{\v{c}}}, and {Wieser}}}]{Nicolaou2017}
{Nicolaou} G., {Behar} E., {Nilsson} H. et~al. (2017) \emph{{Energy-angle
  dispersion of accelerated heavy ions at 67P/Churyumov-Gerasimenko:
  implication in the mass-loading mechanism}}, \emph{\mnras}, \emph{469},
  S339--S345.

\bibitem[{\emph{Niedner}(1981)}]{nied81}
Niedner M.~B. (1981) \emph{Interplanetary gas. {XXVII} - a catalog of
  disconnection events in cometary plasma tails}, \emph{Astrophys. J. Suppl.
  Ser.}, \emph{46}, 141 -- 157.

\bibitem[{\emph{Niedner and Brandt}(1978)}]{Niedner1978}
Niedner M.~B. and Brandt J.~C. (1978) \emph{Interplanetary gas. {XXII} - plasma
  tail disconnection events in comets - evidence for magnetic field line
  reconnection at interplanetary sector boundaries}, \emph{Astrophys. J.},
  \emph{223}, 655--670.

\bibitem[{\emph{Nilsson et~al.}(2018)\emph{Nilsson, Gunell, Karlsson, Brenning,
  Henri, Goetz, Eriksson, Behar, Stenberg{ }Wieser, and
  Valli{\`e}res}}]{Nilsson2018}
Nilsson H., Gunell H., Karlsson T. et~al. (2018) \emph{Size of a plasma cloud
  matters: The polarisation electric field of a small-scale comet ionosphere},
  \emph{Astronomy \& Astrophysics}, \emph{616}, A50.

\bibitem[{\emph{Nilsson et~al.}(2017)\emph{Nilsson, Stenberg{ }Wieser, Behar,
  Gunell, Galand, Simon{ }Wedlund, Alho, Goetz, Yamauchi, Henri, and
  Eriksson}}]{Nilsson2017}
Nilsson H., Stenberg{ }Wieser G., Behar E. et~al. (2017) \emph{Evolution of the
  ion environment of comet 67{P} during the rosetta mission as seen by
  {RPC-ICA}}, \emph{Monthly Notices of the Royal Astronomical Society},
  \emph{469}, S252--S261.

\bibitem[{\emph{Nilsson et~al.}(2015)\emph{Nilsson, Stenberg{ }Wieser, Behar,
  Simon{ }Wedlund, Gunell, Yamauchi, Lundin, Barabash, Wieser, Carr, Cupido,
  Burch, Fedorov, Sauvaud, Koskinen, Kallio, Lebreton, Eriksson, Edberg,
  Goldstein, Henri, Koenders, Mokashi, Nemeth, Richter, Szego, Volwerk, Vallat,
  and Rubin}}]{Nilsson2015a}
Nilsson H., Stenberg{ }Wieser G., Behar E. et~al. (2015) \emph{Birth of a comet
  magnetosphere: A spring of water ions}, \emph{Science}, \emph{347}.

\bibitem[{\emph{Nilsson et~al.}(2020)\emph{Nilsson, Williamson, Bergman,
  Stenberg~Wieser, Wieser, Behar, Eriksson, Johansson, Richter, and
  Goetz}}]{Nilsson2020}
Nilsson H., Williamson H., Bergman S. et~al. (2020) \emph{{Average cometary ion
  flow pattern in the vicinity of comet 67P from moment data}}, \emph{Monthly
  Notices of the Royal Astronomical Society}, \emph{498}, 5263--5272.

\bibitem[{\emph{{Nistic\`o} et~al.}(2018)\emph{{Nistic\`o}, Valdimirov,
  Nakariakov, Battams, and Bothmer}}]{nist18}
{Nistic\`o} G., Valdimirov V., Nakariakov V.~M. et~al. (2018)
  \emph{Oscillations of cometary tails: a vortex shedding phenomenon?},
  \emph{Astron. Astrophys.}, \emph{615}, A143.

\bibitem[{\emph{Odelstad et~al.}(2020)\emph{Odelstad, Eriksson, Andre{\'e},
  Graham, Karlsson, Vaivads, Vigren, Goetz, Nilsson, Henri, and Stenberg{
  }Wieser}}]{Odelstad2020}
Odelstad E., Eriksson A.~I., Andre{\'e} M. et~al. (2020) \emph{Plasma density
  and magnetic field fluctuations in the ion gyro-frequency range near the
  diamagnetic cavity of comet {67P}}, \emph{\jgr}, p. e2020JA028592.

\bibitem[{\emph{Odelstad et~al.}(2018)\emph{Odelstad, Eriksson, Johansson,
  Vigren, Henri, Gilet, Heritier, Vallières, Rubin, and
  André}}]{Odelstad2018}
Odelstad E., Eriksson A.~I., Johansson F.~L. et~al. (2018) \emph{Ion velocity
  and electron temperature inside and around the diamagnetic cavity of comet
  {67P}}, \emph{Journal of Geophysical Research: Space Physics}, \emph{123},
  5870--5893.

\bibitem[{\emph{{\"Opik}}(1964)}]{opik64}
{\"Opik} E.~J. (1964) \emph{The motion of the condensation in the tail of
  {Halley's} comet {June} 5-8, 1910}, \emph{Zeitsschr. Astrophys.}, \emph{58},
  192 -- 201.

\bibitem[{\emph{Ostaszewski et~al.}(2020)\emph{Ostaszewski, Glassmeier, Goetz,
  Heinisch, Henri, Ranocha, Richter, Rubin, and Tsurutani}}]{Ostaszewski2021}
Ostaszewski K., Glassmeier K.-H., Goetz C. et~al. (2020) \emph{Steepening of
  magnetosonic waves in the inner coma of comet 67p/churyumov-gerasimenko},
  \emph{Annales Geophysicae Discussions}, \emph{2020}, 1--37.

\bibitem[{\emph{{Ostaszewski} et~al.}(2020)\emph{{Ostaszewski}, {Heinisch},
  {Richter}, {Kroll}, {Balke}, {Fraga}, and {Glassmeier}}}]{Ostaszewski2020}
{Ostaszewski} K., {Heinisch} P., {Richter} I. et~al. (2020) \emph{{Pattern
  recognition in time series for space missions: A rosetta magnetic field case
  study}}, \emph{Acta Astronautica}, \emph{168}, 123--129.

\bibitem[{\emph{Oya et~al.}(1986)\emph{Oya, Morioka, Miyake, Smith, and
  Tsurutani}}]{Oya1986}
Oya H., Morioka A., Miyake W. et~al. (1986) \emph{Discovery of cometary
  kilometric radiations and plasma waves at comet {H}alley}, \emph{Nature},
  \emph{321}, 307--310.

\bibitem[{\emph{Parker}(1957)}]{park57}
Parker E.~N. (1957) \emph{Sweet's mechanism for merging magnetic fields in
  conducting fluids}, \emph{J. Geophys. Res.}, \emph{62}, 509 -- 520.

\bibitem[{\emph{{Parker}}(1958)}]{Parker1958}
{Parker} E.~N. (1958) \emph{{Dynamics of the Interplanetary Gas and Magnetic
  Fields.}}, \emph{\apj}, \emph{128}, 664.

\bibitem[{\emph{Petschek}(1964)}]{pets64}
Petschek H.~E. (1964) \emph{Magnetic field annihilation}, \emph{NASA Special
  Publication}, \emph{50}, 425 -- 439.

\bibitem[{\emph{{Pierrard} et~al.}(2016)\emph{{Pierrard}, {Lazar}, {Poedts},
  {{\v{S}}tver{\'a}k}, {Maksimovic}, and
  {Tr{\'a}vn{\'\i}{\v{c}}ek}}}]{Pierrard2016}
{Pierrard} V., {Lazar} M., {Poedts} S. et~al. (2016) \emph{{The Electron
  Temperature and Anisotropy in the Solar Wind. Comparison of the Core and Halo
  Populations}}, \emph{\solphys}, \emph{291}, 2165--2179.

\bibitem[{\emph{{Plaschke} et~al.}(2018)\emph{{Plaschke}, {Karlsson},
  {G{\"o}tz}, {M{\"o}stl}, {Richter}, {Volwerk}, {Eriksson}, {Behar}, and
  {Goldstein}}}]{Plaschke2018}
{Plaschke} F., {Karlsson} T., {G{\"o}tz} C. et~al. (2018) \emph{{First
  observations of magnetic holes deep within the coma of a comet}},
  \emph{\aap}, \emph{618}, A114.

\bibitem[{\emph{{Price} et~al.}(2019)\emph{{Price}, {Jones}, {Morrill},
  {Owens}, {Battams}, {Morgan}, {Dr{\"u}ckmuller}, and {Deiries}}}]{Price2019}
{Price} O., {Jones} G.~H., {Morrill} J. et~al. (2019) \emph{{Fine-scale
  structure in cometary dust tails I: Analysis of striae in Comet C/2006 P1
  (McNaught) through temporal mapping}}, \emph{\icarus}, \emph{319}, 540--557.

\bibitem[{\emph{Raeder et~al.}(1987)\emph{Raeder, Neubauer, Ness, and
  Burlaga}}]{Raeder1987}
Raeder J., Neubauer F.~M., Ness N.~F. et~al. (1987) \emph{Macroscopic
  perturbations of the {IMF by P/Halley} as seen by the {Giotto} magnetometer},
  \emph{Astron. Astrophys.}, \emph{187}, 61--64.

\bibitem[{\emph{Rahe}(1968)}]{rahe68}
Rahe J. (1968) \emph{The structure of tail rays in the coma region of comets},
  \emph{Zeitschr. Astrophys.}, \emph{68}, 208 -- 213.

\bibitem[{\emph{Rahe and Donn}(1969)}]{Rahe1969}
Rahe J. and Donn B. (1969) \emph{Ionization and ray formation in comets},
  \emph{Astron. J.}, \emph{74}, 256--258.

\bibitem[{\emph{Ramanjooloo}(2014)}]{rama14}
Ramanjooloo Y. (2014) \emph{How comets reveal structure of the inner
  heliosphere}, \emph{Astron. Geophys.}, \emph{55}, 1.32 -- 1.36.

\bibitem[{\emph{Ramanjooloo and Jones}(2022)}]{rama22}
Ramanjooloo Y. and Jones G.~H. (2022) \emph{Solar wind velocities at comets
  c/2011 l4 pan-starrs and c/2013 r1 lovejoy derived using a new image analysis
  technique}, \emph{J. Geophys. Res.}, \emph{in press}.

\bibitem[{\emph{{Rauer} and {Jockers}}(1993)}]{Rauer1993}
{Rauer} H. and {Jockers} K. (1993) \emph{{Doppler Measurements of the H $_{2}$O
  $^{+}$ Ion Velocity in the Plasma Tail of Comet Levy 1990c}}, \emph{\icarus},
  \emph{102}, 117--133.

\bibitem[{\emph{R{\`e}me et~al.}(1994)\emph{R{\`e}me, Mazelle, {D'Uston},
  {Korth}, {Lin}, and {Chaizy}}}]{Reme1994}
R{\`e}me H., Mazelle C., {D'Uston} C. et~al. (1994) \emph{There is no
  'cometopause' at comet {H}alley}, \emph{J.\ Geophys.\ Res.}, \emph{99},
  2301--2308.

\bibitem[{\emph{{Richter} et~al.}(2016)\emph{{Richter}, {Auster}, {Berghofer},
  {Carr}, {Cupido}, {Forna{\c c}on}, {Goetz}, {Heinisch}, {Koenders}, {Stoll},
  {Tsurutani}, {Vallat}, {Volwerk}, and {Glassmeier}}}]{Richter2016}
{Richter} I., {Auster} H.-U., {Berghofer} G. et~al. (2016) \emph{{Two-point
  observations of low-frequency waves at 67P/Churyumov-Gerasimenko during the
  descent of PHILAE: comparison of RPCMAG and ROMAP}}, \emph{Annales
  Geophysicae}, \emph{34}, 609--622.

\bibitem[{\emph{{Richter} et~al.}(2015)\emph{{Richter}, {Koenders}, {Auster},
  {Fr{\"u}hauff}, {G{\"o}tz}, {Heinisch}, {Perschke}, {Motschmann}, {Stoll},
  {Altwegg}, {Burch}, {Carr}, {Cupido}, {Eriksson}, {Henri}, {Goldstein},
  {Lebreton}, {Mokashi}, {Nemeth}, {Nilsson}, {Rubin}, {Szeg{\"o}},
  {Tsurutani}, {Vallat}, {Volwerk}, and {Glassmeier}}}]{Richter2015}
{Richter} I., {Koenders} C., {Auster} H.-U. et~al. (2015) \emph{{Observation of
  a new type of low-frequency waves at comet 67P/Churyumov-Gerasimenko}},
  \emph{Annales Geophysicae}, \emph{33}, 1031--1036.

\bibitem[{\emph{Riedler et~al.}(1986)\emph{Riedler, Schwingenschuh, Yeroshenko,
  Styashkin, and Russell}}]{Riedler1986}
Riedler W., Schwingenschuh K., Yeroshenko Y.~E. et~al. (1986) \emph{Magnetic
  field observations in comet {Halley's} coma}, \emph{Nature}, \emph{321},
  288--289.

\bibitem[{\emph{Rubin et~al.}(2014)\emph{Rubin, Fougere, Altwegg, Combi,
  Le~Roy, Tenishev, and Thomas}}]{Rubin2014}
Rubin M., Fougere N., Altwegg K. et~al. (2014) \emph{Mass transport around
  comets and its impact on the seasonal differences in water production rates},
  \emph{Astrophysical Journal}, \emph{788}, 168.

\bibitem[{\emph{{Saito} et~al.}(1986)\emph{{Saito}, {Saito}, {Aoki}, and
  {Yumoto}}}]{sait86}
{Saito} K., {Saito} T., {Aoki} T. et~al. (1986) in \emph{ESLAB Symposium on the
  Exploration of Halley's Comet} (B.~{Battrick}, E.~J. {Rolfe}, and
  R.~{Reinhard}, eds.), vol. 250 of \emph{ESA Special Publication}, p. 155.

\bibitem[{\emph{Saito et~al.}(1987)\emph{Saito, Yumoto, Hirao, Minami, Saito,
  and Smith}}]{sait87}
Saito T., Yumoto K., Hirao K. et~al. (1987) \emph{Structure and dynamics of the
  plasma tail of comet {P/Halley. I} - knot event on december 31, 1985},
  \emph{Astron. Astrophys.}, \emph{187}, 209 -- 214.

\bibitem[{\emph{Sauer et~al.}(1995)\emph{Sauer, Bogdanov, and
  Baumg{\"a}rtel}}]{Sauer1995}
Sauer K., Bogdanov A., and Baumg{\"a}rtel K. (1995) \emph{The protonopause—an
  ion composition boundary in the magnetosheath of comets, venus and mars},
  \emph{Advances in Space Research}, \emph{16}, 153--158.

\bibitem[{\emph{Schmid et~al.}(2014)\emph{Schmid, Volwerk, Plaschke,
  {V\"or\"os}, Zhang, Baumjohann, and Narita}}]{Schmid2014}
Schmid D., Volwerk M., Plaschke F. et~al. (2014) \emph{Mirror mode structures
  near venus and comet p/halley}, \emph{Ann. Geophys.}, \emph{32}, 651 -- 657.

\bibitem[{\emph{{Schmidt} and {Wegmann}}(1982)}]{SchmidtWegmann1982}
{Schmidt} H.~U. and {Wegmann} R. (1982) in \emph{IAU Colloq. 61: Comet
  Discoveries, Statistics, and Observational Selection} (L.~L. {Wilkening},
  ed.), p. 538.

\bibitem[{\emph{Schwenn et~al.}(1988)\emph{Schwenn, Ip, Rosenbauer, Balsiger,
  B{\"u}hler, Goldstein, Meier, and Shelley}}]{Schwenn1988}
Schwenn R., Ip W.-H., Rosenbauer H. et~al. (1988) in \emph{Exploration of
  Halley’s Comet}, pp. 160--162, Springer.

\bibitem[{\emph{Silverman and Limor}(2021)}]{silv21}
Silverman S.~M. and Limor E. (2021) \emph{The great comet of 1577: a
  {Palistinian} observation}, \emph{Hist. Geo. Space Sci.}, \emph{12}, 111 --
  114.

\bibitem[{\emph{{Simon Wedlund} et~al.}(2017)\emph{{Simon Wedlund}, {Alho},
  {Gronoff}, {Kallio}, {Gunell}, {Nilsson}, {Lindkvist}, {Behar}, {Stenberg
  Wieser}, and {Miloch}}}]{SimonWedlund2017}
{Simon Wedlund} C., {Alho} M., {Gronoff} G. et~al. (2017) \emph{{Hybrid
  modelling of cometary plasma environments. I. Impact of photoionisation,
  charge exchange, and electron ionisation on bow shock and cometopause at
  67P/Churyumov-Gerasimenko}}, \emph{\aap}, \emph{604}, A73.

\bibitem[{\emph{{Simon Wedlund} et~al.}(2019{\natexlab{a}})\emph{{Simon
  Wedlund}, {Behar}, {Nilsson}, {Alho}, {Kallio}, {Gunell}, {Bodewits},
  {Heritier}, {Galand}, {Beth}, {Rubin}, {Altwegg}, {Volwerk}, {Gronoff}, and
  {Hoekstra}}}]{SimonWedlund2019c}
{Simon Wedlund} C., {Behar} E., {Nilsson} H. et~al. (2019{\natexlab{a}})
  \emph{{Solar wind charge exchange in cometary atmospheres {III}. {R}esults
  from the {R}osetta mission to comet {67P/C}huryumov-{G}erasimenko}},
  \emph{\aap}.

\bibitem[{\emph{{Simon Wedlund} et~al.}(2019{\natexlab{b}})\emph{{Simon
  Wedlund}, {Bodewits}, {Alho}, {Hoekstra}, {Behar}, {Gronoff}, {Gunell},
  {Nilsson}, {Kallio}, and {Beth}}}]{SimonWedlund2019a}
{Simon Wedlund} C., {Bodewits} D., {Alho} M. et~al. (2019{\natexlab{b}})
  \emph{{Solar wind charge exchange in cometary atmospheres. {I}.
  {C}harge-changing and ionization cross sections for He and H particles in
  {H}$_{2}${O}}}, \emph{\aap}.

\bibitem[{\emph{Sishtla et~al.}(2019)\emph{Sishtla, Divin, Deca, Olshevsky, and
  Markidis}}]{Sishtla2019}
Sishtla C.~P., Divin A., Deca J. et~al. (2019) \emph{Electron trapping in the
  coma of a weakly outgassing comet}, \emph{Physics of Plasmas}, \emph{26},
  102904.

\bibitem[{\emph{Slavin et~al.}(1986{\natexlab{a}})\emph{Slavin, Golberg, Smith,
  {McComas}, Bame, Strauss, and Spinrad}}]{Slavin1986b}
Slavin J.~A., Golberg B.~A., Smith E.~J. et~al. (1986{\natexlab{a}}) \emph{The
  structure of a cometary type {I} tail - {Ground-based and ICE} observations
  of {P/Giacobini-Zinner}}, \emph{Geophys. Res. Lett.}, \emph{13}, 1085--1088.

\bibitem[{\emph{Slavin et~al.}(1986{\natexlab{b}})\emph{Slavin, Smith,
  Tsurutani, Sicsoe, Jones, and Mendis}}]{Slavin1986a}
Slavin J.~A., Smith E.~J., Tsurutani B.~T. et~al. (1986{\natexlab{b}})
  \emph{{Giacobini-Zinner} magnetotail - {ICE} magnetic field observations},
  \emph{Geophys. Res. Lett.}, \emph{13}, 283--286.

\bibitem[{\emph{Smith et~al.}(1986)\emph{Smith, Tsurutani, Slavin, Jones,
  Siscoe, and Mendis}}]{Smith1986}
Smith E.~J., Tsurutani B.~T., Slavin J.~A. et~al. (1986) \emph{International
  cometary explorer encounter with {G}iacobini-{Z}inner: Magnetic field
  observations}, \emph{Science}, \emph{232}, 382--385.

\bibitem[{\emph{Snodgrass and Jones}(2019)}]{Snodgrass2019}
Snodgrass C. and Jones G.~H. (2019) \emph{The {E}uropean {S}pace {A}gency's
  {C}omet {I}nterceptor lies in wait}, \emph{Nature Communications}, \emph{10},
  5418.

\bibitem[{\emph{{Stenberg Wieser} et~al.}(2017)\emph{{Stenberg Wieser},
  {Odelstad}, {Wieser}, {Nilsson}, {Goetz}, {Karlsson}, {Andr{\'e}}, {Kalla},
  {Eriksson}, {Nicolaou}, {Simon Wedlund}, {Richter}, and
  {Gunell}}}]{StenbergWieser2017}
{Stenberg Wieser} G., {Odelstad} E., {Wieser} M. et~al. (2017)
  \emph{{Investigating short-time-scale variations in cometary ions around
  comet 67P}}, \emph{\mnras}, \emph{469}, S522--S534.

\bibitem[{\emph{Sweet}(1956)}]{swee56}
Sweet P.~A. (1956) in \emph{Proceedings of the International Astronomican Union
  Symposium on Electromagnetic Phenomena in Cosmical Physics}, p. 123, Kluwer,
  Dordrecht.

\bibitem[{\emph{{Szeg{\"o}} et~al.}(2000)\emph{{Szeg{\"o}}, {Glassmeier},
  {Bingham}, {Bogdanov}, {Fischer}, {Haerendel}, {Brinca}, {Cravens},
  {Dubinin}, {Sauer}, {Fisk}, {Gombosi}, {Schwadron}, {Isenberg}, {Lee},
  {Mazelle}, {M{\"o}bius}, {Motschmann}, {Shapiro}, {Tsurutani}, and
  {Zank}}}]{Szego2000}
{Szeg{\"o}} K., {Glassmeier} K.-H., {Bingham} R. et~al. (2000) \emph{{Physics
  of Mass Loaded Plasmas}}, \emph{Space Science Reviews}, \emph{94}, 429--671.

\bibitem[{\emph{{Tao} et~al.}(2005)\emph{{Tao}, {Kataoka}, {Fukunishi},
  {Takahashi}, and {Yokoyama}}}]{Tao2005}
{Tao} C., {Kataoka} R., {Fukunishi} H. et~al. (2005) \emph{{Magnetic field
  variations in the Jovian magnetotail induced by solar wind dynamic pressure
  enhancements}}, \emph{Journal of Geophysical Research (Space Physics)},
  \emph{110}, A11208.

\bibitem[{\emph{Perez-de Tejada}(1989)}]{Perez1989}
Perez-de Tejada H. (1989) \emph{Viscous flow interpretation of comet halley's
  mystery transition}, \emph{Journal of Geophysical Research: Space Physics},
  \emph{94}, 10131--10136.

\bibitem[{\emph{{Thomsen} et~al.}(1987)\emph{{Thomsen}, {Feldman}, {Wilken},
  {Jockers}, and {Stuedemann}}}]{Thomsen1987}
{Thomsen} M.~F., {Feldman} W.~C., {Wilken} B. et~al. (1987) \emph{{In-situ
  observations of a bi-modal ion distribution in the outer coma of comet
  P/Halley}}, \emph{\aap}, \emph{187}, 141--148.

\bibitem[{\emph{{Timar} et~al.}(2019)\emph{{Timar}, {Nemeth}, {Szego},
  {D{\'o}sa}, {Opitz}, and {Madanian}}}]{Timar2019}
{Timar} A., {Nemeth} Z., {Szego} K. et~al. (2019) \emph{{Estimating the solar
  wind pressure at comet 67P from Rosetta magnetic field measurements}},
  \emph{Journal of Space Weather and Space Climate}, \emph{9}, A3.

\bibitem[{\emph{Verigin et~al.}(1987)\emph{Verigin, Axford, Gringauz, and
  Richter}}]{veri87}
Verigin M.~I., Axford W.~I., Gringauz K.~I. et~al. (1987) \emph{Acceleration of
  cometary plasma in the vicinity of comet {Halley} associated with an
  interplanetary magnetic field polarity change}, \emph{Geophys. Res. Lett.},
  \emph{14}, 987 -- 990.

\bibitem[{\emph{{Vigren} and {Eriksson}}(2019)}]{Vigren2019}
{Vigren} E. and {Eriksson} A.~I. (2019) \emph{{On the ion-neutral coupling in
  cometary comae}}, \emph{\mnras}, \emph{482}, 1937--1941.

\bibitem[{\emph{{Vigren} and {Galand}}(2013)}]{VigrenGaland2013}
{Vigren} E. and {Galand} M. (2013) \emph{{Predictions of Ion Production Rates
  and Ion Number Densities within the Diamagnetic Cavity of Comet
  67P/Churyumov-Gerasimenko at Perihelion}}, \emph{\apj}, \emph{772}, 33.

\bibitem[{\emph{{Volwerk} et~al.}(2014)\emph{{Volwerk}, {Glassmeier}, {Delva},
  {Schmid}, {Koenders}, {Richter}, and {Szeg{\"o}}}}]{Volwerk2014}
{Volwerk} M., {Glassmeier} K.~H., {Delva} M. et~al. (2014) \emph{{A comparison
  between VEGA 1, 2 and Giotto flybys of comet 1P/Halley: implications for
  Rosetta}}, \emph{Annales Geophysicae}, \emph{32}, 1441--1453.

\bibitem[{\emph{{Volwerk} et~al.}(2019)\emph{{Volwerk}, {Goetz}, {Behar},
  {Delva}, {Edberg}, {Eriksson}, {Henri}, {Llera}, {Nilsson}, {Richter},
  {Stenberg Wieser}, and {Glassmeier}}}]{Volwerk2019}
{Volwerk} M., {Goetz} C., {Behar} E. et~al. (2019) \emph{{Dynamic field line
  draping at comet 67P/Churyumov-Gerasimenko during the Rosetta dayside
  excursion}}, \emph{\aap}, \emph{630}, A44.

\bibitem[{\emph{{Volwerk} et~al.}(2020)\emph{{Volwerk}, {Goetz}, {Plaschke},
  {Karlsson}, {Heyner}, and {Anderson}}}]{Volwerk2020}
{Volwerk} M., {Goetz} C., {Plaschke} F. et~al. (2020) \emph{{On the magnetic
  characteristics of magnetic holes in the solar wind between Mercury and
  Venus}}, \emph{Annales Geophysicae}, \emph{38}, 51--60.

\bibitem[{\emph{Volwerk et~al.}(2018)\emph{Volwerk, Goetz, Richter, Delva,
  Ostaszewski, Schwingenschuh, and Glassmeier}}]{Volwerk2018a}
Volwerk M., Goetz C., Richter I. et~al. (2018) \emph{A tail like no other:
  {RPC-MAG's} view of {Rosetta's} tail excursion at comet
  {67P/Churyumov-Gerasimenko}}, \emph{Astron. Astrophys.}, \emph{614}, A10.

\bibitem[{\emph{{Volwerk} et~al.}(2017)\emph{{Volwerk}, {Jones}, {Broiles},
  {Burch}, {Carr}, {Coates}, {Cupido}, {Delva}, {Edberg}, {Eriksson}, {Goetz},
  {Goldstein}, {Henri}, {Madanian}, {Nilsson}, {Richter}, {Schwingenschuh},
  {Stenberg Wieser}, and {Glassmeier}}}]{Volwerk2017}
{Volwerk} M., {Jones} G.~H., {Broiles} T. et~al. (2017) \emph{{Current sheets
  in comet 67P/Churyumov-Gerasimenko's coma}}, \emph{Journal of Geophysical
  Research (Space Physics)}, \emph{122}, 3308--3321.

\bibitem[{\emph{{Volwerk} et~al.}(2021)\emph{{Volwerk}, {Mautner}, {Wedlund},
  {Goetz}, {Plaschke}, {Karlsson}, {Schmid}, {Rojas-Castillo}, {Roberts}, and
  {Varsani}}}]{Volwerk2021}
{Volwerk} M., {Mautner} D., {Wedlund} C.~S. et~al. (2021) \emph{{Statistical
  study of linear magnetic hole structures near Earth}}, \emph{Annales
  Geophysicae}, \emph{39}, 239--253.

\bibitem[{\emph{{Volwerk} et~al.}(2016)\emph{{Volwerk}, {Richter}, {Tsurutani},
  {G{\"o}tz}, {Altwegg}, {Broiles}, {Burch}, {Carr}, {Cupido}, {Delva},
  {D{\'o}sa}, {Edberg}, {Eriksson}, {Henri}, {Koenders}, {Lebreton}, {Mandt},
  {Nilsson}, {Opitz}, {Rubin}, {Schwingenschuh}, {Stenberg Wieser},
  {Szeg{\"o}}, {Vallat}, {Vallieres}, and {Glassmeier}}}]{Volwerk2016}
{Volwerk} M., {Richter} I., {Tsurutani} B. et~al. (2016) \emph{{Mass-loading,
  pile-up, and mirror-mode waves at comet 67P/Churyumov-Gerasimenko}},
  \emph{Annales Geophysicae}, \emph{34}, 1--15.

\bibitem[{\emph{Vourlidas et~al.}(2007)\emph{Vourlidas, Davis, Eyles, Crothers,
  Harrison, Howard, Moses, and Socker}}]{Vourlidas2007}
Vourlidas A., Davis C.~J., Eyles C.~J. et~al. (2007) \emph{First direct
  observation of the interaction between a comet and a coronal mass ejection
  leading to a complete plasma tail disconnection}, \emph{\apjl}, \emph{668},
  L79--L82.

\bibitem[{\emph{{Wattieaux} et~al.}(2020)\emph{{Wattieaux}, {Henri}, {Gilet},
  {Valli{\`e}res}, and {Deca}}}]{Wattieaux2020}
{Wattieaux} G., {Henri} P., {Gilet} N. et~al. (2020) \emph{{Plasma
  characterization at comet 67P between 2 and 4 AU from the Sun with the
  RPC-MIP instrument}}, \emph{\aap}, \emph{638}, A124.

\bibitem[{\emph{{Wellbrock} et~al.}(2018)\emph{{Wellbrock}, {Jones}, {Coates},
  {Simon Wedlund}, {Goetz}, {Dresing}, {Nordheim}, {Mandt}, {Hajra}, {Myllys},
  {Henri}, and {Nilsson}}}]{Wellbrock_etal_EPSC_2018}
{Wellbrock} A., {Jones} G., {Coates} A. et~al. (2018) in \emph{European
  Planetary Science Congress}, pp. EPSC2018--964.

\bibitem[{\emph{Williamson et~al.}(2020)\emph{Williamson, Nilsson,
  Stenberg~Wieser, Eriksson, Richter, and Goetz}}]{Williamson2020}
Williamson H.~N., Nilsson H., Stenberg~Wieser G. et~al. (2020) \emph{Momentum
  and pressure balance of a comet ionosphere}, \emph{Geophysical Research
  Letters}, \emph{47}, e2020GL088666, e2020GL088666 10.1029/2020GL088666.

\bibitem[{\emph{{Winterhalter} et~al.}(2000)\emph{{Winterhalter}, {Smith},
  {Neugebauer}, {Goldstein}, and {Tsurutani}}}]{Winterhalter2000}
{Winterhalter} D., {Smith} E.~J., {Neugebauer} M. et~al. (2000) \emph{{The
  latitudinal distribution of solar wind magnetic holes}}, \emph{\grl},
  \emph{27}, 1615--1618.

\bibitem[{\emph{Witasse et~al.}(2017)\emph{Witasse, Sánchez-Cano, Mays,
  Kajdič, Opgenoorth, Elliott, Richardson, Zouganelis, Zender,
  Wimmer-Schweingruber, Turc, Taylor, Roussos, Rouillard, Richter, Richardson,
  Ramstad, Provan, Posner, Plaut, Odstrcil, Nilsson, Niemenen, Milan, Mandt,
  Lohf, Lester, Lebreton, Kuulkers, Krupp, Koenders, James, Intzekara,
  Holmstrom, Hassler, Hall, Guo, Goldstein, Goetz, Glassmeier, Génot, Evans,
  Espley, Edberg, Dougherty, Cowley, Burch, Behar, Barabash, Andrews, and
  Altobelli}}]{Witasse2017}
Witasse O., Sánchez-Cano B., Mays M.~L. et~al. (2017) \emph{{Interplanetary
  coronal mass ejection observed at STEREO-A, Mars, comet
  67P/Churyumov-Gerasimenko, Saturn, and New Horizons en route to Pluto:
  Comparison of its Forbush decreases at 1.4, 3.1, and 9.9 AU}}, \emph{Journal
  of Geophysical Research (Space Physics)}, pp. n/a--n/a, 2017JA023884.

\bibitem[{\emph{Wolff et~al.}(1985)\emph{Wolff, Siscoe, Sibeck, and
  Neugebauer}}]{wolf85}
Wolff R.~S., Siscoe G.~L., Sibeck D.~G. et~al. (1985) \emph{Cometary rays:
  magnetically channeled outflow}, \emph{Geophys. Res. Lett.}, \emph{12}, 749
  -- 752.

\bibitem[{\emph{{Womack} et~al.}(1997)\emph{{Womack}, {Homich}, {Festou},
  {Mangum}, {Uhl}, and {Stern}}}]{Womack1997}
{Womack} M., {Homich} A., {Festou} M.~C. et~al. (1997) \emph{{Maps of hco+
  Emission in c/1995 o1 (Hale-Bopp)}}, \emph{Earth Moon and Planets},
  \emph{77}, 259--264.

\bibitem[{\emph{Wurm}(1943)}]{wurm43}
Wurm K. (1943) \emph{Die {Natur} der {Kometen}}, \emph{Mitt. Hamburger
  Sternw.}, \emph{8}, Nr. 51.

\bibitem[{\emph{Wurm}(1961)}]{wurm61}
Wurm K. (1961) \emph{Structure and development of the gas tails of comets},
  \emph{Astron. J.}, \emph{66}, 362 -- 367.

\bibitem[{\emph{Wurm}(1963)}]{wurm63}
Wurm K. (1963) in \emph{The moon, meteorites and comets} (G.~P. Kuiper and
  B.~Middlehurts, eds.), pp. 573 -- 617, Univ. Chicago Press, Chicago.

\bibitem[{\emph{Wurz et~al.}(2015)\emph{Wurz, Rubin, Altwegg, Balsiger,
  Berthelier, Bieler, Calmonte, De{ }Keyser, Fiethe, Fuselier, Galli, Gasc,
  Gombosi, J\"ackel, Le~Roy, Mall, R\`eme, Tenishev, and Tzou}}]{Wurz2015}
Wurz P., Rubin M., Altwegg K. et~al. (2015) \emph{Solar wind sputtering of dust
  on the surface of 67p/churyumov-gerasimenko}, \emph{A\&A}, \emph{583}, A22.

\bibitem[{\emph{Yagi et~al.}(2015)\emph{Yagi, Koda, Terai, Fujiwara, and
  Watanabe}}]{yagi15}
Yagi M., Koda J., Terai T. et~al. (2015) \emph{Initial speed of knots in the
  plasma tail of {C/2013 R1(Lovejoy)}}, \emph{Astron. J.}, \emph{149}, 97.

\bibitem[{\emph{{Yang} et~al.}(2016)\emph{{Yang}, {Paulsson}, {Wedlund},
  {Odelstad}, {Edberg}, {Koenders}, {Eriksson}, and {Miloch}}}]{Yang2016}
{Yang} L., {Paulsson} J.~J.~P., {Wedlund} C.~S. et~al. (2016)
  \emph{{Observations of high-plasma density region in the inner coma of
  67P/\linebreak[0]{}Chury\-umov-Gerasimenko during early activity}},
  \emph{\mnras}, \emph{462}, S33--S44.

\bibitem[{\emph{Young et~al.}(2004)\emph{Young, Crary, Nordholt, Bagenal,
  Boice, Burch, Eviatar, Goldstein, Hanley, Lawrence et~al.}}]{Young2004}
Young D., Crary F., Nordholt J. et~al. (2004) \emph{Solar wind interactions
  with comet 19p/borrelly}, \emph{Icarus}, \emph{167}, 80--88.

\bibitem[{\emph{{Zhao} et~al.}(2017)\emph{{Zhao}, {Liu}, {Hu}, and
  {Wang}}}]{Zhao_etal_ApJ_2017}
{Zhao} X., {Liu} Y.~D., {Hu} H. et~al. (2017) \emph{{Propagation
  Characteristics of Two Coronal Mass Ejections from the Sun Far into
  Interplanetary Space}}, \emph{\apj}, \emph{837}, 4.

\bibitem[{\emph{Zolotova et~al.}(2018)\emph{Zolotova, Sizonenko, Mokhmyanin,
  and Veselovsky}}]{zolo18}
Zolotova N., Sizonenko Y., Mokhmyanin M. et~al. (2018) \emph{Indirect solar
  wind measurements using archival cometary tail observations}, \emph{Sol.
  Phys.}, \emph{293}, 85.

\end{thebibliography}

\end{document}